\documentclass[aps,prx,twocolumn,notitlepage,superscriptaddress,nofootinbib]{revtex4-1}
\usepackage{amsmath} % AMS Math Package
\usepackage{amsthm} % Theorem Formatting
\usepackage{bm} % Theorem Formatting
\usepackage{amssymb}	% Math symbols such as \mathbb
\usepackage{graphicx} % Allows for eps images
\usepackage{subfigure}
\usepackage[table]{xcolor}
\usepackage{comment}
\usepackage{accents}
\usepackage{makecell}
\usepackage{dsfont}
\usepackage{booktabs}
\usepackage{bbm}
\usepackage{hyperref}
\usepackage{mathtools}
\usepackage{wasysym}

\hypersetup{
	bookmarks=false,         % show bookmarks bar?
	unicode=false,          % non-Latin characters in Acrobat?s bookmarks
	pdftoolbar=false,        % show Acrobat?s toolbar?
	pdfmenubar=true,        % show Acrobat?s menu?
	pdffitwindow=false,     % window fit to page when opened
	pdfstartview={FitH},    % fits the width of the page to the window
	pdftitle={},    % title
	pdfauthor={Authors},     % author
	pdfsubject={},   % subject of the document
	pdfcreator={},   % creator of the document
	pdfproducer={}, % producer of the document
	pdfnewwindow=true,      % links in new window
	colorlinks=true,       % false: boxed links; true: colored links
	linkcolor=blue,          % color of internal links (change box color with linkbordercolor)
	citecolor=blue,        % color of links to bibliography
	filecolor=magenta,      % color of file links
	urlcolor=blue           % color of external links
}

\usepackage{xcolor}

\renewcommand{\theequation}{\arabic{equation}}

\newcommand{\bea}{\begin{eqnarray}}  
\newcommand{\eea}{\end{eqnarray}}
\newcommand{\ben}{\begin{enumerate}}
\newcommand{\een}{\end{enumerate}}
\newcommand{\be}{\begin{equation}}
\newcommand{\ee}{\end{equation}}

\newcommand{\BigO}{\mathrm{O}}
\newcommand{\smallo}{\mathrm{o}}

\usepackage{xparse}

\newcommand{\ket}[1]{\left| #1 \right>} % for Dirac bras
\newcommand{\bra}[1]{\left< #1 \right|} % for Dirac kets
\newcommand{\braket}[2]{\left< #1 \vphantom{#2} \right|
\left. #2 \vphantom{#1} \right>} % for Dirac brackets
\newcommand{\abs}[1]{\ensuremath{\vert #1 \vert}}

\DeclareMathOperator{\tr}{tr}
\DeclareMathOperator{\Tr}{Tr}

\renewcommand{\theequation}{\arabic{equation}}

\renewcommand{\thesection}{\Roman{section}}

\newcommand{\rd}{\mathrm{d}}

\makeatletter
\newsavebox{\@brx}
\newcommand{\llangle}[1][]{\savebox{\@brx}{\(\m@th{#1\langle}\)}%
  \mathopen{\copy\@brx\mkern2mu\kern-0.9\wd\@brx\usebox{\@brx}}}
\newcommand{\rrangle}[1][]{\savebox{\@brx}{\(\m@th{#1\rangle}\)}%
  \mathclose{\copy\@brx\mkern2mu\kern-0.9\wd\@brx\usebox{\@brx}}}
\makeatother

\begin{document}
\title{
Theory of the phase transition in random unitary circuits with measurements
}
\date{\today}

\author{Yimu Bao}
\thanks{YB and SC contributed equally to this work.}
\affiliation{Department of Physics, University of California, Berkeley, California 94720, USA}
\author{Soonwon Choi}
\thanks{YB and SC contributed equally to this work.}
\affiliation{Department of Physics, University of California, Berkeley, California 94720, USA}
\author{Ehud Altman}
\affiliation{Department of Physics, University of California, Berkeley, California 94720, USA}
\affiliation{Materials Science Division, Lawrence Berkeley National Laboratory, Berkeley, CA 94720, USA}

\begin{abstract}
We present a theory of the entanglement transition tuned by measurement strength in qudit chains evolved by random unitary circuits and subject to either weak or random projective measurements. The transition can be understood as a nonanalytic change in the amount of information extracted by the measurements about the initial state of the system, quantified by the Fisher information. To compute the von~Neumann entanglement entropy $S$ and the Fisher information $\mathcal{F}$, we apply a replica method based on a sequence of quantities  $\tilde{S}^{(n)}$ and $\mathcal{F}^{(n)}$ that depend on the $n$-th moments of density matrices and reduce to $S$ and $\mathcal{F}$ in the limit $n\to 1$.
These quantities with $n\ge 2$ are mapped to free energies of a classical spin model with $n!$ internal states in two dimensions with specific boundary conditions. In particular, $\tilde{S}^{(n)}$ is the excess free energy of a domain wall terminating on the top boundary, and $\mathcal{F}^{(n)}$ is related to the magnetization on the bottom boundary. Phase transitions occur as the spin models undergo ordering transitions in the bulk. Taking the limit of large local Hilbert space dimension $q$ followed by the replica limit $n\to 1$, we obtain the critical measurement probability $p_c=1/2$ and identify the transition as a bond percolation in the 2D square lattice in this limit.  Finally, we show there is no phase transition if the measurements are allowed in an arbitrary nonlocal basis, thereby highlighting the relation between the phase transition and information scrambling. We establish an explicit connection between the entanglement phase transition and the purification dynamics of a mixed state evolution and discuss implications of our results to experimental observations of the transition and simulability of quantum dynamics.
\end{abstract}

\maketitle

%\tableofcontents

\section{Introduction}
Quantum states with high degree of entanglement cannot be efficiently emulated by classical computers and may serve as a resource for quantum science~\cite{nielsen2002quantum}.
A natural way to generate such states is to evolve a quantum system under generic unitary dynamics, which gives rise to extensive entanglement entropy for any subsystems~\cite{nahum2017quantum,von2018operator,nahum2018operator,khemani2018operator,rakovszky2018diffusive}.
However, in realistic situations, the dynamics is also subject to nonunitary evolution owing to local measurements of system degrees of freedom (qudits), performed either by an observer or spuriously by environment. 
Local measurements generally destroy entanglement within a quantum system as they project the measured qudits onto a definite state, disentangled from the rest.
This naturally raises a question: how robust is entanglement generation by unitary evolution against measurements?

Recent works presented numerical results as well as general arguments suggesting that volume-law entanglement in steady states may persist if the rate of measurements is sufficiently small, while frequent enough measurements above a critical rate drive a transition to steady states with area-law entanglement entropy~\cite{li2018quantum,skinner2018measurement,li2019measurement,choi2019quantum,chan2018weak,szyniszewski2019entanglement,gullans2019dynamical}.
Despite this progress, an analytic framework to describe the phase transition is still lacking.
Such a framework is needed in order to understand universal properties of the two distinct dynamical phases and the nature of the transition between them.
Does the phase transition belong to a known universality class?
How does the transition affects physical properties that are more accessible and relevant to experiments than entanglement entropy such as probability distributions of certain measurement outcomes? 
These questions are particularly important in the context of current experimental efforts to demonstrate computational advantages of quantum devices over classical computers. 
In such efforts, quantum systems are interrogated both by a controlled measurement device and  by an uncontrolled environment,
and the entanglement transition may imply a change in the complexity of the quantum dynamics they attempt to simulate.

In this paper, we address the above questions by developing a theoretical framework, in which an ensemble of quantum circuits with measurements are exactly mapped to a series of classical statistical mechanics models.
Our construction generalizes previous works on emergent classical spin models in random tensor networks and random unitary circuits (RUCs)~\cite{hayden2016holographic,qi2017holographic,vasseur2018entanglement,nahum2018operator,zhou2019emergent,hunter2019unitary}.
In particular, mapping to a statistical mechanics model has allowed to analyze entanglement phase transitions in random-tensor networks, which as we will see, are similar to the measurement induced entanglement transition in quantum circuits \cite{vasseur2018entanglement}.
In the framework of random unitary circuits, such mappings allowed understanding the universal features of entanglement growth and operator spreading in a generic unitary time evolution~\cite{nahum2017quantum,nahum2018operator,von2018operator,khemani2018operator,rakovszky2018diffusive}. 
The essential difficulties in generalizing the existing approach to include measurements are related to the nonlinearity of the process; one needs to normalize a many-body wavefunction after each projective measurement.
Furthermore, it is challenging to consider a typical behavior averaged over different measurement outcomes, since their probabilities are not independent of each other and depend on the state history.

\begin{figure}
	\includegraphics[width=0.48\textwidth]{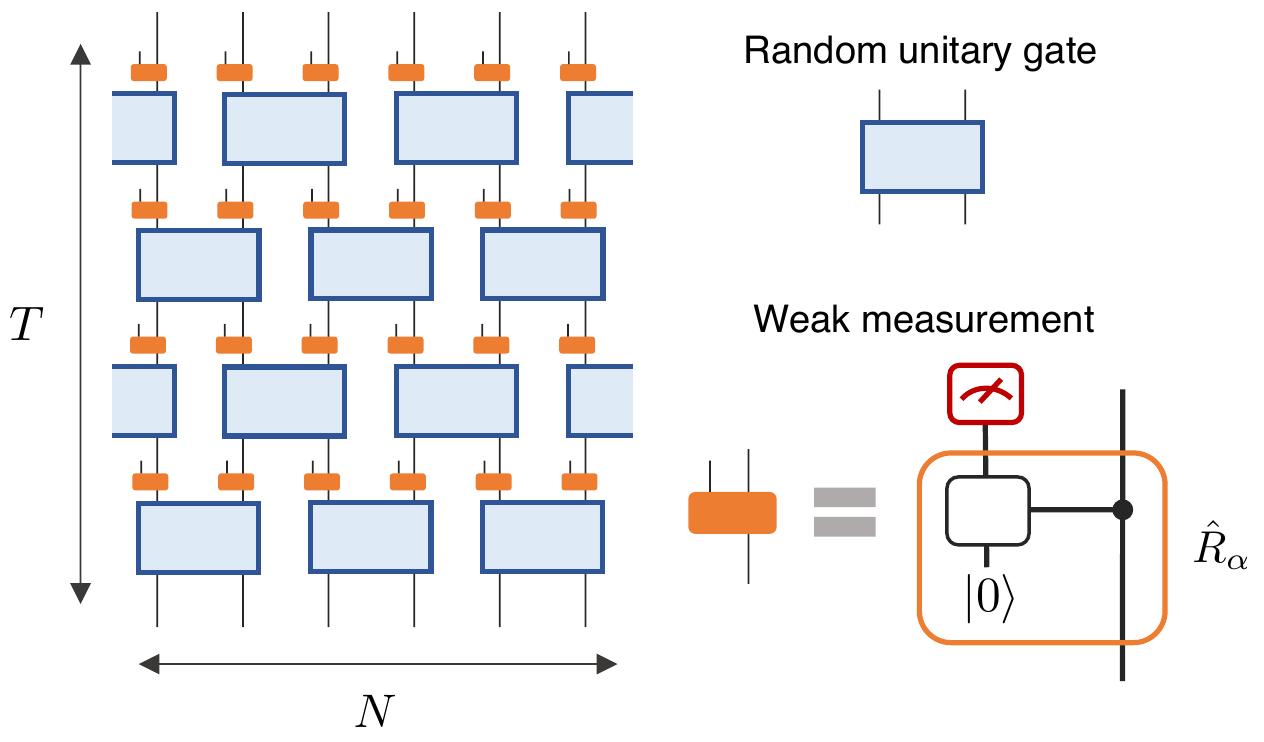}
	\caption{Random unitary circuit with weak measurements. After each layer of unitary gates (blue boxes), a weak measurement (orange box) is performed on every system qudit. The red box indicates a strong projective measurement of an ancilla qudit performed after an entangling unitary $\hat{R}_\alpha$.}
	\label{fig:weak_meas}
\end{figure}

In order to address these challenges, we find that it is conceptually simpler to consider quantum circuits in which every qudit is subject to weak measurement at every time step, instead of ones with a set of randomly distributed strong projective measurements.
The weak measurements are implemented by coupling the system qudits to a set of ancilla qudits that are subsequently measured by strong projective measurements as illustrated in Fig.~\ref{fig:weak_meas}.
The strength of measurement is tuned by the strength of the coupling between the system and the ancilla qudits.
Later, we show that the case of random projective measurements can be implemented as a special case of this circuit.
Hence, our results pertain to random projective measurements studied in Refs.~\cite{li2018quantum,skinner2018measurement,chan2018weak,li2019measurement,choi2019quantum}, while they also generalize to the case of constant weak measurement.
Our main results are insensitive to this detail.

\section{Overview}
\begin{figure}
    \centering
    \includegraphics[width=0.49\textwidth]{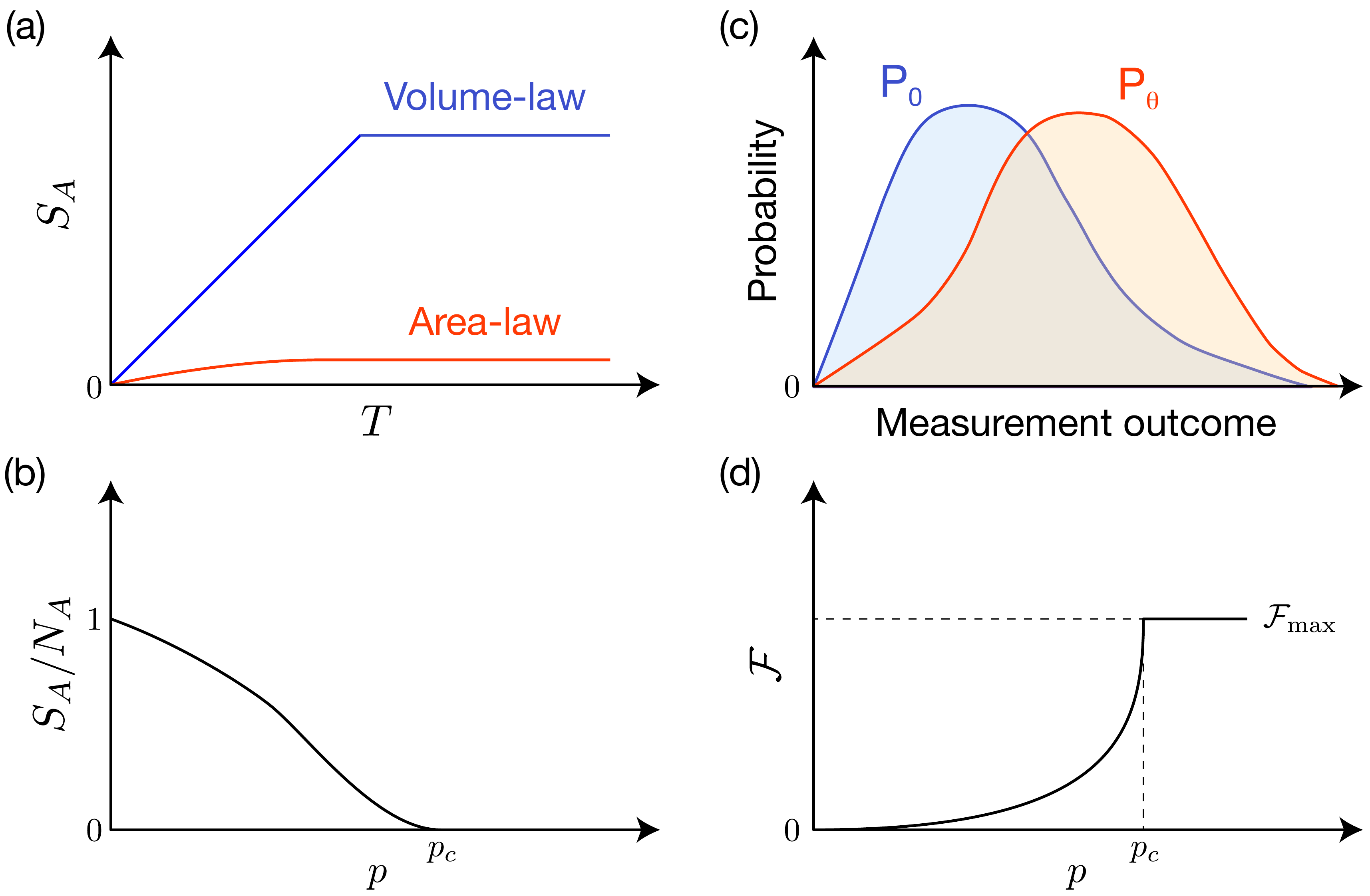}
    \caption{Signatures of the phase transition. 
    (a) Growth and saturation of entanglement entropy $S_A$ of a subsystem $A$ (of size $N_A$).
    (b) Phase transition in the saturation value of $S_A$ from volume- to area-law scaling. (c) Probability distributions $P_0$ and $P_\theta$ of measurement outcomes arising from two distinct initial states. 
    (d) The phase transition is also seen in the Fisher information, quantifying the amount of information that the measurements carry about the initial state of the system. 
    }
    \label{fig:fig1}
\end{figure}

Before proceeding, we provide a brief overview of the main ideas and results of this paper.
First, we present a simple expression of the subsystem entanglement entropy averaged over different measurement outcomes, which is the quantity considered in previous works~\cite{skinner2018measurement,li2018quantum,chan2018weak,li2019measurement,choi2019quantum}.
We show how this quantity can be mapped to the excess free energy originating from a domain wall in a classical spin model,  generalizing the analytic approaches introduced in Refs.~\cite{hayden2016holographic,qi2017holographic,nahum2018operator,vasseur2018entanglement,zhou2019emergent,hunter2019unitary},
and explain the new insights and results obtained from this mapping.

Next, we identify another signature, or an ``order parameter,'' of the same transition, that is more relevant for experiments.
The new quantity offers an alternative interpretation of the phase transition: a sharp change in the amount of information that can be extracted by the measurements about the initial state of the system (see Fig.~\ref{fig:fig1}). 

\subsection{Replica method: conditional entropy}
One of the main quantities considered in previous works is the entanglement entropy of a subsystem $A$, averaged over measurement outcomes and realizations of the RUC:
\be
\overline{\llangle S_A(\mathcal{U})\rrangle}
\equiv
\overline{
\sum_{\bm{i}_M} p_{\bm{i}_M} (\mathcal{U})S\left[\rho_A(\mathcal{U},\bm{i}_M)\right]
}.
\label{eq:cond1}
\ee
Here, the index $\bm{i}_M$ runs over all possible measurement outcomes, $p_{\bm{i}_M}(\mathcal{U})$ is the probability for the measurement outcome $\bm{i}_M$, $S\left[\rho_A(\mathcal{U}, \bm{i}_M)\right]$ is von~Neumann entanglement entropy of $A$ for a given measurement outcome $\bm{i}_M$ in a particular realization $\mathcal{U}$ of RUCs, and the overline denotes averaging over the RUC $\mathcal{U}$.
By using weak measurement formalism, we can rewrite this quantity in a simpler form by extending the Hilbert space to include the subspace $M$ for ancilla qudits.
That is, it exactly equals the conditional entropy:
\begin{align}
\overline{\llangle S_A(\mathcal{U})\rrangle} = \overline{\tilde{S}(A|M)} \equiv  \overline{S[\tilde{\rho}_{AM}]} -\overline{S[\tilde{\rho}_{M}]}, 
\label{eq:cond2}
\end{align}
where $\tilde{\rho}_X$ denotes the reduced density matrix of subsystem $X$ after ancilla qudits are projected onto their diagonal elements, i.e., measured in their computational basis $\ket{\bm{i}_M}$:  $\tilde{\rho}=\sum_{\bm{i}} \ket{\bm{i}_M}\bra{\bm{i}_M} \rho \ket{\bm{i}_M}\bra{\bm{i}_M}  $.

We employ a replica method in order to map the conditional entropy in Eq.~\eqref{eq:cond2} to a classical spin model.
Let us define a series of objects enumerated by an integer $n$,
\begin{align}
\tilde{S}^{(n)}(A|M)\equiv \frac{\log\left(\, \overline{\tr\tilde{\rho}_{AM}^{\,n}}\,\right) -\log \left(\,\overline{ \tr\tilde{\rho}_M^{\,n}}\,\right)}{1-n},
\label{eq:SnA}
\end{align}
from which the von~Neumann conditional entropy $\overline{\tilde{S}(A|M)}$ is recovered in the limit $n\to 1$. It is important to note that $\tilde{S}^{(n)}(A|M)$ is not itself an average conditional (R\'enyi) entropy, but it recovers its meaning only in the replica limit.
Note also that the $n$-th moments of the replicated density matrices are averaged over $\mathcal{U}$ now inside the logarithms.
We will show $\tilde{S}^{(n)}(A|M)$ can be interpreted in terms of free energies of classical spin models in two dimension.
The two dimensions arise from the time, which in our convention flows from bottom to top, and the space dimension on the original qudit chain.
Each classical ``spin'' degree of freedom arises from a random unitary gate and may take as its value any one of the $n!$ elements of the permutation group $\mathcal{P}_n$.
For example, the case $n=2$ corresponds to Ising spins, where an up/down spin maps to a trivial/swap element.
Increasing the measurement strength weakens the couplings between the spins, 
leading to a transition from a ferromagnetic to a paramagnetic phase at a critical measurement strength. 

The generalized entropy $\tilde{S}^{(n)}(A|M)$ corresponds to the difference between free energies that result from different top boundary conditions.
These boundary conditions are arranged such that $\tilde{S}^{(n)}(A|M)$ describes the excess free energy associated with a domain wall that runs in the bulk and connects the two edges of the subsystem $A$ at the top boundary.
The domain wall free energy scales with the length of $A$ in the ferromagnetic phase, which is therefore identified with the volume-law phase, whereas the free energy is of order one in the paramagnetic phase, identified with the area-law phase. We note that entanglement phase transitions in random tensor networks were similarly interpreted in Ref.~\cite{vasseur2018entanglement} as a change of domain wall free energies in an emergent statistical mechanics model. 

The classical spin model description allows infering  crucial information about the transition.
First, we obtain an analytic approximation for the critical measurement strength (or the measurement probability in the case of random projective measurements).
The spin model description greatly simplifies in the limit of a large local Hilbert space dimension $q$ for a qudit, where our model maps to the $n!$-state standard Potts model on the square lattice. On this lattice, we can obtain the critical coupling strength analytically as a function of $n$ and $q$ using the Kramers-Wannier duality.
Taking the replica limit $n\to 1$, we find a critical measurement probability $p_c = 1/2$ in the large-$q$ limit. 
% We note that the value of $p_c$ predicted in this large-$q$ approximation becomes independent of $q$ in the replica limit $n\to 1$, the framework itself  only holds for large $q$.
For a small local Hilbert space dimension  $q=2$, we perform exact numerical simulations for up to $N=30$ qubits and obtain $p_c=0.26\pm 0.02$.

Furthermore, the mapping to an $n!$-state Potts model allows one to infer the
universality class of the entanglement transition in the limit $q\to\infty$. 
Specifically, the partition function of the $Q$-state standard Potts model approaches that of bond percolation in the physical limit $Q\to 1$. This is similar to the prediction made in Ref.~\cite{skinner2018measurement} that the special R\'enyi-$0$ entropy undergoes a percolation transition. However, in our case, note that the bond percolation transition in the von Neumann entanglement entropy is limited to the case of infinite $q$.

\subsection{Alternative signature of the transition}
It is important to find the signatures of the phase transition that are more accessible and relevant for experiments than the conditional entanglement entropy.
In particular, measuring the conditional entanglement entropy is extremely challenging, as it requires  postselecting a particular combination of measurement outcomes with occurrence probability exponentially small in system sizes (in both space and time).
Furthermore, for each measurement outcome, estimating entanglement entropy for the quantum state  necessitates additional exponentially many repetitions of an experiment~\cite{o2015quantum,li2018quantum_query_complexity}.
Consequently, the operational meaning of the conditional entanglement entropy is \emph{a priori} not clear from an experimental point of view.
A quantity with transparent physical meaning is therefore needed.

The quantum circuit we consider involves measurements. Hence, it is natural to seek the signatures of the transition in the probability distribution (or histogram) of measurement outcomes.
Indeed, we show that there is a sharp transition in the amount of information these measurements contain on the initial state of the system, quantified by the Fisher information.
The Fisher information can be derived from a closely related quantity, the Kullback-Leibler (KL) divergence (also known as the relative entropy).
Given the two distributions $P_0(x)$ and $P_\theta(x)$ of measurement outcomes obtained for two close initial states of the system, the KL divergence averaged over the RUC is
\be
D_{\textrm{KL}}(P_0||P_\theta)=\overline{\sum_x P_0(x)\log\left({P_0(x)\over P_\theta(x)}\right)}.
\ee
Here, $\theta$ parameterizes the distance between two initial states.
The variable $x$ runs over all possible ancilla measurement outcomes.
The Fisher information is given by the second order derivative:
\begin{align}
    \mathcal{F} \equiv \left. \partial^2_\theta D_{\textrm{KL}}(P_0||P_\theta) \right|_{\theta = 0}.
\end{align}
Note that the first order derivative vanishes due to the positivity of the KL divergence and the fact that it vanishes for $\theta=0$.

Using the replica method, we recover the KL divergence and the Fisher information in the limit $n\to 1$ from a series of auxilliary functions:
\begin{align}
D^{(n)}(P_0||P_\theta) &\equiv \frac{\log\overline{\tr [\tilde{\rho}_{M,0}\,\tilde{\rho}^{\, n-1}_{M,\theta}]} - \log\overline{\tr[\tilde{\rho}^{\,n}_{M,0}]}}{1-n},\\
\mathcal{F}^{(n)} &\equiv \left. \partial^2_\theta D^{(n)}(P_0||P_\theta) \right|_{\theta = 0}.
\end{align}
Note that $\tilde{\rho}_{M,\theta}$ denotes the diagonal density matrix of ancilla qudits composed of the probability distribution $P_\theta(x)$. 
We show that $D^{(n)}$ maps to the excess free energy associated with applying a field on the bottom boundary in the same classical spin model derived for $\tilde{S}^{(n)}$. $\mathcal{F}^{(n)}$ is proportional to the probability of finding a certain subset of spin values (out of the $n!$ possibilities) on the bottom boundary sites.
Hence, the Fisher information undergoes the same transition as the entanglement entropy.
Above the critical measurement rate, $\mathcal{F}$ reaches a maximal value, reflecting the fact that the measurement device can obtain, over a long enough measurement time, maximal classical information on the initial state.
Below $p_c$, the value of $\mathcal{F}$ starts to deviate from its maximum, reflecting incomplete information on the initial state [see Fig.~\ref{fig:fig1}(d)].

The transition in information flow can be understood in terms of the natural quantum error correction implemented by the unitary components of the quantum circuit~\cite{choi2019quantum}. 
In the volume-law phase, information scrambling by the unitary evolution hides information in nonlocal degrees of freedom, thus protecting it from being revealed by local measurements.
The error correction becomes ineffective if the measurement rate exceeds a threshold, or if the measurements are capable of decoding nonlocal correlations, as we show in this work.

The rest of this paper is organized in the following order.
In Sec.~\ref{sec:setup}, we introduce our RUC model with weak measurements and elaborate on the generalized quantities $\tilde{S}^{(n)}$ and $\mathcal{F}^{(n)}$ as well as their relations to the von~Neumann entropy and the Fisher information.
In Sec.~\ref{sec:sm}, we derive the mapping between a RUC with weak measurements and a series of classical statistical mechanics models with various boundary conditions. 
Section~\ref{sec:pt} is dedicated to discuss the nature of the phase transition for different $n$. We analytically compute the critical/threshold measurement strength in the large $q$ limit and identify the universality class of the phase transition.
Then, in Sec.~\ref{sec:quantum}, we slightly digress from our main topic and discuss the absence of the phase transition in the presence of measurements on ancilla qudits in an arbitrary nonlocal basis.
Finally, we conclude with discussions and outlook in Sec.~\ref{sec:discussion}.

\section{Random unitary circuits with weak measurements}\label{sec:setup}
We consider an 1D array of $N$ qudits undergoing random unitary circuit evolution and weak measurements.
The unitary circuit consists of independent Haar random $q^2\times q^2$ unitary gates acting on the nearest neighboring qudits, each with local Hilbert space dimension $q$.
While not very important for our results, we assume a periodic boundary condition for concreteness.
The depth of the circuit $T$ corresponds to the discrete time of our model.
We are generally interested in the regime $N, T \gg 1$.
The layout of the unitary gates is illustrated in Fig.~\ref{fig:weak_meas}.

After each layer of unitary gates, every qudit is weakly measured.
For each weak measurement, an ancilla qudit with local Hilbert space dimension $q'$ is newly introduced and coupled to the system qudit via an entangling unitary gate $\hat{R}_\alpha$. 
While a specific choice of $q'$ or $\hat{R}_\alpha$ is not very important for most of our results, we focus on a particular example, where an ancilla has $q'=q+1$ internal states (enumerated by $\ket{i}_m$ with $i\in \{0, 1, \dots, q\}$), and the coupling $\hat{R}_\alpha$ takes the form of a controlled rotation:  
\begin{align}
\hat{R}_\alpha &= \sum_{i=1}^{q} \hat{P}_i \otimes e^{-i \alpha \hat{X}_i }.
\label{eq:ctrl_rot}
\end{align}
Here, $\hat{P}_i = \ket{i}_s\bra{i}_s$ is the projector onto one of the system qudit states $\ket{i}_s$ (enumerated with $i\in \{1, 2, \dots, q\}$), $\alpha \in [0,\pi/2]$ is a tunable parameter that controls the strength of the weak measurement, and $\hat{X}_i = \ket{i}_m\bra{0}_m + \ket{0}_m\bra{i}_m$ is a generalized Pauli matrix that rotates the quantum state of an ancilla between $\ket{0}_m$ and $\ket{i}_m$.
We assume that every ancilla qudit is initially prepared in $\ket{0}_m$.
Figure~\ref{fig:weak_meas} shows diagrammatic representations of weak measurements.

The entangling gate $\hat{R}_\alpha$ correlates the quantum state of a system qudit with that of an ancilla qudit.
For example, after applying $\hat{R}_\alpha$ with $\alpha = \pi/2$, the ancilla qudit state becomes $\ket{i}_m$ if and only if the system qudit is in $\ket{i}_s$. Therefore, any strong, projective measurements of the ancilla qudit in the computational basis $\ket{i}_m$ reveals the quantum state of the system qudit.
When $\alpha <\pi/2$, the correlation between the system and ancilla qudit becomes weaker, thereby a projective measurement of the ancilla constitutes a weak measurement of the system qudit.

The weak measurement strictly generalizes  projective measurements. In fact, it can be easily shown that the weak measurement with our choice of $\hat{R}_\alpha$ (followed by a projection in the computational basis) is equivalent to a probabilistic, projective measurement.
While the measurement of $\ket{i}_m$ with $i\neq0$ necessarily implies the system qudit in $\ket{i}_s$, the ancilla in $\ket{0}_m$ for any $\alpha < \pi/2$ reveals no information about the system qudit.
Therefore, the measurement of $\ket{0}_m$ state corresponds to not performing a projective measurement. 
This relation between weak and projective measurements can be made precise by explicitly computing the density matrix of a system-ancilla qudit pair after applying $\hat{R}_\alpha$ followed by projective measurements of the ancilla in the computational basis:
\begin{align}
    \mathcal{N}_\phi & \left[ \hat{R}_\alpha \left(\rho_{\textrm{in}} \otimes \ket{0}_m\bra{0}_m\right) \hat{R}^\dagger_\alpha \right] = \nonumber \\
    & (1-p) \rho_{\textrm{in}} \otimes \ket{0}_m\bra{0}_m + p \sum_i \hat{P}_i \rho_{\textrm{in}} \hat{P}_i \otimes \ket{i}_m\bra{i}_m,
    \label{eqn:proj_meas_channel}
\end{align}
where $\mathcal{N}_\phi[\rho] = \sum_i \ket{i}_m\bra{i}_m \rho \ket{i}_m\bra{i}_m$ is the dephasing channel acting on ancilla that corresponds to the measurement in the computational basis, $\rho_{\textrm{in}}$ is the density matrix of the system qudit before the weak measurement, and $p = \sin^2\alpha$.
We find that the resultant quantum channel is equivalent to that of performing a projective measurement with the probability $p$.
Furthermore, by tracing over system degrees of freedom, one can obtain the probability distribution of different measurement outcomes (including whether or not the measurement has been performed) from the diagonal components of the ancilla density matrix.

When more than one system/ancilla qudits are considered, one can in principle apply a (potentially nonlocal) unitary among the ancilla qudits before performing projective measurements. Such a procedure amounts to probabilistically measuring multi-qudit correlations of physical degrees of freedom. In our work, we first focus on the simple local measurements in computational basis in Secs.~\ref{sec:sm} and~\ref{sec:pt}, until we lift this condition later in Sec.~\ref{sec:quantum}.

\subsection{Entanglement entropy within the system}
One of the central quantities of interest is entanglement entropy within the system.
More specifically, we are interested in the entanglement between two complementary subsystems $A$ and $B$, averaged over all possible measurement outcomes from the collection of ancilla qudit $M$.
Given a particular set of unitary gates $\mathcal{U}$, and measurement outcomes $\bm{i}_M$, the system remains in a pure state after time evolution, and the von~Neumann entropy $S[\rho] \equiv -\tr{[ \rho \log\rho ]}$ of a subsystem characterizes the amount of entanglement.
Here, we consider its average behavior:
\begin{align}
 \overline{\llangle S_A \rrangle} &= \overline{\sum_{\bm{i}_M} p_{\bm{i}_M}(\mathcal{U})S[\rho_{A}(\mathcal{U},\bm{i}_M)]}, \label{eq:average_entropy}
\end{align} 
where  $\overline{\cdot}$ and $\llangle\cdot\rrangle$ denote the averaging over  $\mathcal{U}$ and over measurements $\bm{i}_M$, respectively; $\rho_X(\mathcal{U},\bm{i}_M)$ is the normalized reduced density matrix of the subsystem $X$ for a given pair of $\mathcal{U}$ and $\bm{i}_M$; and $p_{\bm{i}_M}(\mathcal{U})$ is the probability associated with a particular outcome $\bm{i}_M$ for a given $\mathcal{U}$.
For a system of $N$ qudits with RUC of depth $T$, the measurement outcome $\bm{i}_M$ from a set of ancilla qudits is enumerated by a string of $q'$ outcomes of length $NT$, i.e., $\bm{i}_M \in \left(\mathds{Z}_{q'}\right)^{\otimes NT}$, which encodes both different positioning of projective measurements and their outcomes.
Here and below, we simplify our notations by omitting the explicit dependence of $\rho_A$ and $p_{\bm{i}_M}$ on $\mathcal{U}$ whenever unambiguous, using $S_X$ to denote $S[\rho_X]$ for any subsystem $X$, and adapting a bold symbol for any indices running over exponentially many possibilities, e.g. basis states $\ket{\bm{i}_M}$, $\ket{\bm{j}_A}$ or $\ket{\bm{k}_B}$ for subsystems $M$, $A$, or $B$, respectively.

One of the nice properties of the weak measurement formalism is that it allows us to rewrite the average over measurements $\llangle S_A\rrangle$ in a simple form that depends only on the global density matrix of the system-environment combined without any averaging.
Consider the global density matrix $\tilde{\rho}_{ABM}$ in an extended Hilbert space that results from the RUC with given $\mathcal{U}$ and a set of couplings $\hat{R}_\alpha$ followed by the dephasing $\mathcal{N}_\phi $ \emph{without} projecting onto any particular outcome,
\begin{align}
\nonumber
    \tilde{\rho}_{ABM} = \sum_{\bm{i}_M} p_{\bm{i}_M} \ket{\Psi(\mathcal{U},\bm{i}_M)} \bra{\Psi(\mathcal{U},\bm{i}_M)} \otimes \ket{\bm{i}_M}\bra{\bm{i}_M},
\end{align}
where $\ket{\Psi(\mathcal{U},\bm{i}_M)}$ is a normalized pure state of the system for a given pair $\mathcal{U}$ and $\bm{i}_M$.
Then, $\llangle S_A \rrangle$ is exactly the conditional entropy of $A$ on $M$:
\begin{align}
\label{eq:cond_etrp}
% \langle S_A \rangle = \tilde{S}(A|M) \equiv \tilde{S}_{AM} - \tilde{S}_{M},
\llangle S_A \rrangle = \tilde{S}(A|M) \equiv \tilde{S}_{AM} - \tilde{S}_{M},
\end{align}
where $\tilde{S}_X$ indicates that the corresponding entropy is computed from a reduced density matrix $\tilde{\rho}_X$ in the extended Hilbert space, i.e., $\tilde{S}_X \equiv S[\tilde{\rho}_X]$. We note that $\tilde{\rho}_X$ is diagonal in the measurement basis of the ancilla qudits.
This relation is valid for any $\mathcal{U}$ (before averaging) and can be easily derived from explicit expressions:
\begin{align}
    \tilde{S}_{AM} &= -\sum_{\bm{i}_M} \tr \left(p_{\bm{i}_M} \rho_{A,\bm{i}_M} \log p_{\bm{i}_M} \rho_{A,\bm{i}_M}\right), \\
    \tilde{S}_{M} &= -\sum_{\bm{i}_M} p_{\bm{i}_M} \log p_{\bm{i}_M},
\end{align}
where we use $\rho_{A,\bm{i}_M}$ to denote $\rho_{A}(\mathcal{U}, \bm{i}_M)$.
We note that Eq.~\eqref{eq:cond_etrp} holds only when von~Neumann (or Shannon) entropy is used as the entropy measure, and the reduced density matrix $\tilde{\rho}_{AM}$ is block-diagonal in the basis $\ket{\bm{i}_M}$. In our case, the latter is satisfied owing to the dephasing channel $\mathcal{N}_\phi$, which enforces the measurement in the computational basis.

Equation~\eqref{eq:cond_etrp} greatly simplifies our problem since the conditional entropy $\tilde{S}(A|M)$ can be evaluated without explicitly computing entanglement entropies for different measurement outcomes. 
However, exact computation of $\tilde{S}(A|M)$ for any given $\mathcal{U}$ is still a formidable task.
For this reason, we introduce a series of closely related quantities, the $n$-th conditional entropies, that are more accessible for analytic calculation and recover $\tilde{S}(A|M)$ averaged over $\mathcal{U}$ in the limit $n\to 1$.
For any $n\geq 1$, these objects are defined as
\begin{align}
\tilde{S}^{(n)}(A|M) \equiv \tilde{S}^{(n)}_{AM} -\tilde{S}^{(n)}_{M}, \label{eq:cond_etrp_n}
\end{align}
with
\begin{align}
\label{eq:our_renyi}
    \tilde{S}^{(n)}_X \equiv \frac{\log \overline{ \tr\tilde\rho_X^n }}{1-n}.
\end{align}
While $\tilde{S}^{(n)}(A|M)$ quantifies the amount of the entanglement between subsystems $A$ and $B$, it does not correspond to the R\'enyi conditional entropy. Instead, this object measures the $n$-th moment of the entanglement spectrum, weighted by the $n$-th power of the measurement outcome probability:
\begin{align}
    \tilde{S}^{(n)}(A|M)
    = \frac{1}{1-n} \log \left(\; \frac{\overline{ \sum_{\bm{i}_M} p_{\bm{i}_M}^n \tr \rho_{A,\bm{i}_M}^n}}
    { \overline{ \sum_{\bm{i}_M} p_{\bm{i}_M}^n}}\;\right).
    \label{eq:Sn}
\end{align}
For a larger $n$, the averaging is more strongly weighted by relatively more likely measurement outcomes. 
A sharp change in the behavior of $\tilde{S}^{(n)}(A|M) $ signifies a qualitative change in the entanglement spectrum in its $n$-th moment. It is also worth noting that analytically evaluating $\tilde{S}^{(n)}(A|M) $ is much easier than $\overline{\tilde{S}(A|M)}$ as the average over $\mathcal{U}$ is taken inside the logarithm and $\tilde{S}^{(n)}(A|M)$ depends only on the $n$-th moment of the system-ancilla density matrix.

Crucially, the average von~Neumann conditional entropy is recovered in the limit $n\to 1$  
\begin{align}
    \overline{\tilde{S}(A|M)} = \lim_{n\rightarrow 1} \tilde{S}^{(n)}(A|M).
\end{align}
This follows from the analytic relation $\overline{\tr [\rho \log \rho]} = \lim_{n\rightarrow 1} (n-1)^{-1}\log \overline{ \tr \rho^n }$.
In Sec.~\ref{sec:sm}, we show that both
$\tilde{S}^{(n)}_{AM}$ and $\tilde{S}^{(n)}_{M}$  (with $n\ge 2)$ exactly map to the free energies of classical spin models with $n!$ internal states with different boundary conditions (up to a constant factor $n-1$), which reduce to the $n!$-state standard Potts models in the limit $q\to\infty$.
We investigate their behaviors in various parameter regimes and show their implications to the average conditional entropy $\overline{\tilde{S}(A|M)}$ in the interesting physical limit $n\to 1$.

\subsection{Kullback-Leibler divergence and Fisher information from measurement outcomes}
\label{sec:KL_div}
We now introduce another signature of the phase transition in the quantum circuit, which is motivated by the interpretation given in Ref.~\cite{choi2019quantum}. There, it was argued that the entanglement phase transition can be understood from the perspective of quantum error correction: in the volume-law phase, the scrambling of information by the unitary gates protects quantum correlations from sparse local measurements, thereby providing a natural error correction mechanism. At higher rate of measurement, this mechanism fails, and information encoded in the circuit is no longer protected from measurements.
Thus, we expect that there is a sharp difference across the phase transition in the amount of information that can be gained from local measurements about the state of the system. 

Here, we utilize the Fisher information~\cite{van2000asymptotic} as a measure of how much information about a system flows into the ancilla degrees of freedom.
Fisher information characterizes how much information a set of observables carry about an unknown parameter, and it can be derived from the closely related Kullback-Leibler (KL) divergence.
For example, consider an observable $X$ drawn from a probability distribution $P(x;\theta)$ that depends on an unknown parameter $\theta$.
If $\theta$ is perturbed away from its original value $\theta_0$ (without loss of generality, we set $\theta_0 = 0$), the probability distribution for $X$ also changes.
Then, one can quantify the amount of the information about $\theta$ carried in $X$ by measuring how distinct the new distribution $P_\theta=P(x;\theta)$ is from its original $P_0=P(x;0)$, using the KL divergence 
\begin{align}
    D_{\textrm{KL}}(P_0||P_\theta) \equiv \sum_x P_0(x) \log\left( \frac{P_0(x)}{P_\theta(x)} \right).
\end{align}
The Fisher information is defined as the second order derivative 
\begin{align}
    \mathcal{F} = \left. \partial_{\theta}^2 D_{\textrm{KL}}(P_0||P_\theta) \right|_{\theta=0}.
\end{align}
The KL divergence cannot be negative and vanishes if and only if $P_0=P_\theta$, which implies that the first order derivative generally vanishes at $\theta = 0$.

In our model, we are interested in the Fisher information carried by a set of measurement outcomes $\bm{i}_M$ about the initial quantum state of the system.
For simplicity, we assume that the system is initialized in either one of two nearby product states, $\ket{\Psi_0}=\ket{\psi_0}^{\otimes N}$ and 
$\ket{\Psi_\theta}=\delta U(\theta)\ket{\psi_0}^{\otimes N}$ with local perturbation $\delta U(\theta)$.
The parameter $\theta$ characterizes the strength of the perturbation.
For concreteness (and without loss of generality), we consider the perturbation $\delta U(\theta) = e^{-i \hat{X}_1 \theta}$ with $\theta \ll 1$ applied to a qudit at position $x_0$, where $\hat{X}_1$ is the generalized Pauli matrix in Eq.~\eqref{eq:ctrl_rot}.

Consider two systems with initial states $\ket{\Psi_0}$ and $\ket{\Psi_\theta}$ that evolve with identical circuits.
We examine the KL divergence averaged over $\mathcal{U}$:
\begin{align}
    D_{\textrm{KL}}(P_0||P_\theta) &\equiv  \overline{ \sum_{\bm{i}_M} p_{0, \bm{i}_M}(T) \log \left( \frac{p_{0, \bm{i}_M}(T)}{p_{\theta, \bm{i}_M}(T)} \right) }\\
    &=\overline{ \tr \left[ \tilde{\rho}_{M,0} \left(\log \tilde{\rho}_{M,0} - \log \tilde{\rho}_{M,\theta} \right) \right] },
\end{align}
where $\tilde{\rho}_{M,\theta}$ is the reduced density matrix of ancilla qudits for a given $\mathcal{U}$ and the initial state $\ket{\Psi_\theta}$.
Note that $\tilde{\rho}_{M,\theta}$ is a diagonal matrix, whose elements denote the probabilities for different outcomes $p_{\theta,\bm{i}_M}$.

A few remarks are in order.
First, $D_{\textrm{KL}}(P_0||P_\theta)$ is only a function of 
the local perturbation strength $\theta$ and time $T$ due to the averaging over random unitary gates.
Hence, we simplify our notation by using $D_{\textrm{KL}}(\theta, T)$ below.
Second, when $\theta =0$, the two initial states coincide, so  $D_{\textrm{KL}}(0, T) = 0$ at all time.
However, for any $\theta \neq 0$, one expects that $D_{\textrm{KL}}(\theta, T)$ would generally grow over time as two initial states should be better distinguished by more accumulated measurement outcomes. 
Indeed, it can be shown that $D_{\textrm{KL}}(\theta, T)$ is a nondecreasing function in time, using the monotonicity of relative entropy~\cite{araki1976relative,lindblad1975completely,uhlmann1977relative}.
Interestingly, we will show in Sec.~\ref{sec:sm} that in our model $D_{\textrm{KL}}(\theta, T)$ cannot grow indefinitely, but rather approaches a finite saturation value; a nonanalytic change in the saturation value at long time signifies the phase transition.
Finally, we note that the Fisher information can be directly extracted by computing the second order derivative in $\theta$.

Similar to the case of the average von~Neumann entropy, direct calculations of $D_{\textrm{KL}}(\theta, T)$ is a formidable task.
Once again, we use the replica method and introduce closely related quantities, the $n$-th divergence and the $n$-th Fisher information, based on the $n$-th moment of $\tilde{\rho}_{M,\theta}$:
\begin{align}
    D^{(n)}(\theta,T) &\equiv  \frac{\log \overline{ \tr [\tilde{\rho}_{M,0} \tilde{\rho}_{M,\theta}^{n-1} ]} - \log \overline{  \tr [ \tilde{\rho}_{M,0}^n ]} }{1-n}
    \label{eq:n_divergence}\\
    \mathcal{F}^{(n)}(T) &\equiv \left. \partial_\theta^2 D^{(n)}(\theta,T)\right|_{\theta = 0}.
\end{align}
The KL divergence and the Fisher information can be obtained via the analytic relations:
\begin{align}
    D_{\textrm{KL}}(\theta,T) = \lim_{n\rightarrow 1} D^{(n)}(\theta,T),\;\;
    \mathcal{F}(T) = \lim_{n \rightarrow 1} \mathcal{F}^{(n)}(T).\label{eq:div_anal}
\end{align}
We will show that $D^{(n)}$  also maps to the difference of free energies of the same classical spin models as those for $S^{(n)}$ with two different bottom boundary conditions, and $\mathcal{F}^{(n)}$ is proportional to the density of spins coupled to a boundary field (closely related to boundary magnetization) at the bottom boundary.

\section{Mapping to spin models}\label{sec:sm}
In this section, we develop a mapping between the RUC with weak measurements and a series of classical spin models.
Within this mapping, the generalized quantities $\tilde{S}^{(n)}$ and $D^{(n)}$ for integer $n\geq 2$ are related to free energies of 2D classical spin models. The distinction between the two is only in the boundary conditions imposed on the top and bottom, while the bulk spin model remains the same.
This implies that the two quantities detect the same phase transition.

Our mapping builds on earlier works in which the $n$-th moment of a density matrix evolved by a random unitary circuit is mapped to the partition function of a classical spin model with $n!$ states~\cite{nahum2017quantum,zhou2019emergent,hunter2019unitary}.
Similar emergent classical spin models have been also studied in random tensor networks~\cite{hayden2016holographic,qi2017holographic,vasseur2018entanglement}. Our mapping generalizes these works by incorporating projective measurements of ancilla qudits. 

An essential building block for calculating both the generalized entropies $\tilde{S}^{(n)}$ in Eq.~\eqref{eq:cond_etrp_n} and the generalized KL divergences $D^{(n)}$ in Eq.~\eqref{eq:n_divergence}
is the $n$-th moment of a density matrix: 
\begin{equation}
\label{eqn:moments}
    \mu^{(n)}_{AM} = \tr\left[\tilde{\rho}_{AM, 1} \tilde{\rho}_{AM, 2} \cdots \tilde{\rho}_{AM, n} \right].
\end{equation}
Here, $\tilde{\rho}_{AM,i}$ is the reduced density matrix of the ancilla qudits $M$ together with a subsystem $A$ obtained from the evolution of an initial state $\ket{\Psi_i}$ of the system.
The form of the quantity in Eq.~\eqref{eqn:moments} is sufficiently general to cover the calculation of both $\tilde{S}^{(n)}$ and $D^{(n)}$. In the former, we take all $\ket{\Psi_i}$ to be identical. 
In the latter, we take the subsystem $A$ to be an empty set, while the $\tilde{\rho}_{AM, i}$ are evolved from the initial state $\ket{\Psi_0}$ or $\ket{\Psi_\theta}$ as dictated by Eq.~\eqref{eq:n_divergence}.

The essence of our mapping is the identification of the $n$-th moment with a classical partition function
\begin{equation}
    \overline{\mu^{(n)}_{AM}} = \mathcal{Z}^{(n)}_{AM},
\end{equation}
where $\mathcal{Z}^{(n)}_{AM}$ is the partition function of a classical spin model with $n!$ internal states:
\begin{align}
    \mathcal{Z}^{(n)}_{AM} = \sum_{\{\sigma_{\bm{r}}\}} W(\{\sigma_{\bm{r}}\}).
\end{align}
Here, $\sigma_{\bm{r}}$ is a classical variable at $\bm{r} = (x,t)$ that may take $n!$ different values, $W(\{\sigma_{\bm{r}}\})$ is the Boltzmann weight given the spin configuration $\{\sigma_{\bm{r}}\}$.
Depending on the choice of the $n$-th moment $\mu^{(n)}_{AM}$, the partition function $\mathcal{Z}^{(n)}_{AM}$ is evaluated with different boundary conditions.
The top boundary condition of the spin model is determined by the choice of subsystem $A$. The bottom boundary condition depends on initial states of the density matrices. 
For example, when $\tilde{\rho}_{AM,i}$ have the same initial states, i.e., $\ket{\Psi_i} = \ket{\Psi_j}$ for any $i,j$, the spin model has an open boundary condition at the bottom. When $\tilde{\rho}_{AM,i}$ have the distinct initial states, the spin model exhibits a boundary ``magnetic" field at the bottom boundary.
Using this mapping, $\tilde{S}^{(n)}$ and $D^{(n)}$ are mapped to the free energy cost of exciting a domain wall at the top boundary and applying a magnetic field at the bottom boundary, respectively. 
A dictionary of the mapping is provided in Table~\ref{tab:dict}.

\begin{table}
\begin{tabular}{|c|c|}
    \hline
    Random unitary circuits & Classical spin models \\
    \hline
     $\overline{\tr\rho_{AM}^n}$ & $\mathcal{Z}^{(n)}_{AM}$ \\
    $-\log\overline{\tr\rho_{AM}^n}$ & $F^{(n)}_{AM}$ \\
    $\tilde{S}^{(n)}(A|M)$ & $(F^{(n)}_{AM} - F^{(n)}_{M})/(n-1)$\\
    $D^{(n)}(P_0||P_\theta)$ & $(F^{(n)}_{M,\theta} - F^{(n)}_{M})/(n-1)$\\
    $\mathcal{F}^{(n)}$ & $2 \langle m^\downarrow_{1}\rangle/(n-1)$\\
    \hline
\end{tabular}
\caption{Dictionary of the mapping between the RUC with weak measurements and classical spin models.
On the left column, $\tilde{S}^{(n)}$, $D^{(n)}$, and $\mathcal{F}^{(n)}$ are generalized entropy, the divergence of probability distributions, and Fisher information that identify the phase transition, respectively. 
On the right column, $\mathcal{Z}^{(n)}$ and $F^{(n)}$ denote the partition function and the free energy of a classical spin model under a certain boundary condition specified by their subscripts. $\langle m^{\downarrow}_1\rangle $ denotes the probability that a spin at position $\bm{r} = (x_0,1)$ belongs to the down-type (see the main text). }
\label{tab:dict}
\end{table}

Before we present any technical details, we provide the outline of the mapping. 
Our mapping consists of two steps. In the first step, we consider the $n$-th moment that involves $n$ copies of density matrix $\tilde{\rho}_{AM, i}$.
We note that the $n$ copies of the density matrix is an $n$-th order monomial of random unitary gate $U$ and its conjugate $U^\dagger$.
By averaging over the unitary gates, we rewrite $\overline{\mu_{AM}^{(n)}}$ as a sum of terms labeled by two emergent spin variables $\sigma$ and $\tau$, each living on one of the two honeycomb sublattices:
\begin{align}
\overline{\mu_{AM}^{(n)}}
 =
\sum_{\{\sigma_{\bm{r}},\tau_{\bm{r'}}\}}
W(\{\sigma_{\bm{r}}, \tau_{\bm{r'}}\}),\label{eqn:partition_func}
\end{align}
where $W(\{\sigma_{\bm{r}},\tau_{\bm{r'}}\})$ is the weight associated with a given spin configuration $\{\sigma_{\bm{r}},\tau_{\bm{r'}}\}$.
We perform this calculation by utilizing a tensor network representation of $\overline{\mu_{AM}^{(n)}}$ and present an explicit expression for $W$.
At this point, the weight $W(\{\sigma_{\bm{r}},\tau_{\bm{r'}}\})$ is not always positive, and the r.h.s. of Eq.~\eqref{eqn:partition_func} cannot be interpreted as the partition function of a spin model.
In the second step of the mapping, we resolve the negative weight in two different cases.
In the case of $n = 2$, the negative weight $W(\{\sigma_{\bm{r}}, \tau_{\bm{r'}}\})$ can be removed by integrating out $\tau$ variables (or alternatively $\sigma$ variables).
The resultant expression can be interpreted as the partition function of a $2$D classical Ising model on a triangular lattice.
In the case of $n \geq 3$, integrating out $\tau$ variables is not sufficient, and we further consider the limit of large Hilbert space dimension $q$.
For a sufficiently large but finite $q$, the weights in Eq.~\eqref{eqn:partition_func} become positive, and $\overline{\mu_{AM}^{(n)}}$ can be interpreted as the partition function of a spin model with $n!$ internal states on a $2$D triangular lattice.
In the limit $q\to \infty$, this model further reduces to the $n!$-state standard Potts model on a square lattice.

In what follows, we provide the details of the above procedure, starting from the simplest example $\tilde{S}^{(n)}$ with $n=2$. Its modification for $D^{(n)}$ and the generalization for $n\geq 3$ are straightforward and will be discussed in detail later in this section.

\subsection{Mapping between purity and partition function of classical Ising model}\label{sec:sm_mapping}

\begin{figure*}[t!]
\includegraphics[width=\textwidth]{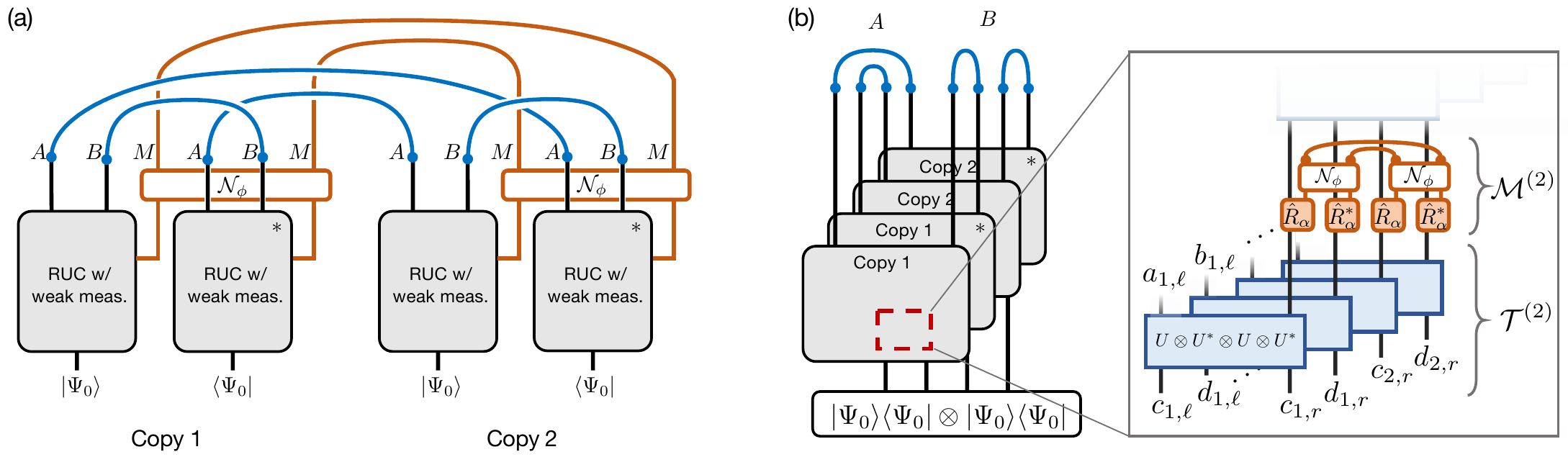}
	\caption{Tensor network representation of the purity $\tr{[\tilde{\rho}_{AM}^2]}$.
	(a) Two copies of the initial states $\ket{\Psi_0}\bra{\Psi_0}$ are evolved under a RUC with weak measurements, before jointly contracted (or traced out).
	The asterisk ($^*$) indicates the complex conjugate for the corresponding tensor.
	Notice the different connectivities for contracting the subsystems $A$, $B$, and $M$ (blue and orange lines). 
	(b) The network can be rearranged in such a way that the top boundary contractions are simplified and that the identical unitary tensors are brought together. 
	Inset: Four copies of a random unitary gate $U\otimes U^*\otimes U\otimes U^*$ form a tensor $\mathcal{T}^{(2)}$.
	The combination of the coupling unitary $\hat{R}_\alpha$, dephasing gates $\mathcal{N}_\phi$, and the contraction of the ancilla qudits is denoted by $\mathcal{M}^{(2)}$.
	}
	\label{fig:ising_wide_01}
\end{figure*}

Here, we explicitly show that the average purity $\overline{\mu_{AM}^{(2)}} = \overline{\tr \tilde{\rho}_{AM}^2}$ for a subsystem $AM$ maps to the partition of a 2D classical Ising model under a certain boundary condition.
We will pay close attention to how the partition function of the classical Ising model depends on various parameters of the original model such as the local Hilbert space dimension $q$ of the system qudits and the measurement strength $\alpha \in [0, \pi/2]$.
We first discuss the mapping of $\tilde{S}^{(2)}$ and elaborate how the boundary conditions for the spin model depend on the choice of the subsystem $A$.
Once the relationship between the purity and the partition function is established, we will discuss how the boundary conditions are modified for the calculation of $D^{(2)}$.
Recall that we choose $A$ to be the empty set when computing $D^{(2)}$.
Generalization of the mapping to $\tilde{S}^{(n)}$ and $D^{(n)}$ for  $n\geq2$ is deferred to Secs.~\ref{sec:sm_n} and~\ref{sec:sm_op}.

The first step of our derivation is to rewrite the second moment (purity) using the swap technique:
\begin{align}
\label{eqn:swap_technique}
\mu_{AM}^{(2)} = \mathrm{Tr}
\left[\left( \mathcal{C}_A^{(2)} \otimes \mathcal{I}_B^{(2)} \otimes \mathcal{C}_M^{(2)} \right)\left(\tilde{\rho}_{ABM} \otimes \tilde{\rho}_{ABM} \right) \right],
\end{align}
where $\mathrm{Tr}$ represents tracing over a twofold replicated Hilbert space for $ABM$, 
$\mathcal{C}_X^{(2)} = \sum_{\bm{i}_X,\bm{i}_X'} \ket{\bm{i}_X}\otimes\ket{\bm{i}_X'}\bra{\bm{i}_X'}\otimes\bra{\bm{i}_X}$ is an operator that swaps the quantum states of the subspace $X$ between two copies, and $\mathcal{I}^{(2)}$ is the identity operator.
Each copy of the quantum state $\tilde{\rho}_{ABM}$ is generated from an initial product state $\ket{\Psi_0}=\ket{\psi_0}^{\otimes N}$ by evolving it under an identical RUC and weak measurements (i.e., applying the coupling unitaries $\hat{R}_\alpha$ followed by dephasing of every ancilla qudit).
Equation~\eqref{eqn:swap_technique} rewrites a nonlinear function of the density matrix (on the left-hand side) as a linear expectation value of an operator in an extended Hilbert space (on the right-hand side).
In turn, this allows the tensor network (TN) representation of $\mu_{AM}^{(2)}$ [see Fig.~\ref{fig:ising_wide_01}(a)].
Notice that the tensors are contracted across the two different copies for the subsystems $A$ and $M$, reflecting the swap operators, while they are connected within each copy for the subsystem $B$.
For reasons that will become clear below, we call the operators $\mathcal{O}_{\textrm{top}}^{(2)} = \mathcal{C}_A^{(2)}\otimes \mathcal{I}_B^{(2)} \otimes \mathcal{C}_M^{(2)}$ the top boundary operator.

\begin{figure*}[t!]
\includegraphics[width=\textwidth]{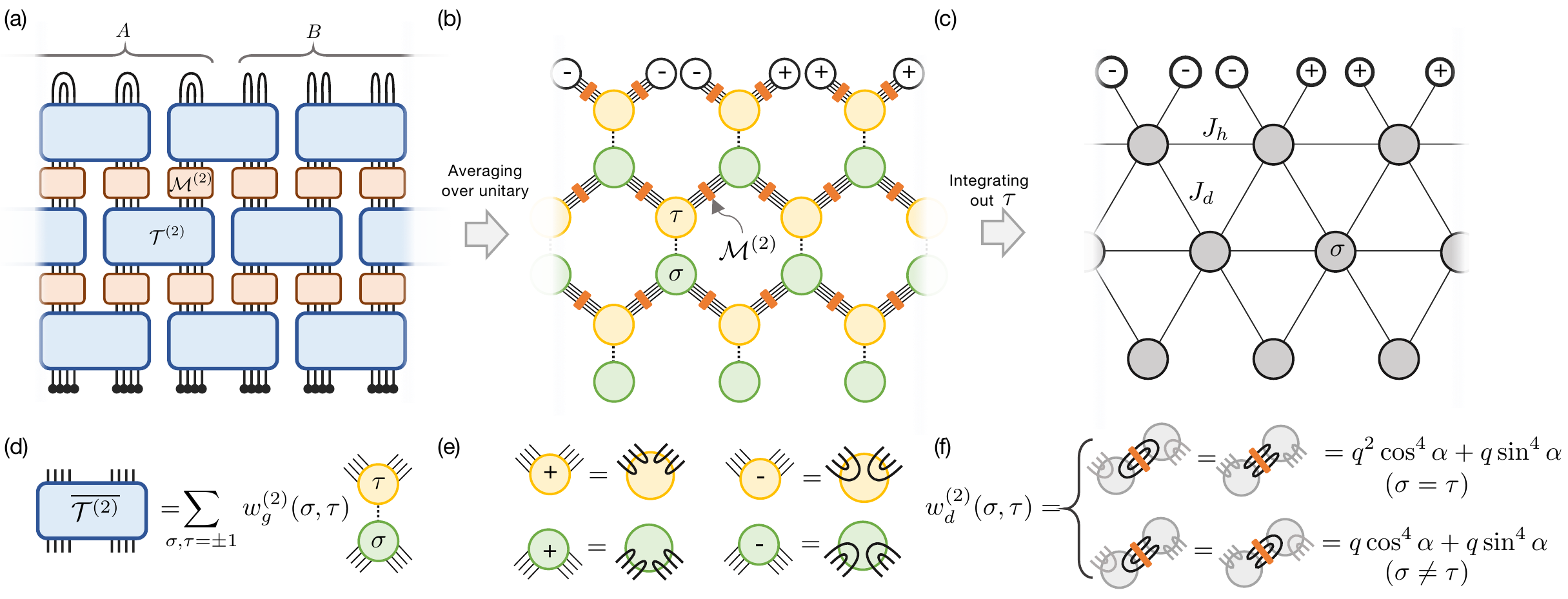}
	\caption{Mapping between a RUC with weak measurements and classical Ising model.
	(a) Tensor network representation of the RUC with weak measurements after the rearrangement [see also Fig.~\ref{fig:ising_wide_01}(b)].
	Each $\mathcal{T}^{(2)}$ represents four identical unitary tensors, and $\mathcal{M}^{(2)}$ represents a weak coupling $\hat{R}_\alpha$ followed by the contraction of ancilla qudits [Fig.~\ref{fig:ising_wide_01}(b) inset].
	(b) After averaging over unitary, the TN reduces to a network on a honeycomb lattice. Each site of the lattice hosts a classical Ising variable $\sigma$ or $\tau$.
	(c) Integrating out $\tau$ variables leads to the 2D classical Ising model on a triangular lattice with nearest-neighbor interactions $J_h$ and $J_d$.
	(d) Averaging $\mathcal{T}^{(2)}$ over Haar random unitary allows one to replace the tensor as a sum of simple diagrams labeled by a pair of classical variables $\sigma,\tau \in \{ \pm 1\}$.
	(e) Diagrammatic representations of the tensors $\hat{\tau}$ and $\hat{\sigma}$.
	(f) Contracting a pair of diagonally neighboring $\hat{\sigma}$ and $\hat{\tau}$ tensors leads to a closed loop diagram, whose value depends only on the relative sign between $\sigma$ and $\tau$.}
	\label{fig:ising_wide_02}
\end{figure*}

Our next step is to perform averaging over random unitary gates.
Owing to their statistical independence, we can average unitary gates separately.
In our diagrammatic representation, a tensor corresponding to each gate $U$ appears exactly four times: $U$ and $U^\dagger$ applied to the bra and ket vectors of the system in two copies.
We rearrange the TN such that four identical tensors are placed together [see Figs.~\ref{fig:ising_wide_01}(b) and \ref{fig:ising_wide_02}(a)] and consider their joint tensor $\mathcal{T}^{(2)}$ defined as
\begin{align}
\label{eqn:2-design}
    \mathcal{T}^{(2)}_{\mathbf{ab};\mathbf{cd}} \equiv U_{a_1c_1}\otimes U^*_{b_1d_1} \otimes U_{a_2c_2} \otimes U^*_{b_2d_2},
\end{align}
where each index on the right-hand side represents a pair of input or output qudits and therefore runs over $d = q^2$ possible quantum states. Specifically, $\mathbf{a} \equiv (a_1,a_2)\equiv (a_{1,\ell},a_{1,r},a_{2,\ell},a_{2,r})$.
Figure~\ref{fig:ising_wide_01}(b) inset shows the diagrammatic representation of Eq.~\eqref{eqn:2-design}.

For Haar random unitary gates, the average of $\mathcal{T}^{(2)}$ can be exactly computed using a property of unitary $2$-design:
\begin{align}
\overline{\mathcal{T}^{(2)}_{\mathbf{ab;cd}}}
=
\sum_{\sigma,\tau =\pm1} w_g^{(2)}(\sigma,\tau) \hat{\tau}_{\mathbf{ab}}
\hat{\sigma}_{\mathbf{cd}}, \label{eqn:2-design_avg}
\end{align}
where $\sigma, \tau \in \{\pm1\}$ are classical Ising variables, 
\begin{align}
\label{eqn:weingarten_coeff}
w_g^{(2)}(\sigma,\tau) = 
\frac{\delta_{\sigma,\tau}}{d^2-1}  
-\frac{1-\delta_{\sigma,\tau}}{d(d^2-1)} 
\end{align}
is a coefficient called the Weingarten function~\cite{collins2003moments}, and $\hat{\sigma}$ and $\hat{\tau}$
are tensors associated with the variables $\sigma$ and $\tau$ defined as 
\begin{align}
\label{eqn:tensor_content}
    \hat{\xi}_{\mathbf{ab}} 
    =\left\{
\begin{array}{ll}
\delta_{a_1b_1} \delta_{a_2b_2} & \textrm{if  } \xi = +1\\
\delta_{a_1b_2} \delta_{a_2b_1} & \textrm{if  } \xi = -1
\end{array}
\right.
.
\end{align}
The diagrammatic representations of Eqs.~\eqref{eqn:2-design_avg} and \eqref{eqn:tensor_content} are given in Figs.~\ref{fig:ising_wide_02}(d) and~\ref{fig:ising_wide_02}(e).
Equation~\eqref{eqn:2-design_avg} plays the central role in our mapping; practically, it implies that, after averaging, each unitary gate in our TN can be replaced by a sum of simpler tensors labeled by the Ising variables $\sigma, \tau \in \{\pm1\}$.
The corresponding tensors $\hat{\sigma}, \hat{\tau}$ with $\sigma, \tau \in \{\pm1\}$ in Eq.~\eqref{eqn:tensor_content} describe different ways to pair up indices associated with bra/ket vectors.

Now, it remains to contract these tensors on the entire network [Fig.~\ref{fig:ising_wide_02}(b)]. 
In particular, we need to incorporate the effects of weak measurements by contracting tensors associated with ancilla qudits $\mathcal{M}^{(2)}$ [Fig.~\ref{fig:ising_wide_01}(b) inset] with a pair of diagonally neighboring $\hat{\sigma}$ and $\hat{\tau}$ tensors [Fig.~\ref{fig:ising_wide_02}(f)].
As shown in Appendix~\ref{app:weights}, these contractions lead to the weight 
\begin{align}
w_d^{(2)}(\sigma,\tau)
= \left\{\begin{array}{ll}
q^2\cos^4\alpha + q \sin^4\alpha & \textrm{if  }  \sigma = \tau \\
q\cos^4\alpha + q \sin^4\alpha & \textrm{if  } \sigma \neq \tau
\end{array}
\right.
,\label{eqn:w_d_2}
\end{align}
where we recall that $q$ and $\alpha$ denote the local Hilbert space dimension and the measurement strength, respectively.
The dependence of $w_d^{(2)}(\sigma,\tau)$ on $\alpha$ distinguishes our spin model from the previous result described in Refs.~\cite{nahum2017quantum,zhou2019emergent,hunter2019unitary}, which is recovered when $\alpha = 0$.

By applying Eq.~\eqref{eqn:2-design_avg} to every gate and Eq.~\eqref{eqn:w_d_2} to the contraction of the diagonally neighboring tensors, we obtain an expression for the purity
\begin{align}
\label{eqn:honeycomb_model}
    \overline{\mu_{AM}^{(2)}} =\sum_{\{\sigma_r,\tau_{r}\}} W_{\text{top}}W_{\text{bottom}}\prod_{\langle {\bm r},{\bm r}'\rangle} w^{(2)}(\sigma_{\bm r},\tau_{{\bm r}'}) \,  
\end{align}
that indeed looks like the partition function of an Ising spin model. The sum is on all the configurations of the pairs $(\sigma_{\bm{r}},\tau_{\bm{r}})$ of Ising variables arising from the unitary gates at $\bm{r} = (x, t)$ in two dimension. The weights $w^{(2)}(\sigma_{\bm r},\tau_{{\bm r}'})$ are functions of nearest neighbor spin variables at $\langle {\bm r},{\bm r}'\rangle$ in the network given by $w^{(2)}=w_d^{(2)}$ or $w_g^{(2)}$ depending on the orientation of the bond, i.e., the Weingarten function $w_g^{(2)}(\sigma, \tau)$ is treated as the weight for a vertical bond.
The weights $W_{\textrm{top}}$ and $W_{\textrm{bottom}}$ are the contributions from boundary conditions.

At the top boundary, closed diagrams arise from contracting $\hat{\tau}$ tensors according to their appropriate boundary conditions: qudit indices are connected across two copies in the subsystem $A$ and within each copy in the subsystem $B$.
According to Eq.~\eqref{eqn:tensor_content}, these contractions are equivalent to introducing of a set of additional $\hat{\sigma}$ tensors at $t = T + 1$ with fixed values $\sigma_{x,T+1} = -1$ for $x\in A$ and $+1$ for $x\in B$ [Fig.~\ref{fig:ising_wide_01}(b)], leading to the top boundary contribution~\footnote{When the position $x$ is at the interface between $A$ and $B$, one needs to first decompose the index $\ket{\mathbf{a}}=\ket{a_{1,\ell},a_{2,\ell}}\otimes \ket{a_{1,r},a_{2,r}}$ and use a  different boundary condition for each qudit degree of freedom.}
\begin{align}
    W_\textrm{top} = \prod_{x\in A} w_d^{(2)}(-1,\tau_{x,T}) \prod_{x\in B} w_d^{(2)}(+1,\tau_{x,T}).
\end{align}
In fact, this equivalence is natural since a contraction with $\hat{\sigma}$ represents evaluating the expectation value of an identity $\mathcal{I}^{(2)}$ ($\sigma = 1$) or swap $\mathcal{C}^{(2)}$ ($\sigma = -1$) operator for any state in a two-fold replicated Hilbert space:
\begin{align}
\label{eqn:c_tensor}
\hat{\sigma}_{\mathbf{ab}} = 
\left\{
\begin{array}{cc}
\tr\left(\mathcal{I}^{(2)}\ket{\mathbf{a}}\bra{\mathbf{b}}\right) & \textrm{if } \sigma = +1\\
\tr\left(\mathcal{C}^{(2)}\ket{\mathbf{a}}\bra{\mathbf{b}}\right) & \textrm{if } \sigma = -1
\end{array}
\right.
,
\end{align}
and similarly for $\hat{\tau}$.

At the bottom boundary, $\hat{\sigma}$ variables are contracted with tensors representing the initial state of the system.
Here, both copies are initialized in the same product state $\ket{\Psi_0}=\ket{\psi_0}^{\otimes N}$.
In such a case, tensor contractions lead to unity, independent of the Ising variable since $\tr{\left( \mathcal{I}^{(2)} \ket{\psi_0}\bra{\psi_0} \otimes \ket{\psi_0}\bra{\psi_0} \right)}=\tr{\left(\mathcal{C}^{(2)} \ket{\psi_0}\bra{\psi_0} \otimes \ket{\psi_0}\bra{\psi_0}\right)} = 1$ from Eq.~\eqref{eqn:tensor_content}. As a result, $W_{\textrm{bottom}} = 1$.

Combining these results, we obtain an intermediate expression
\begin{align}
    \overline{\mu_{AM}^{(2)}} = 
    \sum_{ \{\sigma_{\bm{r}}, \tau_{\bm{r}}\} }
    \sideset{}{'}\prod_{\langle \bm{r},\bm{r'} \rangle} w^{(2)}(\sigma_{\bm{r}},\tau_{\bm{r'}}), 
\end{align}
where the summation runs over all possible assignment of bulk Ising variables while the product with prime runs over every neighboring pairs on a honeycomb lattice including the extra boundary spins at $t=T+1$ with fixed $\sigma_{x,T+1}$ values.
We emphasize that the boundary conditions are implicit and depend on the the choice of the subsystem $A$ [Fig.~\ref{fig:ising_wide_02}(b)].
While the above expression already resembles the partition function of a classical Ising model on a honeycomb lattice, there is still a problem that the  Weingarten function contributes negative weights.
This sign problem can be fixed by explicitly integrating out all $\tau$ variables, which leads to a model defined on a triangular lattice
\begin{align}
\label{eqn:Ising_spin_model}
    \overline{\mu_{AM}^{(2)}} = 
    \mathcal{Z}^{(2)}_{AM} =
    \sum_{ \{\sigma\} }
    \sideset{}{'}\prod_{\langle \sigma_1,\sigma_2,\sigma_3 \rangle} \bar{w}^{(2)}(\sigma_1,\sigma_2,\sigma_3), 
\end{align}
where $\left<\sigma_1, \sigma_2, \sigma_3\right>$ denotes a lower-facing triangle with three neighboring vertices $\sigma_1, \sigma_2, \sigma_3$ (in clockwise order starting from top left), and $\bar{w}^{(2)}$ is a three-body weight that is nonnegative for arbitrary choice of $q$ and $\alpha$.
The explicit expression of $\bar{w}^{(2)}$ is given in Appendix~\ref{app:three-body_weight}.
Apparently, the right-hand side of Eq.~\eqref{eqn:Ising_spin_model} is a partition function of a classical Ising spin model on the 2D triangular lattice [Fig.~\ref{fig:ising_wide_02}(c)].

Crucially, except for boundary contributions, the model given by Eq.~\eqref{eqn:Ising_spin_model} preserves the Ising $\mathbb{Z}_2$ symmetry associated with the global transformation $\sigma \mapsto -\sigma$ since both $w_g^{(2)}(\sigma,\tau)$ and $w_d^{(2)}(\sigma,\tau)$ only depend on the relative sign of two arguments.
Consequently, the bulk three-body weight decomposes into pairwise contributions $\bar{w}^{(2)}(\sigma_1, \sigma_2, \sigma_3) \propto e^{-J_h \sigma_1 \sigma_2 - J_d \sigma_1 \sigma_3 - J_d \sigma_2 \sigma_3}$. The couplings, tuned by the measurement strength, are ferromagnetic $J_d\leq 0$ along diagonal bonds and antiferromagnetic $J_h\geq 0$ along horizontal bonds (Appendix~\ref{app:Ising_coupling}).
It is interesting to note that this classical Ising model with anisotropic pairwise interaction on the triangular lattice is exactly solvable~\cite{eggarter1975triangular,tanaka1978triangular}.
In our case, $J_h + J_d \leq 0$, hence there is a ferromagnetic-paramagnetic phase transition at a critical value of those parameters tuned by the measurement strength.
A large value of $q\gg 1$ and weak measurement $\alpha \ll 1$ correspond to the low-temperature limit of the spin model resulting in the ferromagnetic phase.
In the limit of vanishing measurement strength $\alpha = 0$, the model reduces to the previous result discussed in Ref.~\cite{nahum2017quantum}.
On the other hand, the limit of strong measurement $\alpha\to \pi/2$ maps to high temperature, giving rise to the paramagnetic phase.
We will elaborate on this phase transition in Sec.~\ref{sec:pt}.

Having established the mapping in Eq.~\eqref{eqn:Ising_spin_model}, the second conditional entropy in Eq.~\eqref{eq:Sn} can be expressed as $\tilde{S}^{(2)} = F_{AM}^{(2)} - F_{M}^{(2)}$, where $F_X^{(2)} \equiv -\log\mathcal{Z}^{(2)}_X$ with $X = AM, M$ is the free energy of the spin model.
In the case of $\mathcal{Z}_{AM}^{(2)}$, the Ising symmetry is explicitly broken on the top boundary; the spin variables on the top boundary ($t = T+1$) are pinned to $\sigma = -1$ and $+1$ in subsystem $A$ and $B$, respectively.
$\mathcal{Z}_{M}^{(2)}$ corresponds to the special case when the subsystem $A$ is an empty set.
Therefore, the second conditional entropy is given by the excess free energy of an Ising domain wall terminating on the edges of region $A$ on the top boundary.

Finally, we discuss the important modification of our mapping that allows analyzing the second KL divergence $D^{(2)}$. 
Calculation of this object involves
comparing two probability distributions (diagonal density matrices) $\tilde{\rho}_{M,0}$ and $\tilde{\rho}_{M,\theta}$ obtained from the evolution of two distinct initial states with the same quantum circuit. This requires the computation of
\begin{align}
    \mu_{M,\theta}^{(2)} &\equiv \tr{\left[\tilde{\rho}_{M,0} \tilde{\rho}_{M,\theta}\right]} \nonumber\\
    &= 
    \Tr{\left[\left( \mathcal{I}_{AB}^{(2)} \otimes \mathcal{C}_M^{(2)}\right) \tilde{\rho}_{ABM,0} \otimes \tilde{\rho}_{ABM,\theta} \right]}.
\end{align}
This equation is related to Eq.~\eqref{eqn:swap_technique} when the subsystem $A$ is an empty set.
The average of $\mu_{M,\theta}^{(2)}$ maps to the same Ising model as Eq.~\eqref{eqn:Ising_spin_model} in the bulk.
A nontrivial modification arises only in the bottom boundary condition, where density matrices for two initial states are contracted with the $\hat{\sigma}$ tensors in the bottom layer.
For product states  $\ket{\Psi_0}=\ket{\psi_0}^{\otimes N}$ and $\ket{\Psi_\theta}=\delta U(\theta)\ket{\psi_0}^{\otimes N}$, this leads to nontrivial additional weights on the bottom boundary
\begin{align}
    W_\textrm{bottom} = \left|\bra{\psi_0}\delta U(\theta)\ket{\psi_0}\right|^{(1-\sigma_{x_0,1})}.
\end{align}
In terms of the Ising spin model description, the extra weight appears as a ``magnetic field'' term on the pertubed site
\begin{align}
\label{eqn:Ising_spin_model_with_boundary}
    \mathcal{Z}^{(2)}_{M,\theta}=
    \sum_{ \{\sigma\} }\;
    e^{-h(1-\sigma_{x_0,1})}
    \sideset{}{'}\prod_{\langle \sigma_1,\sigma_2,\sigma_3 \rangle} \bar{w}^{(2)}(\sigma_1,\sigma_2,\sigma_3).
\end{align}
Here, $h = - \log\cos\theta $ with $\cos\theta \equiv |\bra{\psi_0}\delta U(\theta)\ket{\psi_0}|$.
We also note that the top boundary boundary condition is $\sigma_{x,T+1}=+1$ everywhere owing to the trivial choice of $A$.
The generalization of this result for any pair of inhomogeneous product states, e.g. $\ket{\Psi} = \otimes_k \ket{\psi_k}$ is straightforward.

In the Ising model, the second KL divergence $D^{(2)} = F_{M, \theta}^{(2)} - F_{M}^{(2)}$ with $F_{M, \theta}^{(2)} = -\log\mathcal{Z}_{M,\theta}^{(2)}$ is identified as the free energy cost of applying a boundary magnetic field on the perturbed site. The corresponding second Fisher information is given by
\begin{align}
    \mathcal{F}^{(2)} = 1 - \left< \sigma_{x_0, 1} \right>,
\end{align}
where $\langle \sigma_{x_0, 1} \rangle$ is the magnetization at $\bm{r} = (x_0, 1)$.

The boundary magnetization $\langle \sigma_{x_0, 1} \rangle$ exhibits distinct behaviors in the two phases of the Ising model.
In the ferromagnetic phase (weak or sparse measurements), the  top boundary condition ($\sigma = +1$) induces a nonvanishing magnetization on the bottom boundary so that $\langle \sigma_{x_0, 1} \rangle>0$ and $\mathcal{F}^{(2)}<1$.
In the paramagnetic phase (relatively strong or frequent measurements), the bottom boundary magnetization vanishes in the thermodynamic limit $T\to \infty$. We conclude that the second Fisher information $\mathcal{F}^{(2)}$ is less than unity in the volume-law phase, increases continuously with measurement strength, and saturates, in a nonanalytic way, to $\mathcal{F}^{(2)}=1$ at and beyond the critical point.

\subsection{Generalization to the $n$-th moment}\label{sec:sm_n}
\begin{figure}
    \centering
    \includegraphics[width=0.48\textwidth]{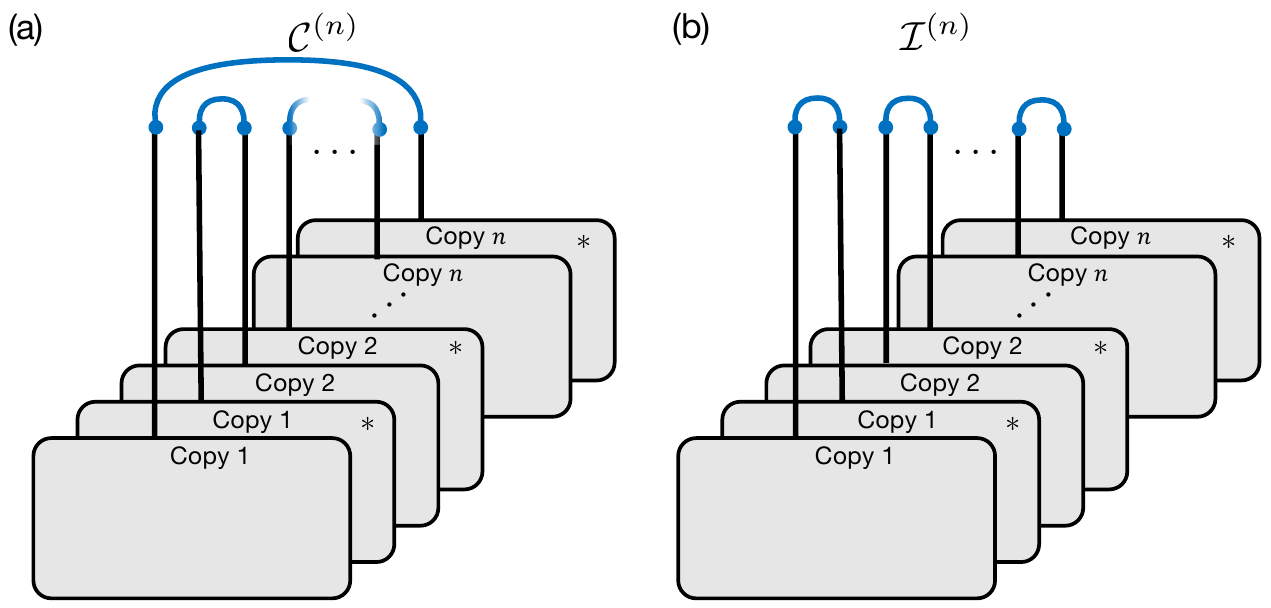}
    \caption{TN representations of the top boundary contractions for an $n$-th moment.
    (a) For subsystem $A$ and $M$, $n$ pairs of indices are cyclically shifted before being contracted.
    (b) For subsystem $B$, indices are contracted within each copy.}
    \label{fig:top_boundary_cond}
\end{figure}

We now generalize the mapping discussed in the previous section to arbitrary $n\geq 2$.
To this end, consider a generic $n$-th moment $\mu_{AM}^{(n)}$ that involves $n$ copies of density matrices: 
\begin{align}
    \mu_{AM}^{(n)} = \tr\left[ \tilde{\rho}_{AM, 1} \tilde{\rho}_{AM, 2} \cdots \tilde{\rho}_{AM, n} \right].
\end{align}
All density matrices $\tilde{\rho}_{AM,i}$ are obtained from the same RUC with weak measurements, but starting from a potentially different initial state of the system $\ket{\Psi_i}$.
Similar to the swap trick in Sec.~\ref{sec:sm_mapping}, $\mu^{(n)}_{AM}$ can be written as
\begin{align}
    \mu_{AM}^{(n)} =\mathrm{Tr} \Big[&\Big(\mathcal{C}_A^{(n)} \otimes \mathcal{I}_B^{(n)} \otimes \mathcal{C}_M^{(n)}\Big) \nonumber \\
    &\Big(\tilde{\rho}_{ABM, 1}\otimes \tilde{\rho}_{ABM, 2}\otimes \cdots \otimes\tilde{\rho}_{ABM, n}\Big) \Big],
\end{align}
where $\mathcal{I}_B^{(n)}$ is the identity operator and $\mathcal{C}_X^{(n)}$ is a  cyclic permutation of $n$ copies of the Hilbert space of subsystem $X$:
\begin{align}
\mathcal{C}_X^{(n)} = \sum_{\{\bm{i_k}\}} \ket{\bm{i_n,i_1,i_2,\cdots,i_{n-1}}}\bra{\bm{i_1,i_2,\cdots,i_n}},
\end{align}
where $\bm{i_k}$ $(\bm{k = 1, 2, \cdots, n})$ enumerates quantum states of $X$ in the $k$-th copy.
Below we omit the subscript and simply use $\mathcal{I}^{(n)}$ and $\mathcal{C}^{(n)}$ to denote the identity and cyclic permutation operators acting on $n$ copies of an appropriate subsystem.
The cyclic permutation $\mathcal{C}^{(n)}$ generalizes the swap operation $\mathcal{C}^{(2)}$ between two copies. 
The TN representations of the boundary contractions with $\mathcal{C}^{(n)}$ and $\mathcal{I}^{(n)}$ are shown in Figs.~\ref{fig:top_boundary_cond}(a) and~\ref{fig:top_boundary_cond}(b), respectively.

We now average $\mu_{AM}^{(n)}$ over individual random unitary gate $U$.
Since $\mu_{AM}^{(n)}$ is an $n$-th order monomial of $U$ and $U^\dagger$, the joint tensor $\mathcal{T}^{(n)}$ takes the form
\begin{align}
    \mathcal{T}^{(n)}_{\mathbf{ab;cd}} \equiv U_{a_1c_1} \otimes U^*_{b_1d_1} \otimes \cdots \otimes U_{a_nc_n} \otimes U^*_{b_nd_n},
\end{align}
where we used the composite indices, e.g. $\mathbf{a} \equiv (a_1,a_2,\cdots, a_n)$, for multiple copies.
For unitary gates drawn from the Haar measure, the average $\mathcal{T}^{(n)}$ can be written as
\begin{align}
    \overline{\mathcal{T}^{(n)}_{\mathbf{ab;cd}}} = \sum_{\sigma,\tau \in \mathcal{P}_n} w_g^{(n)}(\sigma,\tau) \hat{\tau}_{\mathbf{ab}} \hat{\sigma}_{\mathbf{cd}},
    \label{eqn:unitary_n_design}
\end{align}
where the emergent spin degrees of freedom $\sigma$ and $\tau$ can be any of the $n!$ members of the permutation group defined over $n$ elements, $\mathcal{P}_n$.
The tensor $\hat{\sigma}$ describes different ways (permutations) to contract indices
\begin{align}
    \hat{\sigma}_{\mathbf{ab}} = \prod_{k=1}^n \delta_{a_k,b_{\sigma(k)}}.\label{eqn:permutation_tensor}
\end{align}
The coefficient $w_g^{(n)}(\sigma,\tau)$ is the Weingarten function for unitary group $U(q^2)$ that depends only on $\sigma \tau^{-1}$.
An exact expression for $w_g^{(n)}$ is known~\cite{collins2003moments,collins2006integration,zinn2010jucys,novak2010jucys,matsumoto2013weingarten,novaes2014elementary}, and
in the limit of a large $q$~\cite{collins2003moments}, it reduces to
\begin{align}
w_g^{(n)}(\sigma, \tau) = \left\{ \begin{array}{cc}
    1/q^{2n} & \textrm{if }  \sigma = \tau \\
    \smallo(1/q^{2n+1}) & \textrm{if } \sigma \neq \tau 
\end{array}\right. . \label{eqn:weingarten_n}
\end{align}
Analogous to the case of $n = 2$, the application of Eq.~\eqref{eqn:unitary_n_design} 
to every gate in the RUC leads to an intermediate ``spin model'' defined on a honeycomb lattice:
\begin{align}
    \overline{\mu_{AM}^{(n)}} = \sum_{\{\sigma_{\bm{r}}, \tau_{\bm{r}}\}} W(\{\sigma_{\bm{r}}, \tau_{\bm{r}}\}),
    \label{eqn:partition_func_n}
\end{align}
where the weight $W(\{\sigma_{\bm{r}}, \tau_{\bm{r}}\})$ can be evaluated by decomposing the tensor network into smaller diagrams.
In the bulk, closed diagrams associated with diagonally neighboring spin pairs take the value (see  Appendix~\ref{app:weights})
\begin{equation}
    w_d^{(n)}(\sigma, \tau) = q^{\textrm{\#cycle}(\sigma\tau^{-1})}\cos^{2n}\alpha + q \sin^{2n}\alpha,
\label{eq:weight_diag_n}
\end{equation}
where $\textrm{\#cycle}(\xi)$ denotes the number of cycles in the cyclic representation of a permutation $\xi \in \mathcal{P}_n$~\footnote{For example, when $\sigma = \tau$, $\sigma\tau^{-1} = (1)(2)\cdots(n)$ in the cyclic representation, hence $\textrm{\#cycle}(\sigma\tau^{-1}) = n$.}.
The top boundary contractions are equivalent to introducing an additional layer of tensors $\hat{\sigma}$ at time $T+1$.
The values of the variables $\sigma_{x,T+1}$ are fixed according to the top boundary operator.
For example, $\mu_{AM}^{(n)}$ requires $\hat{\sigma}_{x,T+1}$ to be the cyclic permutation and identity operators in the subsystem $A$ and $B$, respectively.
At the bottom boundary, extra weights arise from the overlap between different initial states $\prod_{k = 1}^n \braket{ \psi_{\sigma(k), x}}{\psi_{k, x}}$ at each site $x$.

Integrating out $\tau$ variables leads to a spin model on a triangular lattice:
\begin{align}
\label{eqn:Potts_model_n}
    \overline{\mu_{AM}^{(n)}} = 
    \mathcal{Z}^{(n)}_{AM} =
    \sum_{ \{\sigma\} }
    \sideset{}{'}\prod_{\langle \sigma_1,\sigma_2,\sigma_3 \rangle} \bar{w}^{(n)}(\sigma_1,\sigma_2,\sigma_3).
\end{align}
Unlike the case of $n = 2$, the three-body weight $\bar{w}^{(n)}$ may still be negative for a certain $q$ and $\alpha$.
Nevertheless, we find that for any given $n$, there exists a finite range of $(q, \alpha)$ for which the weight is nonnegative.
More specifically, we introduce $\kappa = q^{n-1}\cot^{2n}(\alpha)$ and show that the weights are all positive provided $q^2 > n$ and 
\begin{align}
    \frac{1}{n!}\left(\frac{1}{1+\kappa}\right)^2 > \left(\frac{q^2 + n}{q^2 - n} \right)^n - 1.\label{eqn:validity}
\end{align}
The proof is given in Appendix~\ref{app:proof_validity}.
For a given $n$, the inequality is generally satisfied when fixing $\kappa$ and considering a large $q$ regime, where the right-hand side can be arbitrarily small.

Our spin model exhibits the $\mathcal{P}_n \times \mathcal{P}_n$ symmetry --- the system is invariant under the transformation $\sigma \mapsto \xi_1 \circ \sigma \circ \xi_2 $ for any pair $(\xi_1,\xi_2) \in \mathcal{P}_n \times \mathcal{P}_n$. This can be already seen from $w_g^{(n)}$ and $w_d^{(n)}$, which depends only on the conjugacy class of $\sigma\tau^{-1}$.
At a sufficiently high temperature, e.g. $\pi/2-\alpha \ll 1$, the model is in the disordered (paramagnetic) phase.
As the effective temperature is lowered by varying $q$ and $\alpha$, the system may spontaneously breaks the symmetry and undergoes a transition into a ferromagnetic phase.
In the limit $q\to \infty$, the ordering transition can be quantitatively investigated as the symmetry of the spin model extends to the Potts symmetry $\mathcal{P}_{n!}$ for $n\geq 3$, for which the ordering transition has been well studied.
Crucially, the phase transition point lies within the validity regime of our mapping in Eq.~\eqref{eqn:validity}.

\subsection{\texorpdfstring{$\tilde{S}^{(n)}$}{tilde $S^{(n)}$} and $D^{(n)}$ in spin model descriptions}
\label{sec:sm_op}
We now present the explicit spin model description of our generalized quantities, $\tilde{S}^{(n)}$ and $D^{(n)}$ (see Table~\ref{tab:dict} for a summary of results). We also discuss their qualitative behaviors in the two different phases.
\subsubsection{The $n$-th generalized conditional entropy} \label{sec:sm_bdry_cond_etrp}

The $n$-th generalized conditional entropy $\tilde{S}^{(n)}$ in  Eq.~\eqref{eq:cond_etrp_n} can be rewritten in a more suggestive form:
\begin{align}
    \tilde{S}^{(n)}(A|M) = \frac{1}{n-1} \left( F_{AM}^{(n)} - F_{M}^{(n)} \right),
\end{align}
where $F_{X}^{(n)} \equiv -\log \mathcal{Z}_{X}^{(n)}$ is the free energy of the classical spin model with the appropriate boundary condition for subsystem $X$.
Specifically, in the first term $F^{(n)}_{AM}$, the top boundary variables are fixed to one of the two distinct elements of $\mathcal{P}_n$ -- a cyclic permutation for spins positioned inside subsystem $A$ and an identity element for those in $B$. In the second term, all variables on the top boundary are uniformly set to the identity element.
Therefore, $\tilde{S}^{(n)}$ corresponds to the excess  free energy  associated with a domain wall terminating on the top boundary at the interfaces between regions $A$ and $B$ (up to a constant factor, $n-1$). 
%We note that the specific choice of the boundary variables is not important as the system exhibits the permutation symmetry.  

The free energy cost for having a domain wall is qualitatively different in the two phases of the spin model.
In the ferromagnetic phase, the excitation of a domain wall requires an energy proportional to its length.
Pinning inhomogeneous boundary conditions results in two ordered domains of spin variables, whose interface scales with the linear size of a subsystem.
In this phase, $\tilde{S}^{(n)}$ thus also grows linearly with the subsystem size, corresponding to the volume-law entangling phase.
In contrast, domain wall excitations are condensed in the paramagnetic phase and do not cost extensive free energies.
Hence, in-homogeneous boundary conditions lead to a free energy change of at most order one, leading to the area-law scaling of $\tilde{S}^{(n)}$. 

It is interesting to note the close relation between the statistical mechanics interpretation of the entanglement entropy of the time-evolving state in the circuit to holographic entanglement entropy encoded in a random tensor network. In Ref.~\cite{vasseur2018entanglement}, the latter was similarly mapped to the free energy of a domain wall in a classical spin model.  The entanglement phase transition in both cases corresponds to a ferromagnetic transition affecting a change in the scaling of the domain wall free energy. 

\subsubsection{The $n$-th generalized KL divergence and Fisher~information}
The $n$-th generalized KL divergence $D^{(n)}$ can be rewritten as
\begin{align}
    D^{(n)}(P_0||P_\theta) = \frac{1}{n-1}\left( F^{(n)}_{M,\theta} - F^{(n)}_{M} \right),
\end{align}
where $F^{(n)}_{M,\theta} = -\log \mathcal{Z}_{M,\theta}^{(n)}$ with $\mathcal{Z}^{(n)}_{M,\theta} = \overline{\tr \tilde{\rho}_{M,0}\tilde{\rho}_{M,\theta}^{n-1}}$.
Similar to the case of $\tilde{S}^{(n)}$, $D^{(n)}$ also corresponds to the free energy difference in spin models with two distinct boundary conditions.
In this case, however, the boundary conditions are identical at the top and distinguished only at the bottom, originating from the different initial states of the $n$ density matrices.
More specifically, in order to compute $F_{M,\theta}^{(n)}$, we consider the system initialized in $\ket{\Psi_0}$ for the first copy (or equivalently any one of the $n$ copies) and in $\ket{\Psi_\theta}$ for the rest. 
Following our derivation in the preceding section, this leads to nontrivial weights $W_\textrm{bottom}$ associated with the contractions of $\hat{\sigma}$ at the site being perturbed: for any $\sigma \in \mathcal{P}_n$ that permutes the first copy to itself, the weight is trivial unity, and otherwise it contributes a factor  $|\bra{\psi_0}\delta U(\theta)\ket{\psi_0}|^2=\cos^2\theta$ to $W_\textrm{bottom}$.
Among total $n!$ permutations of $\mathcal{P}_n$, $(n-1)!$ elements belong to the former case (up-type) and the remaining $n!-(n-1)!$ elements constitute the latter case (down-type).
Hence, the bottom boundary weights can be interpreted as an effect of a boundary ``magnetic field'' that distinguishes the down-type spin variables from the up-type ones at site $x_0$. $D^{(n)}$ corresponds to the free energy cost of applying the boundary field at site $x_0$ up to a constant prefactor.

The qualitative behaviors of $D^{(n)}$ in two different phases of the spin model can be intuitively understood by the $n$-th Fisher information:
\begin{align}
    \mathcal{F}^{(n)} = \frac{2}{n-1}\left\langle m_{x_0, 1}^\downarrow \right\rangle,
    \label{eqn:Fn}
\end{align}
where $\langle m_{x_0,1}^\downarrow \rangle$ is the probability that the spin at $(x_0,1)$ belongs to a down-type in the spin model with open bottom boundary conditions.\footnote{We consider the open boundary condition since $\mathcal{F}^{(n)}$ is obtained in the limit $\theta \to 0$.}
Owing to the translation symmetry of our model (after averaging over $\mathcal{U}$) and top boundary conditions, $\langle m_{x_0,1}^\downarrow \rangle$ is independent of $x_0$, hence we use a simpler notation $\langle m_1^\downarrow \rangle$ in the rest of the paper.

For a relatively short evolution time $T$ (i.e., the temporal dimension of the spin system is short), the density of down-type spins, $\langle m_1^\downarrow \rangle$, is smaller than its na\"ive statistical average:
\be
\langle m^{\downarrow}_1\rangle<{n !-(n-1) ! \over n!}=1- 1/n 
\ee
due to the effect of the symmetry-breaking top boundary conditions at $t=T+1$ (see Monte Carlo results in Appendix~\ref{app:mc_distinguishability}).
This memory of the top boundary conditions persists in the ferromagnetic phase even in the long-time limit, so $\lim_{T\rightarrow \infty} \langle m_1^\downarrow \rangle < 1-1/n$ due to the long range order. On the the other hand, in the paramagnetic phase. the state of the bottom boundary becomes uncorrelated with the top boundary; the number of down-type spins saturates to its simple statistical average $1-1/n$.
Therefore, through  Eq.~\eqref{eqn:Fn}, the bulk phase transition of the spin model manifests as a nonanalytic change in the long-time limit of the $n$-th Fisher information at the critical point, which reaches its fully saturated value $\mathcal{F}^{(n)}={2/n}$ only in the paramagnetic phase. In the replica limit $n\to 1$, we get the saturation value $\mathcal{F}=2$ (see Fig.~\ref{fig:fig1}).

The saturation of the Fisher information to its maximal value in the paramagnetic phase (or equivalently area-law phase) indicates that the measurement device (ancilla qudits) obtains \emph{all} the information about the initial state, accessible from a set of local measurements.
In contrast, the saturation to a value less than its maximum in the volume-law phase implies that \emph{some} information cannot be extracted even after infinitely many measurements, owing to the effective quantum error correcting properties of scrambling dynamics~\cite{choi2019quantum}.
We note that in the special limit $n\rightarrow 1$, the nondecreasing nature of $D^{(1)}$ as a function of $T$ can be independently derived from the monotonicity of the relative entropy~\cite{araki1976relative,lindblad1975completely,uhlmann1977relative}.

\section{Phase transitions}\label{sec:pt}

In this section, we provide detailed analysis on the nature of the phase transitions in our models.
The phase transition points in our spin models are analytically solvable in two cases: (1)  $n=2$ with an arbitrary choice $q$, and (2) $n\geq 2$ in the limit $q \to \infty$.
The former case is discussed in Sec.~\ref{sec:pt_2}, where we utilize that the emergent 2D Ising model on a triangular lattice is exactly solvable~\cite{eggarter1975triangular,tanaka1978triangular}.
The phase transition is described by conformal field theory (CFT) with known critical exponents.
The latter case is elaborated in Sec.~\ref{sec:pt_largeq}, where we show that our spin model further reduces to the $n!$-state standard Potts model on a square lattice in the large $q$ limit.
The phase transition of such models can be exactly computed using the Kramers-Wannier duality~\cite{potts1952some,kihara1954statistics} and belongs the first order phase transition for $n\geq 3$.
Then, we perform the analytic continuation in order to identify the universality class and extrapolate the phase transition point for the $n=1$ case in Sec.~\ref{sec:pt_analytic_cont}.
In this limit, our model reduces to a bond percolation problem; the phase transition is described by CFT with well-known critical exponents.
In order to study the phase transition for small local Hilbert space dimensions, we perform exact numerical simulations up to $N=30$ qubits ($q=2$) and extract the critical point.
We also present simulation results for the entanglement phase transition in R\'enyi-$k$ entropies in Sec.~\ref{sec:pt_numerics}, which suggests that the phase transition point may not depend on the R\'enyi order $k$.

\subsection{Classical Ising model ($n =2$ with an arbitrary $q$)}\label{sec:pt_2}
When $n = 2$, the average purity is mapped to the partition function of 2D classical Ising model on a triangular lattice with appropriate boundary conditions.
The bulk Hamiltonian of such system in the unit of the effective temperature is
\begin{align}
    \beta\mathcal{H}_{\textrm{Ising}} = \sum_{\left< \bm{r}, \bm{r'}\right>_d} J_d \sigma_{\bm{r}}\sigma_{\bm{r'}} + \sum_{\left< \bm{r}, \bm{r'} \right>_h} J_h \sigma_{\bm{r}}\sigma_{\bm{r'}},\label{eqn:Ising_Ham}
\end{align}
where $\left< \bm{r}, \bm{r'}\right>_{d/h}$ represents a pair of diagonally/horizontally neighboring sites, and $J_d$ ($J_h$) is the corresponding Ising coupling strength that depends on $q$ and $\alpha$.
Exact expressions for $J_d$ and $J_h$ are provided in Appendix~\ref{app:Ising_coupling}. 
In our model, it can be shown that $J_d \leq 0$ is ferromagnetic, $J_h \geq 0$ is antiferromagnetic, and they satisfy $J_d + J_h \leq 0$ for arbitrary $q$ and $\alpha$.
This model is exactly solvable and exhibits a paramagnetic-to-ferromagnetic phase transition when $J_d$ and $J_h$ satisfy~\cite{eggarter1975triangular,tanaka1978triangular}:
\begin{align}
    2 e^{2 J_h} = e^{-2 J_d} - e^{2 J_d}.
\end{align}
From this condition, the critical measurement strength $\alpha_c^{(2)}$ can be computed for an arbitrary $q$
\begin{align}
    \tan^4\left(\alpha^{(2)}_c\right) = \frac{(q-1)(q^2+1)}{\sqrt{2 + 2q^4} - 2} - 1. \label{eq:alphac_2}
\end{align}
The universality of this phase transition is described by the Ising CFT with the critical exponents $\nu=1$ and $\beta = 1/8$.

\subsection{Standard Potts model ($n \geq 2$ with $q\to \infty$)}\label{sec:pt_largeq}

In the case $n \geq 3$, our interpretation of different $n$-th moments as partition functions of classical spin models is valid only for a finite range of $(q,\alpha)$.
A sufficient condition for the validity is provided in Eq.~\eqref{eqn:validity}, which can be satisfied by a finite $q$ for any given $n$.
In this section, we consider a limiting case $1/q\rightarrow 0 $, where our spin model description becomes valid for every $n$ and is further simplified, allowing extracting exact phase transition points.
We believe this analysis should provide insights to the phase transitions even when $1/q \neq 0$, where we expect small modifications to the critical measurement probability arising from corrections in powers of $1/q$.

We begin our analysis by pointing out that, in the limit $q \to \infty$, the phase transition in the spin model occurs close to $\alpha_c^{(n)} \sim \pi/2$ (or $p_c^{(n)} \sim 1$), i.e., the measurement strength is near its maximum.
This can be easily checked by estimating the couplings in the spin model as a function of $q$ and $\alpha$ (see Appendix~\ref{app:Ising_coupling} for an example of $n = 2$).
Therefore, we consider the limit $q \rightarrow \infty$ together with $\alpha \rightarrow \pi/2$ while keeping $\kappa = q^{n-1}\cot^{2n}(\alpha)$ fixed.
We will see that our model exhibits the phase transition point at $\kappa_c^{(n)} = \sqrt{n!}$, which is consistent with our limit as well as the validity criterion in Eq.~\eqref{eqn:validity}.

\begin{figure}[t!]
	\includegraphics[width=0.48\textwidth]{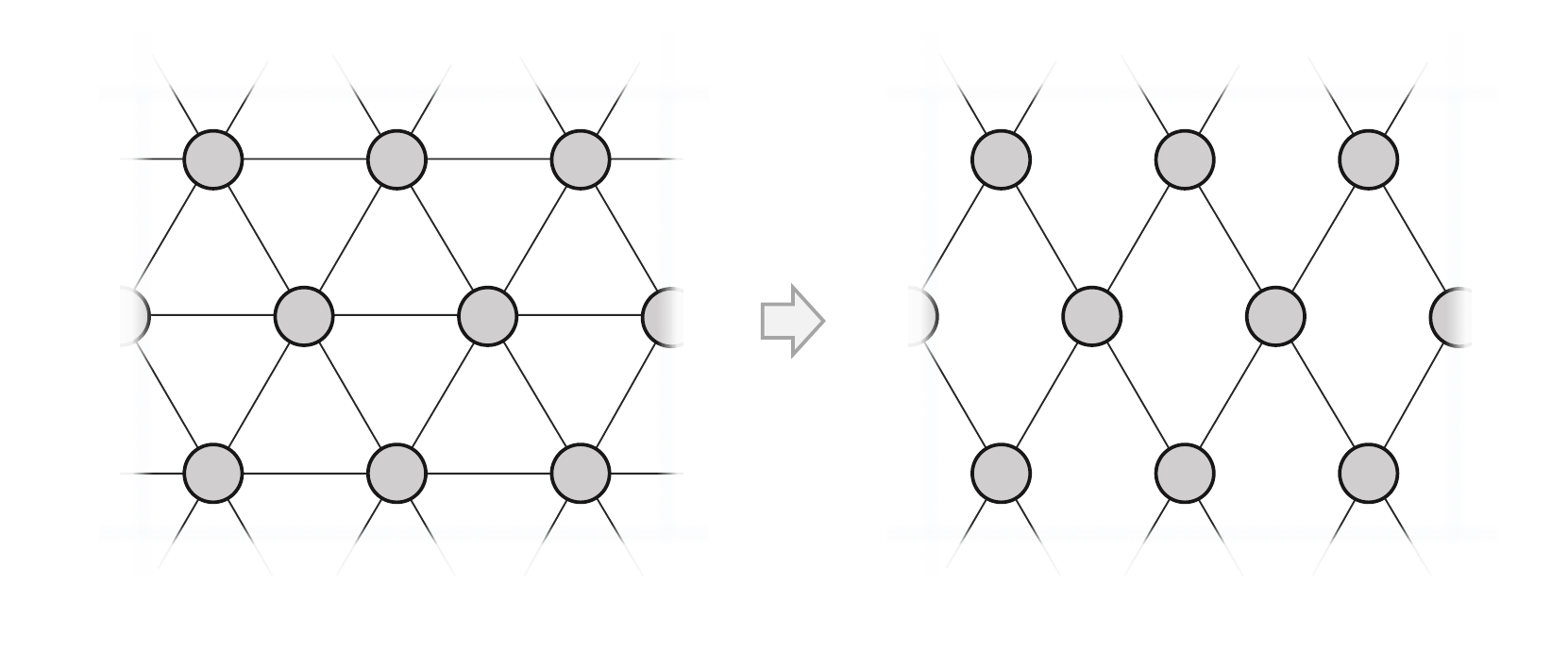}
	\caption{Our spin models reduce to the standard Potts models on a square lattice in the limit $q\rightarrow \infty$, $\alpha \rightarrow \pi/2$ while keeping $\kappa = q^{n-1}\cot^{2n}(\alpha)$ fixed (see the main text). The phase transition occurs at $\kappa_c^{(n)} = \sqrt{n!}$.}
	\label{fig:Potts_reduction}
\end{figure}

An important consequence of the limit $1/q\rightarrow 0$ with a fixed $\kappa$ is the emergence of the Potts symmetry $\mathcal{P}_{n!}$.
More specifically, both weights $w_d^{(n)}$ and $w_g^{(n)}$ become dramatically simplified from  Eqs.~\eqref{eqn:weingarten_n} and \eqref{eq:weight_diag_n}:
\begin{align}
    w_d^{(n)}(\sigma, \tau) &\approx q \left(1 + \kappa\delta_{\sigma\tau}\right), \\
    w_g^{(n)}(\sigma,\tau) &\approx q^{-2n} \delta_{\sigma,\tau},
\end{align}
which is invariant under any permutations of $n!$ elements in $\mathcal{P}_n$, hence the symmetry group is $\mathcal{P}_{n!}$.
By integrating out $\tau$ variables, we obtain the three-body weight and its corresponding effective Hamiltonian for spin variables on a triangular lattice:
\begin{align}
    \beta \mathcal{H}_{\textrm{Potts}} \approx \sum_{\left< \bm{r}, \bm{r'}\right>_d} J \delta_{\sigma_{\bm{r}} \sigma_{\bm{r'}}},\label{eqn:standard_potts}
\end{align}
where $J = -\ln(\kappa + 1) < 0$ is a ferromagnetic coupling between diagonally neighboring sites, and the couplings between horizontally neighboring sites vanish.
Thus, our model simplifies to the standard Potts model defined on a square lattice (Fig.~\ref{fig:Potts_reduction}), for which the phase transition point as a function of $J$ (in units of effective temperature) has been well studied~\cite{potts1952some,kihara1954statistics}.

\begin{figure}[t!]
	\includegraphics[width=0.48\textwidth]{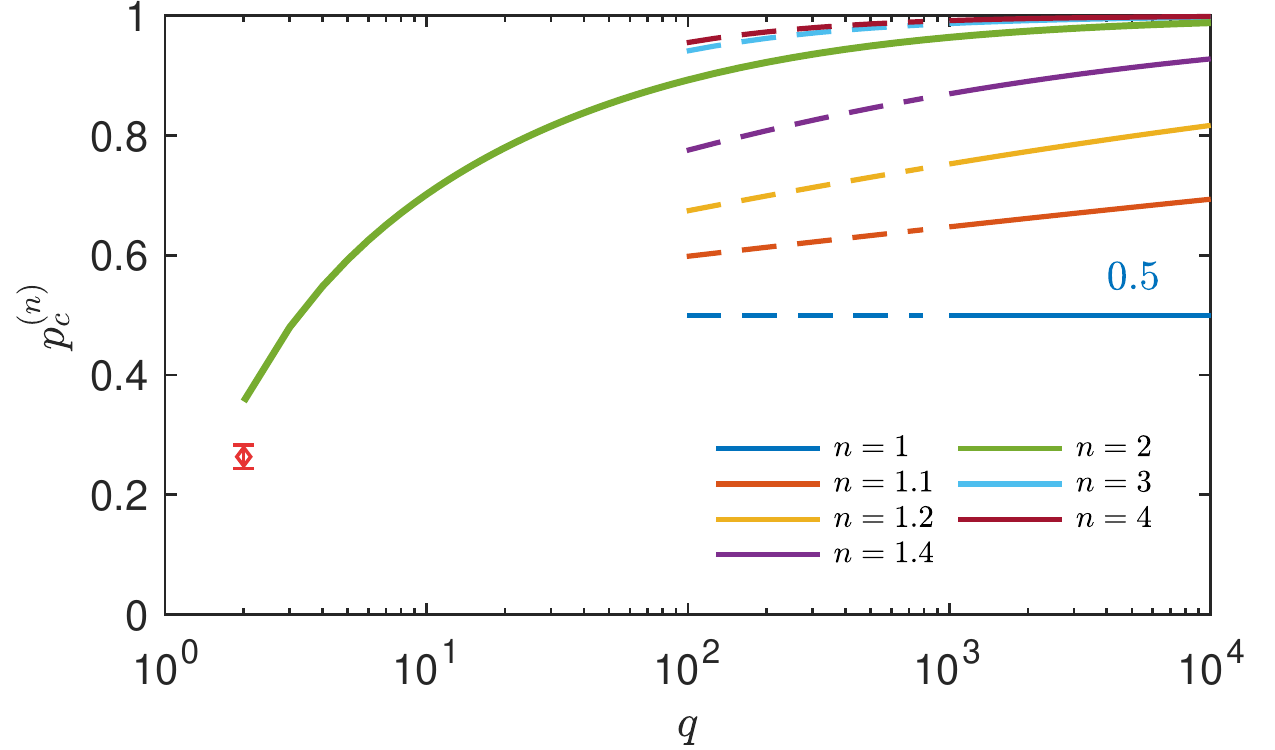}
	\caption{Phase transition points $p_c^{(n)}$ as a function of $q$. 
	Different curves represent various replica index $n$. For $n = 2$, the critical point is given by the exact solution in Eq.~\eqref{eq:alphac_2}.
	For integer $n \geq 3$, $p_c^{(n)}$ is estimated from  Eq.~\eqref{eq:critical_alpha}, which is valid for $q\gg 1$. For noninteger $n$ and $n = 1$, the transition points are extracted from the analytic continuation of Eq.~\eqref{eq:critical_alpha}.
	Solid lines represent the exact value ($n = 2$) or the results in the large $q$ regime ($n \geq 3$ and noninteger $n$) while dashed lines indicate that our approximation needs not be valid as $q$ decreases. 
	The red diamond indicates numerically extracted critical point $p_c = 0.26\pm0.02$ with $q = 2$ for von~Neumann entanglement entropy.
	}
	\label{fig:transition_points}
\end{figure}

The transition point of the standard Potts model is exactly solvable using the Kramers-Wannier duality~\cite{potts1952some,kihara1954statistics}, which gives $J_{c} = -\log(1+\sqrt{n!} )$, or equivalently $\kappa^{(n)}_c = \sqrt{n!}$. For completeness, we review the duality in Appendix~\ref{app:duality}.
This transition point corresponds to the strength $\alpha_c^{(n)}$ and the probability $p_c^{(n)}$:
\begin{align}
\alpha^{(n)}_c = \arctan{\left(\left( \frac{q^{n-1}}{\sqrt{n!}}\right)^{1/2n}\right)},\\
p_c^{(n)} = 1/\left( 1+ \left( \frac{\sqrt{n!}}{q^{n-1}}\right)^{1/n}\right)
\label{eq:critical_alpha}
\end{align}
for weak and projective measurements formalisms, respectively.
In the case $n = 2$, this critical point agrees with the exact solution in Eq.~\eqref{eq:alphac_2} in the large $q$ limit.
It is well known that the $n!$-state Potts model exhibits a first-order phase transition for $n\geq3$~\cite{baxter1973potts,baxter1978triangular}. A summary of $p_c^{(n)}$ as a function of $q$ for various $n$ is presented in Fig.~\ref{fig:transition_points}.

\subsection{Bond Percolation ($n\rightarrow 1$ and $q\to \infty$)}
\label{sec:pt_analytic_cont}
Our exact results when $q \to \infty$ in the preceding section allow one to perform the analytic continuation to study an important limiting case $n\rightarrow 1$, where our generalized quantities $\tilde{S}^{(n)}$ and $\mathcal{F}^{(n)}$ approach the averaged von~Neumann entanglement entropy and the Fisher information in measurement outcomes, respectively.
Here, we compute the critical point and identify the universality of class of the phase transition.

In the limit $q\rightarrow \infty$, we have seen that our model reduces to the $n!$-state standard Potts model on a 2D square lattice with the coupling $J = -\ln(\kappa + 1)$.
Taking another limit $n\rightarrow 1$, the standard Potts model becomes equivalent to a bond percolation problem~\cite{kasteleyn1969phase,wu1978percolation,cardy1992critical}, where each bond of a square lattice is ``activated'' with the probability 
\begin{align}
\label{eqn:activation_prob}
    f = \frac{\kappa}{1+\kappa} = p.
\end{align}
The set of activated bonds percolates the network of the 2D square lattice when $f$ exceeds the critical point $f_c=1/2$. 
In our model, this criterion leads to $\lim_{n\rightarrow 1}\kappa_c^{(n)}= 1$, corresponding to
\begin{align}
\label{eq:critical_alpha_n_1}
    \lim_{n\rightarrow 1} \alpha_c^{(n)} &= \pi/4,\\
    \lim_{n\rightarrow 1} p_c^{(n)} &= 1/2.
\end{align}
Thus, the universality class of the phase transition for $q\to\infty$ belongs to that of a bond percolation transition in the 2D square lattice.
At the critical point, the 2D model can be described by CFT with critical exponents $\nu = 4/3$ and $\beta = 5/36$~\cite{kesten1982percolation,kesten1987scaling,grimmett1999percolation}.

A few remarks are in order.
First, while the percolation model has been discussed in Ref.~\cite{skinner2018measurement}, we emphasize that the origin of the model in the present work is different.
In Ref.~\cite{skinner2018measurement}, the model has been adapted primarily to explain the phase transition in the R\'enyi-$0$ entropy, relying on simple geometric properties of tensor network representations of many-body wave functions.
In particular, the probability of projective measurement in RUCs has been directly identified with the activation probability in the percolation problem. 
In this paper, however, the percolation problem arises from the analytic continuation of emergent effective spin models that we obtain only after averaging over unitary gates and weak measurements. More importantly, we predict a percolation transition in the von Neumann entanglement entropy {\it only} in the limit $q\to\infty$ (i.e., large local dimension). As discussed in Ref.~\cite{vasseur2018entanglement}, the $1/q$ corrections give rise to a reduction of the Potts symmetry ($\mathcal{P}_{n!}\to \mathcal{P}_{n}\times \mathcal{P}_{n}$) and constitute relevant perturbations at the percolation fixed point. It is an interesting direction for future research to characterize the critical point in a system of qudits with a finite number of internal states and in particular qubits with $q=2$.
Second, in the previous studies~\cite{li2018quantum,li2019measurement,choi2019quantum}, the critical point and critical exponents of the entanglement phase transition in the Clifford circuit are also obtained from numerics. However, the critical point and the universality extracted from analytic continuation here may not be applied to the Clifford circuit owing to the fact that the Clifford group does not form a unitary ${t}$-design for $t \geq 4$~\cite{webb2015clifford}. Thus, Eq.~\eqref{eq:critical_alpha} is not valid for the phase transition in the Clifford circuit when $n \geq 4$.
Finally, we note that, in a strict sense, our analytic continuation is valid only when $q\rightarrow \infty$ since our analytic expressions for $\alpha_c^{(n)}$ (or $p_c^{(n)}$) are exact only in this limit. 
Surprisingly, we find that the resultant critical point does not depend on $q$ in its leading order. A careful quantitative study of potential $1/q$ corrections may be an interesting future direction.

\subsection{Exact Numerical Simulations}
\label{sec:pt_numerics}

\begin{figure}[t!]
    \centering
    \includegraphics[width=0.48\textwidth]{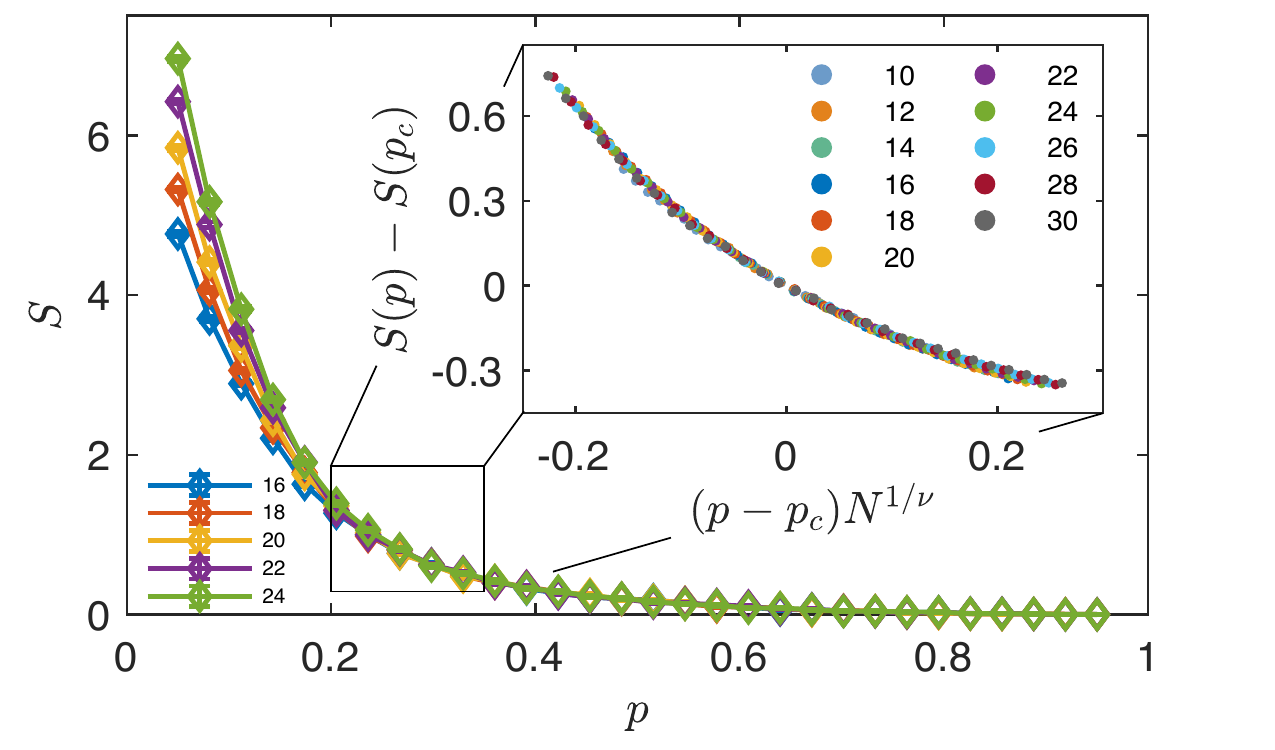}
    \caption{Exact numerical simulations for the entanglement phase transition in a RUC with probabilistic projective measurements. The horizontal and vertical axes represent the measurement probability $p$ and the half-chain von~Neumann entanglement entropy $S$ for $q = 2$ and various system sizes up to $N=24$.
    The inset shows the collapse of data from finite-size scaling analysis based on the scaling formula $S(p, N)-S(p_c, N) = g((p-p_c)N^{1/\nu})$~\cite{skinner2018measurement} with a specific choice $p_c = 0.27$ and $\nu = 2.9$.
    For the scaling analysis, we use the results from system sizes up to $N=30$ with $q = 2$ and a smaller window of $p\in [0.2, 0.35]$.}
    \label{fig:pt_vn}
\end{figure}

In order to see the behavior of the phase transition with a small value of $q$, we perform exact numerical simulations of a RUC with projective measurements for $q=2$ (qubits).
We are able to obtain exact results for up to $N=30$ spins by using a customized algorithm that leverages tensor representations of many-body wave functions,  (see Appendix~\ref{app:ED_numerics} for details).
Based on finite size scaling analysis for $10\leq N\leq 30$, we extract $p_c = 0.26 \pm 0.02$ for von~Neumann entropy, consistent with previous numerical results up to $N = 24$ qubits~\cite{skinner2018measurement} (Fig.~\ref{fig:pt_vn}). The extracted critical point seems to deviate from our analytic prediction for $q\to \infty$ limit in Eq.~\eqref{eq:critical_alpha_n_1}, which indicates that a significant $1/q$ correction is present.

We note that the replica limit introduced in the preceding sections is tailored for von~Neumann entropy.
For many practical applications, it is also important to understand the behaviors of R\'enyi entropies of order $k\neq 1$.
In particular, it is well known that quantum states in 1D with area-law R\'enyi-$k$ entanglement entropy with $k<1$ can be efficiently simulated by using matrix product state representations~\cite{verstraete2006matrix,schuch2008entropy}.
Therefore, the entanglement phase transition in R\'enyi-$k$ entropy with $k<1$ is directly related to the classical simulability of quantum dynamics.

\begin{figure}[t!]
    \centering
    \includegraphics[width=0.48\textwidth]{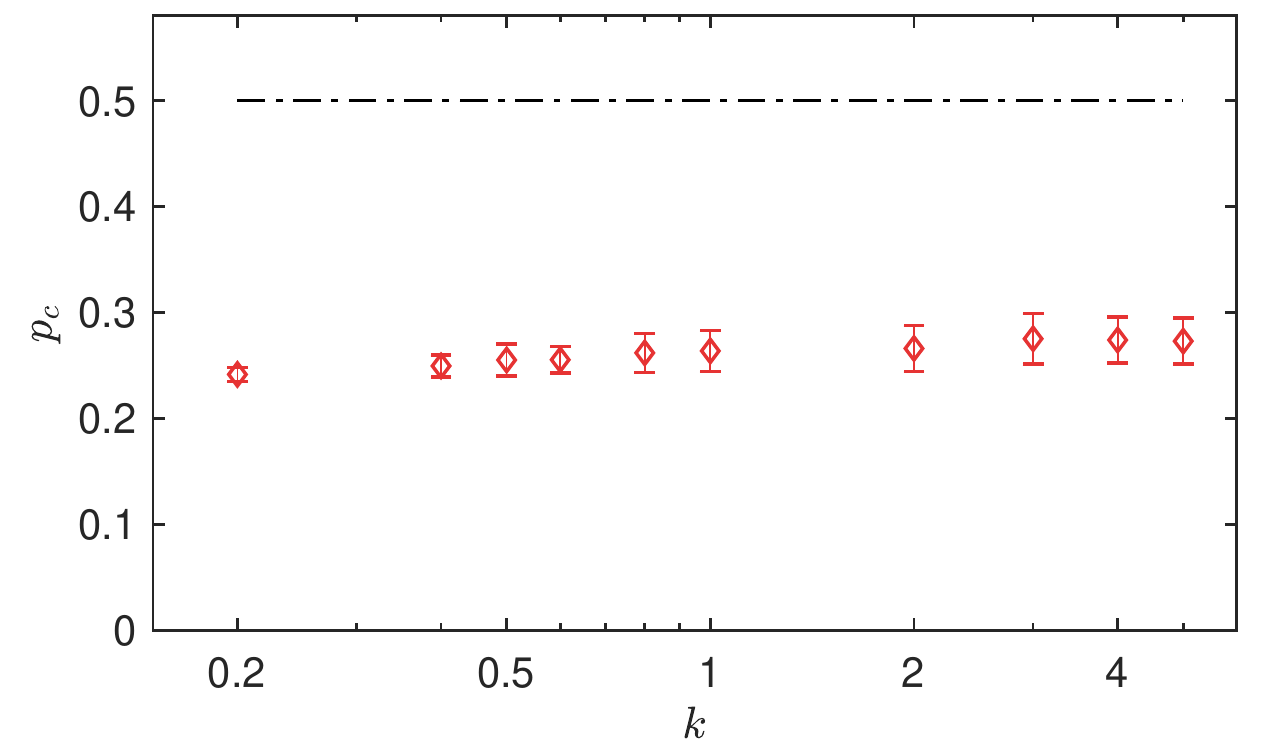}
    \caption{
    Critical measurement probabilities $p_c$ for different orders $k$ of R\'enyi entropy fitted from exact numerical simulations.
    The von~Neumann entropy corresponds to $k=1$.
    Red diamonds represent $p_c$ extracted from finite-size scaling analysis using the results from exact numerical simulation for $q = 2$ up to $N = 30$ and $p \in [0.2, 0.35]$.
    Error bars are estimated from statistical fluctuations of our scaling analysis based on the bootstrapping method and may not represent the accuracy to their true value (see Appendix~\ref{app:ED_numerics_fss}). 
    The dash-dotted line shows the analytic prediction $p_c = 1/2$ in the large $q$ limit as well as the critical point for R\'enyi-$0$ entropy based on the percolation in the unitary network~\cite{skinner2018measurement}.
    }
    \label{fig:crit_pt}
\end{figure}

With this motivation in mind, we numerically study the phase transitions in R\'enyi-$k$ entropy for various $k$ ranging from $0.2$ to $5$ and extract critical measurement probabilities.
The results are summarized in Fig.~\ref{fig:crit_pt}.
Interestingly, $p_c$ does not exhibit a strong dependence on the order $k$.
This suggests that phase transitions for different R\'enyi entropies ($k>0$) occur simultaneously and that the dynamics in the area-law phase for von~Neumann entropy already implies its simulability using a classical algorithm. 

\section{Absence of phase transition for unrestricted measurements}
\label{sec:quantum}
In previous sections, we focused on the RUC with \emph{simple} weak measurements, in which the ancilla qudits are measured only in their local computational basis.
In this section, we lift this constraint, allowing measurements of any complete set of observables (positive-operator valued measure), which could be arbitrarily nonlocal. 
Thus, we must now consider the full quantum density matrix of the ancilla qudits $\rho_M$ without applying the dephasing channel, i.e., without projecting to its diagonal elements in the computational basis.
With the full density matrix in hand, we can in principle apply, for example, an arbitrary unitary $V$ to the entire set of $NT$ ancilla qudits before projecting them in the computational basis.
By appropriately choosing $V$, this setup allows one to extract more information about the system including nonlocal correlations (in both space and time) hidden among multiple system qudits.
We show that there is no phase transition in the presence of such unrestricted (fully quantum) measurements. 
The absence of phase transition under these conditions can be understood from two complementary perspectives.
The first perspective is based on the insights of Ref.~\cite{choi2019quantum}, namely that the volume-law phase remains stable against sufficiently sparse measurements owing to the natural error correction achieved by the information scrambling of unitary gates. The scrambling transforms information into a highly nonlocal form, protecting the information from local measurements.
However, if the measurements are also arbitrarily nonlocal, then information is no longer protected. 
The scrambling becomes meaningless, and the volume-law phase immediately destabilizes.

Another perspective on the problem is afforded by considering the entanglement entropy between the system and ancilla qudits.
Through the coupling gates $\hat{R}_\alpha$, the entanglement entropy generally grows linearly in time as $S[\rho_S] \sim p NT \log q $ and is expected to saturate to its maximal possible value $N\log q$, after a time $T^*\sim 1/p$. 
In this situation, as we show in Sec.~\ref{sec:system_rho} below, the system qudits reach a steady state (at time $T\gg T^*$), which retains no information about their initial state. Therefore, given the unitarity of the combined system, all the information must have gone into the ancilla qudits~\cite{schumacher2002approximate}.
For this reason, we expect no phase transition: the ancilla qudits will eventually attain the full information about the initial state of the system, regardless of the measurement strength.

It is interesting to see how the presence of unrestricted measurements modifies the mapping to the classical spin models. Below, we show that removing the constraint imposed by the dephasing channel results in classical spin models that do not respect the permutation symmetry nor the Potts symmetry when $q\to \infty$, and consequently do not exhibit an ordering phase transition.

\subsection{Dynamics of quantum relative entropy of ancilla qudits}\label{sec:quantum_rl_etrp_no_pt}
To characterize the amount of information stored in the full density matrix of the measurement device (ancilla qudits), 
we generalize the classical KL divergence to the quantum relative entropy
\begin{align}
    D_\mathcal{Q}(\theta, T) \equiv \overline{\tr\rho_{M,0}\left(\log\rho_{M,0} - \log\rho_{M,\theta}\right)}.
\end{align}
Here, there is no dephasing of the ancilla qubits. From the quantum relative entropy, one can readily compute the Kubo-Mori-Bogoliubov (KMB) quantum Fisher information defined as 
\begin{equation}
    \mathcal{F}_{\textrm{KMB}} \equiv \partial_{\theta}^2 D_{\mathcal{Q}}(\theta, T)\Big|_{\theta = 0},
\end{equation}
which is an attainable upper bound to the (conventional) Fisher information, optimized over all possible choices of positive-operator valued measure~\cite{uhlmann1977relative,hiai1991proper,ogawa1999strong,hayashi2001asymptotics,nagaoka2005fisher} (see Appendix~\ref{app:Fisher_info} for a review). Thus, $D_\mathcal{Q}$ is an appropriate quantum generalization of $D_{\textrm{KL}}$.

In order to analyze these quantities in the classical spin model description, we again use the replica method. 
To this end, we introduce the $n$-th quantum relative entropy and the corresponding $n$-th KMB Fisher information:
\begin{align}
    D_{\mathcal{Q}}^{(n)}(\theta, T) &\equiv \frac{1}{1-n} \log \left(\frac{\overline{\tr\rho_{M,0}\rho_{M, \theta}^{n-1}}}{\overline{\tr\rho_{M,0}^n}} \right),\\
    \mathcal{F}_{\textrm{KMB}}^{(n)} &\equiv \partial_\theta^2 D_{\mathcal{Q}}^{(n)}(\theta, T)\Big|_{\theta = 0},
\end{align}
which reduce to $D_\mathcal{Q}$ and $\mathcal{F}_{\textrm{KMB}}$, respectively, in the limit $n \rightarrow 1$.

The mapping to classical spin models described in Sec.~\ref{sec:sm} generalizes to this case in a straightforward way. 
The $n$-th moment $\nu^{(n)}_M = \tr[\rho_{M,1}\rho_{M,2}\cdots\rho_{M,n}]$ is mapped to the partition function of a modified classical spin model after averaging over unitaries. As before, in the limit $\theta^2 \ll 1$, the leading order contribution to $D^{(n)}_\mathcal{Q}$ can be interpreted as the density of down-type spins in the bottom layer $\langle m_1^\downarrow\rangle$ up to a constant prefactor.
The only modification to our spin model originates from the absence of the dephasing channels $\mathcal{N}_\phi$, which modifies the two-body Boltzmann weight $w_d^{(n)}(\sigma,\tau)$.
Crucially, this modification breaks the permutation symmetry (see Appendix~\ref{app:quantum}).
For this reason, the ferromagnetic to paramagnetic phase transition cannot exist.

In order to further develop the intuition, we focus on the second quantum relative entropy $D^{(2)}_\mathcal{Q}$ as a specific example, and explicitly demonstrate the absence of the phase transition verified with numerical simulations.
Similar to the discussion in Sec.~\ref{sec:sm_mapping}, the second moment $\nu_M^{(2)}$ is mapped to the partition function of classical Ising model on a triangular lattice (see Appendix~\ref{app:quantum} for detailed derivations).
In this case, the down-type spin is identified with a spin with $\sigma=-1$, hence $\langle m_1^\downarrow \rangle = \langle m_1^{-} \rangle$.

\begin{figure}[t!]
	\includegraphics[width=0.48\textwidth]{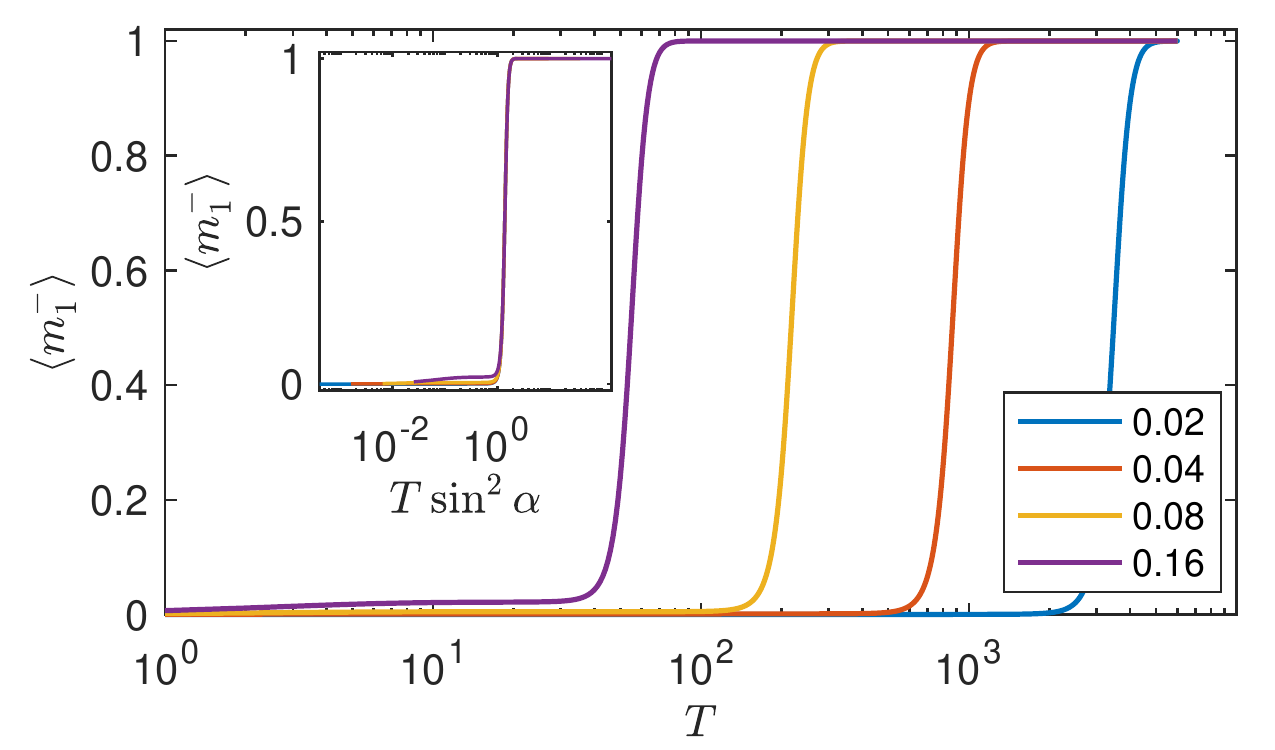}
	\caption{
	The density of spins with $\sigma = -1$ in bottom layer,  $\langle m_1^- \rangle$, as a function of time $T$ when the full density matrix of ancilla qudits is retained, i.e., without projection to diagonal elements.
	Different curves represent various measurement strengths $\alpha$ from $0.02$ to $0.16$ given in the legend.
	The data collapse shown in the inset clearly demonstrates the absence of phase transition. The time dependence of $\langle m_1^-\rangle$ for all measurement strengths collapses onto a single curve when plotted as a function of $T\sin^2\alpha$.
	The results are obtained from exact numerical simulation using the transfer matrix method with $N = 20$. 
% 	Additional numerical results for different $N$ collapse on top of each other and therefore are not presented here.
	}
	\label{fig:quantum_transient}
\end{figure}

Crucially, the Ising symmetry is explicitly broken: 
the three-body weights $\bar{v}^{(2)}(\sigma_1, \sigma_2, \sigma_3)$ are  larger when the majority of spins are $\sigma = -1$ (rather than $\sigma = +1$), i.e.,
$\bar{v}^{(2)}(+,-,-)=\bar{v}^{(2)}(-,+,-) > \bar{v}^{(2)}(+,-,+) = \bar{v}^{(2)}(-,+,+)$ and $\bar{v}^{(2)}(-,-,-) > \bar{v}^{(2)}(+,+,+)$ for an arbitrary $\alpha > 0$.
These imbalances in weights prefer spin variables $\sigma =-1$ over $\sigma=+1$.
Furthermore, $\bar{v}^{(2)}(-,-,+) =0$ independent of $\alpha$ while $\bar{v}^{(2)}(+,+,-) > 0$.
This implies that if $\sigma = -1$ at every position at a certain time $t^*$, then all the spins at $t< t^*$ must also be in the state $\sigma = -1$.
This constraint originates from the unitarity of the dynamics of system and ancilla qudits~\cite{nahum2018operator} (see also Appendix~\ref{app:unitarity}).
Therefore, we expect $\langle m_1^-\rangle$ to saturate to unity in the limit $T\rightarrow \infty$.
Numerics provided in Fig.~\ref{fig:quantum_transient} verifies this statement.

Figure~\ref{fig:quantum_transient} presents the exact simulation of $\langle m_1^- \rangle$ as a function of time $T$ using the transfer matrix method for various measurement strengths $\alpha$ from $0.02$ to $0.16$. 
$\langle m_1^- \rangle$ remains small for a long time and rapidly increases to unity.
%at a sufficiently long time for various measurement strengths $\alpha$ from $0.02$ to $0.16$. 
In the inset, we obtain a collapse of curves for various measurement strengths when $\langle m_1^- \rangle$ is plotted as a function of rescaled time $T\sin^2\alpha$.
% The quadratic scaling agrees with the intuition that the increase of $D_{\mathcal{Q}}^{(2)}$ (or Fisher information) in every time step is proportional to the average number of measured qudits $ N \sin^2(\alpha)\approx N \alpha^2$.
% This scaling also suggests that $\langle m_1^- \rangle$ will eventually saturate to unity at a time scale $T^* \sim 1/\alpha^2$ for $0< \alpha \ll 1$.
The scaling suggests that $\langle m_1^- \rangle$ approaches close to the saturation value at a timescale $T^*$, when every system qudit is measured once on average, i.e., $T^*\sin^2\alpha = T^* p = \BigO (1)$.
% At the time scale $T^*$, the entanglement between the system and ancilla qudits reaches its maximum value and the ancilla qudits obtained the information encoded in the initial state of system qudits.
Note that this timescale $T^*$ coincides with the timescale at which the system qudits achieve maximal entanglement with the ancilla qudits as estimated earlier in this section.
%This quantitatively corroborates our interpretation that 

%Since the time scale $T^*$ is also when the system qudits become maximally entangled with the ancilla qudits, the absence of the phase transition in $\langle m_1^- \rangle$ (or equivalently $D_\mathcal{Q}^{(2)}$ and $\mathcal{F}_\textrm{KMB}^{(2)}$) can be understood from the fact that ancilla qudits always obtain the full information encoded in the initial state of system qudits.
% This indicates that the saturation of $\langle m_1^- \rangle$ happens when the ancilla qubits obtained the information encoded in the initial state of system qudits.

%
The dynamics of $\langle m_1^- \rangle$ is reminiscent of the information dynamics in the Hayden-Preskill protocol in black holes~\cite{hayden2007black}.
The initial state of system qudits (namely, a newly born black hole) encodes Alice's diary, and the ancilla qudits act as the Hawking radiation collected by an external observer, Bob.
Bob cannot decode Alice's diary until the black hole emits sufficiently large amount of Hawking radiations such that the remaining black hole is maximally entangled with the early Hawking radiation, at which point Bob suddenly becomes able to decode the diary by collecting a few more bits of radiation.
%Since the entropy between the system and ancilla qudits eventually grows to its maximum value for any nonvanishing measurement strength, no phase transition happens in $\langle m_1^- \rangle$ (or equivalently $D_\mathcal{Q}^{(2)}$ and $\mathcal{F}_\textrm{KMB}^{(2)}$).
%The absence of phase transition in $D_\mathcal{Q}$ (or equivalently $\mathcal{F}_\textrm{KMB}$) will be further demonstrated after we show the density matrix of system qudits evolves to the maximally mixed state in the next section.

% We note that the absence of phase transition in $D_\mathcal{Q}^{(n)}$ for $n \geq 3$ cannot be rigorously demonstrated following the same argument due to the presence of negative three-body weights. However, it will be clear that no phase transition happens in $D_\mathcal{Q}^{(n)}$ and $D_\mathcal{Q}$ (or equivalently $\mathcal{F}_\textrm{KMB}^{(n)}$ and $\mathcal{F}_\textrm{KMB}$) after we show the density matrix of system qudits evolves to the maximally mixed state in the next section.

% The absence of a phase transition in $D_\mathcal{Q}^{(2)}$ can be generalized to $n \geq 2$. As discussed in Appendix~\ref{app:quantum}, in the spin model description of $D_\mathcal{Q}^{(n)}$, the cyclic permutation is preferred in the spin configuration.
% Moreover, the three-body weight $\bar{v}^{(n)}(\mathcal{C}^{(n)},\mathcal{C}^{(n)},\sigma_3) = 0$ for any $\sigma_3 \neq \mathcal{C}^{(n)}$, where $\mathcal{C}^{(n)}$ is the cyclic permutation [Fig.~\ref{fig:top_boundary_cond}(a)]. 
% Therefore, one layer of $\sigma = \mathcal{C}^{(n)}$ at a certain time $t^*$ will force all the spins at time $t < t^*$ to take the value $\mathcal{C}^{(n)}$.
% For a sufficiently large $T$, spins in the bottom layer are all polarized to $\sigma = \mathcal{C}^{(n)}$, and $\langle m^\downarrow_{1}\rangle = 1$ since $\mathcal{C}^{(n)}$ belongs to the down-type.
% Consequently, there cannot be a phase transition in $D_\mathcal{Q}^{(n)}$ for any integer $n \geq 2$ and nor for $D_\mathcal{Q}$ (or $\mathcal{F}_{\textrm{KMB}}$).

\subsection{Dynamics of the system qudits}\label{sec:system_rho}

In the preceding section, we have shown the absence of a phase transition based on the amount of information accessible to ancilla qudits when the measurement basis is unrestricted and potentially nonlocal.
Here, we explore the complementary perspective by considering the reduced density matrix of the system qudits.
We will see that the system density matrix evolves to a maximally mixed steady state independent of its initial state for any $\alpha > 0$~\cite{li2018quantum,skinner2018measurement}.
This observation, together with results in preceding sections, provides a clear intuitive understanding regarding the flow of quantum information: all information about the initial state of the system is transferred to ancilla qudits at a sufficiently long time for $\alpha >0$.
Furthermore, we point out that the presence or the absence of the phase transition cannot be unambiguously answered by the reduced density matrix of the system alone because it does not depend on the choice of measurement basis for ancilla qudits.
One of the most important implications of our results in this section is that the phase transition is inherent to the data collected on \emph{individual quantum trajectories} of the system, and it goes away when the quantum state of the system is averaged over those trajectories.

We consider the quantum relative entropy 
$\bar{D}_{\mathcal{Q}} \equiv \overline{\tr\rho_{S,0}(\log\rho_{S,0} - \log\rho_{S, \theta})}$
between the reduced density matrices of the system originating from two close initial states $\ket{\Psi_0}$ and $\ket{\Psi_\theta}$.
$\bar{D}_\mathcal{Q}$ can be similarly analyzed using the spin model description, and we show this quantity always decays to zero in the limit $T \rightarrow \infty$.
As before, we define the replicated objects $\bar{D}^{(n)}_{\mathcal{Q}} \equiv (\log\overline{\tr\rho_{S,0}\rho_{S,\theta}^{n-1}} - \log\overline{\tr\rho_{S, 0}^n})/(1-n)$, which is  proportional to the number of down-type spins in the bottom layer. 
An important modification arises from the top boundary conditions: the additional spin degrees of freedom for the system take the value $\mathcal{C}^{(n)}$ (i.e., $\sigma_{x,T+1} = \mathcal{C}^{(n)}$) everywhere, while $n$ copies of ancilla qudits are contracted by the identity operator $\mathcal{I}^{(n)}$ (in contrast to  $\mathcal{C}^{(n)}$ for $D_\mathcal{Q}^{(n)}$).
This leads to modified Boltzmann weights.
Here, we focus on the case of $n = 2$.
Instead of favoring $\mathcal{C}^{(2)}$, the spin model now favors $\mathcal{I}^{(2)}$ in the bulk for an arbitrary $\alpha > 0$.
Consequently, all the spins at the bottom boundary are polarized to $\mathcal{I}^{(2)}$, leading to a vanishing density of down-type spins as $T\rightarrow \infty$.
% for any integer $n \geq 2$.
% From the analytic continuation, we conclude that $\bar{D}_\mathcal{Q}$ vanishes as well.

The vanishing $\bar{D}_\mathcal{Q}^{(2)}$ implies that the density matrix of system qudits evolves to an identical steady state regardless of its initial state.
The steady state is indeed maximally mixed, which can be shown by considering one of the initial states being the maximally mixed state: the bottom boundary condition is modified in a nonperturbative way, but our argument above still holds.
Therefore, $\bar{D}_\mathcal{Q}^{(n)}$ ($n \geq 2$) and $\bar{D}_\mathcal{Q}$ decay to zero as $T \to \infty$.
% the quantum relative entropy $D_\mathcal{Q}$ eventually saturates to its maximal value for any nonvanishing $\alpha$.
% Since the maximally mixed initial state remains maximally mixed, the steady state of the system is maximally mixed regardless of its initial state.
From the information theoretical perspective, the system loses the quantum information of its initial state.
Since the system and ancilla qudits combined undergo a closed unitary time evolution, one can conclude that the full quantum information about the initial state is recoverable from the ancilla qudits~\cite{schumacher2002approximate}.
Finally, we note that the spin model descriptions of $D_\mathcal{Q}$ and $\bar{D}_\mathcal{Q}$ are identical up to the exchange of $\mathcal{C}^{(n)}$ and $\mathcal{I}^{(n)}$ in boundary conditions.
As a result, the saturation of $D_\mathcal{Q}$ and the decay of $\bar{D}_{\mathcal{Q}}$ occur exactly on the same timescale $T^*$.

\section{Discussion and outlook}\label{sec:discussion}
\begin{figure}[t!]
	\includegraphics[width=0.49\textwidth]{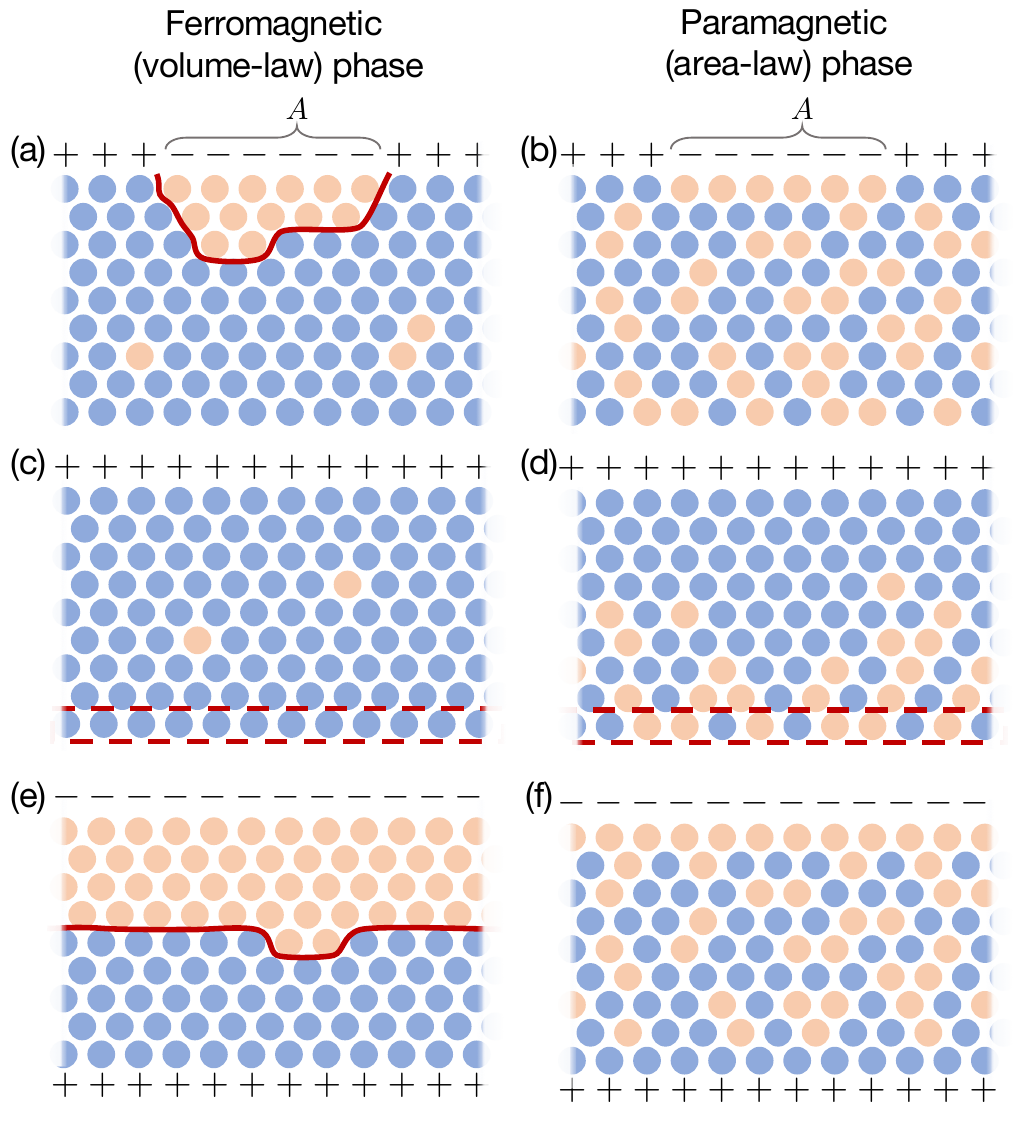}
	\caption{The different signatures of the entanglement phase transition as they manifest in the classical spin model descriptions. For simplicity, we present the descriptions in terms of the Ising spin model with $n=2$. Left and right columns correspond to the ferromagnetic (volume-law) and paramagnetic (area-law) phases.
	(a) and (b) Average entanglement entropy of a subsystem $A$ is related to the excess free energy of a domain wall (red solid line) terminating at the edges of $A$ on the top boundary.
	(c) and (d) Fisher information is related to the average magnetization density at the bottom (red dotted box), via Eq.~\eqref{eqn:Fn}, when the top boundary is fixed at $\sigma_{x,T+1} = +1$.
	(e) and (f) Purification of a mixed state evolution (or equivalently the average entropy of the system in steady states) is related to the excess free energy of a domain wall running across the entire system (red solid line).
	}
	\label{fig:spin_model_summary_diagrams}
\end{figure}

\subsection{Relation to purification phase
transition}\label{sec:purification}
Recently, Ref.~\cite{gullans2019dynamical} has pointed out that the entanglement phase transition occurs concurrently with the change in the purification dynamics of an initially mixed state.
More specifically, Ref.~\cite{gullans2019dynamical} considers the entropy of an initially maximally mixed state undergoing a RUC with projective measurements: when $p>p_c$ the quantum trajectories of the system density matrix rapidly approach pure states, while for $p<p_c$ the system remains in a mixed state with a finite entropy density for an exponentially long time in its system size.
Based on numerical simulations of 1D qubit chains evolved under Clifford gates and projective measurements, it has been observed that the critical measurement probability of the purification phase transition equals that of the entanglement phase transition with high accuracy~\cite{gullans2019dynamical}.

Using our mapping to a series of spin models, we can show that the purification phase transition is indeed identical to the entanglement phase transition for Haar random unitary circuits with projective or weak measurements. As a measure of the purity, we consider the von Neuman entropy of the full system. To this end, we consider the series of generalized conditional entropies  
$\tilde{S}_\text{mix}^{(n)}(A|M)$, with $A$ now identified with the {\it entire system} and the subscript `mix' indicates that the initial state is maximally mixed. As before, the special features of $\tilde{S}_\text{mix}^{(n)}(A|M)$ are enacted through bottom and top boundary conditions. For the spins in the bottom layer, contraction with the maximally mixed initial state  $\rho_\text{max}^{\otimes n} = (\frac{1}{q^{n}} \mathcal{I}^{(n)})^{\otimes N} $ can be accounted by introducing additional spins at $t = 0$ with fixed value $\sigma_{x, 0} = \mathcal{I}^{(n)}$ (identity permutation).  
In this case, since the subsystem $A$ covers the entire system (we are calculating the entropy of the whole system), the top boundary conditions are also homogenous. Similar to the discussion in Sec.~\ref{sec:sm_op}, the conditional entropy now corresponds to the difference between a configuration with all top spins fixed to $\mathcal{C}^{(n)}$ (cyclic permutation) and one with all of them in the $\mathcal{I}^{(n)}$ (identity permutation), while the bottom boundary condition is uniformly $\mathcal{I}^{(n)}$.
In the paramagnetic (area-law) phase, such free energy cost is of order unity independent of the system size. This implies a purified phase.
In the ferromagnetic (volume-law) phase, the excess free energy of a domain wall traversing the the entire system scales as $\sim N$. This corresponds to the mixed phase with total entropy proportional to volume.
The spin model descriptions of the subsystem entanglement entropy, the Fisher information, and the steady state entropy for a mixed initial state are summarized in Fig.~\ref{fig:spin_model_summary_diagrams} for $n=2$.

\subsection{Experimental considerations}\label{sec:exp_protocol}
There are several potential experimental platforms to investigate the phase transition, 
 including superconducting quantum circuits, trapped ions and neutral atoms, and ultracold  atomic systems~\cite{bernien2017probing,zhang2017observation,neill2018blueprint,harris2018phase,brydges2019probing,lukin2019probing}.
For theoretical convenience, we have focused on the entanglement phase transition in a circuit of random unitary gates.
However, we expect that this is not crucial and a system undergoing a chaotic hamiltonian evolution with measurements would also exhibit a similar phase transition.
In experiments, the projective measurements can be directly implemented when  quantum nondemolition (QND) measurements are possible.
Alternatively, one can introduce a set of ancilla qudits coupled to a system, postponing all measurements to the end of each experimental run of quantum dynamics.

For experimental observations of the phase transition, the biggest challenge in our view lies in identifying realistic observables that detect the phase transition.
As we discussed in the introduction, the direct measurement of the (conditional) entanglement entropy is extremely challenging, as it requires  
a large number of experimental repetitions that scales exponentially with the value of the entropy.
This exponential overhead is fundamental, limited by the complexity of estimating an entropy~\cite{o2015quantum,li2018quantum_query_complexity}.
Furthermore, for quantum dynamics with projective measurements, there are additional multiplicative overhead scaling exponentially with $\sim p NT$, associated with the postselection of different measurement outcomes.
This is because the entanglement entropy needs to be evaluated for individual trajectory of an open system dynamics, and, in order to accumulate sufficient statistics for a single trajectory, experiments need to be repeated over at least $\sim q^{pNT}$ times.\footnote{For this conservative estimate, we assumed that different measurement outcomes are not correlated as expected in RUCs.}

The transition in the KL divergence introduced in Sec.~\ref{sec:KL_div} can partially alleviate the exponential overhead since its detection only requires a number of samples sufficient to discriminate two probability distributions without any postselections and/or entanglement measurements.
While precisely computing the KL divergence in general still requires exponentially many samples, we note that one can utilize other empirical methods such as Kolmogorov-Smirnov test~\cite{degroot2012probability} or evaluating an estimator for cross entropy~\cite{boixo2018characterizing}.
It remains open if our phase transition can be faithfully identified by any local observables, or by interferometric methods.

Another important consideration is the effect of imperfections in experiments.
Common sources of the imperfections include imprecise implementations of unitary evolutions, and dephasing or depolarization induced by uncontrolled noise.
In general, these types of errors can be formulated as a quantum channel.
Following the discussion in Sec.~\ref{sec:quantum} and Appendix~\ref{app:quantum}, it is straightforward to see that an uncontrolled quantum channel generally leads to an explicit breaking of the permutation symmetry in the spin model description and hinders the observation of a sharp phase transition.
More specifically, for a system of $N$ qudits evolved for $T$ time steps, the effects of explicit symmetry-breaking perturbations (experimental imperfections) become significant when $\eta NT \gtrsim \BigO(N)$, where $\eta$ is the effective strength of perturbations for a single qudit per time step obtained in the spin model description. 
Under this condition, the strength of symmetry-breaking perturbations arising from experimental imperfections exceeds that of boundary conditions discussed in Sec.~\ref{sec:sm}.

Finally, we note that the effects of experimental imperfections can be relatively more significant near the phase transition point. In particular, the susceptibility to symmetry-breaking perturbations is significantly enhanced near the phase transition in our spin model descriptions with $n=2$ or $n\rightarrow 1$.
This feature suggests a novel approach to characterize the amount of experimental imperfections by studying the phase transition or the lack thereof in quantum systems of finite sizes.

\subsection{Implications to the simulability of open system dynamics}
Quantum dynamics with area-law scaling entanglement entropy can be often simulated using classical computers.
In particular, many-body wave functions of $1$D systems can be efficiently represented by MPS.
Therefore, the entanglement phase transition described in this work may be also interpreted as a phase transition in the simulability of an open system dynamics.
Such interpretation, however, requires additional considerations.
For simulations of quantum dynamics, it is necessary that expectation values of observables can be approximated within a given accuracy using an MPS with the bond dimension that scales polynomially in the system size.
While this is guaranteed from area-law scaling R\'enyi-$k$ entropies with $k < 1$~\cite{verstraete2006matrix,schuch2008entropy}, it is not necessarily the case for quantum states with area-law scaling von~Neumann or R\'enyi entropies with $k \geq 1$.
This is because, under certain circumstances (see Ref.~\cite{schuch2008entropy} for examples), small Schmidt coefficients in the tail of an entanglement spectrum contribute significantly to evaluating an observable.
In many physical quantum states, however, the entanglement spectra often follow well-known distributions, such as the Boltzmann distribution~\cite{DeutschETH,SrednickiETH,rigol2008thermalization,kaufman2016quantum,MBL_review_Huse,MBL_review_Altman},
and an area-law scaling von~Neumann or R\'enyi-$k$ entropy with $k> 1$ already provides pragmatic criteria for classical simulability.
Our numerical simulation results in Sec.~\ref{sec:pt_numerics} suggest that the phase transition in von~Neumann entropy is indeed accompanied by transitions in R\'enyi-$k$ entropies with $k>0$.
More quantitative analysis of the entanglement spectra of quantum states resulting from chaotic dynamics and projective measurements would be an interesting future direction.

We emphasize that while the area-law scaling entanglement is a sufficient condition for efficient simulations, it is not necessary.
In particular, we have seen in Sec.~\ref{sec:quantum} that the phase transition (or its existence) sensitively depends on the type of information extracted by the measurements.
By choosing a different measurement basis (nonlocal or quasilocal), it may be possible to modify the effective phase transition point such that quantum states in typical trajectories exhibit area-law scaling of entanglement even for $p$ less than its critical value based on na\"ive local projections.
Thus, our result does not rule out the possibility that the volume-law phase can also be efficiently simulated by appropriately sampling different trajectories.

\subsection{Outlook}
Our work opens several new directions.
We proposed the KL divergence as a new measure to detect the entanglement phase transition. 
One intriguing future direction is to find local observables that faithfully and efficiently detect the phase transition.
In case such observables do not exist, designing an interferometric method to detect the phase transition or providing a fundamental complexity bound on the hardness of observing the transition would be interesting.

Another intriguing direction is to establish quantitative connections between quantum chaos and the entanglement phase transition.
The analyses in Sec.~\ref{sec:quantum} and Ref.~\cite{choi2019quantum} indicate that the stability of the volume-law entangling phase is intimately related to the effective quantum error corrections arising from information scrambling, which is a generic feature of chaotic many-body dynamics.
Quantifying the rate of information scrambling, for instance, by the critical probability or the rate of quasilocal projective measurements may provide new insight to quantum chaos from the perspective of information theory.
Indeed, it was previously demonstrated that noninteracting particles or nonscrambling dynamics of Bell pairs cannot exhibit stable volume-law entangled phases, i.e., $p_c = 0$~\cite{cao2018entanglement,chan2018weak}.
It remains open whether or not the phase transition can occur in nontrivial integrable systems such as Bethe ansatz solvable models or strongly disordered systems in many-body localized phases~\cite{sutherland2004beautiful,MBL_review_Huse,MBL_review_Altman}.
Alternatively, one may characterize quantum chaos from the Kolmogorov-Sinai entropy~\cite{alicki1994defining,cotler2018superdensity} of measurement outcomes, which has been widely used as a diagnostic of chaotic dynamics in classical settings.

Finally, our mapping technique in Sec.~\ref{sec:sm} may provide a promising framework to analyze recently proposed quantum supremacy test protocols~\cite{boixo2018characterizing,Supremacy_from_RUC}.
We expect that, using classical spin model descriptions, one can efficiently evaluate the average discrepancy between sampling distributions from an ideal RUC and from its experimental realization as a function of the degree of various imperfections.
Such analysis would provide an estimate to the maximum amount of imperfections that are tolerable to demonstrate quantum supremacy under reasonable conditions~\cite{Supremacy_from_RUC}, or allow characterizing near-term quantum devices~\cite{bernien2017probing,zhang2017observation,neill2018blueprint,harris2018phase,brydges2019probing}.

\vspace{3mm}
\noindent{\it Note --} Recently a related work appeared on the arXiv \cite{Jian2019} in parallel to ours. This work uses a different replica scheme to map quantum circuits with projective measurements to statistical mechanics models, which leads to results consistent with ours.

\section*{Acknowledgments}
We thank William Berdanier, Fernando Brand\~ao, Xiangyu Cao, Ignacio Cirac, Jordan Cotler, Michael Gullans, David Huse, Mikhail Lukin, Hannes Pichler, Frank Pollmann, John Preskill, and Shivaji Sondhi for helpful discussions.
We would like to thank especially Xiao-Liang Qi for discussions at the early stage of this work and Yi-Zhuang You for pointing out the connection between the percolation problem and the Potts model in the limit $n \rightarrow 1$.
We are also very grateful to Chao-Ming Jian, Yi-Zhuang You, Romain Vasseur, and Andreas Ludwig for pointing out errors in the previous version of this paper, namely the identification of symmetries in classical spin models with a finite $q$ and the evaluation of the critical measurement probability in the replica limit $n\to 1$. 
SC acknowledges support from the Miller Institute for Basic Research in Science.
EA acknowledges support from the ERC synergy grant UQUAM, the Gyorgy Chair in Physics at UC Berkeley, and the Department of Energy project DE-SC0019380 ``The Geometry and Flow of Quantum Information: From Quantum Gravity to Quantum Technology."
Parts of numeric simulations are performed using the Savio computational cluster resource provided by the Berkeley Research Computing program at the University of California, Berkeley  (supported by the UC Berkeley Chancellor, Vice Chancellor for Research, and Chief Information Officer).
\bibliography{refs}

\appendix
\numberwithin{equation}{section}
\renewcommand{\theequation}{\thesection\arabic{equation}}

\section{Details of the mapping to classical spin models}
\subsection{Derivation of $w_d^{(n)}$}
\label{app:weights}

\begin{figure}[t!]
	\centering
	\includegraphics[width=0.4\textwidth]{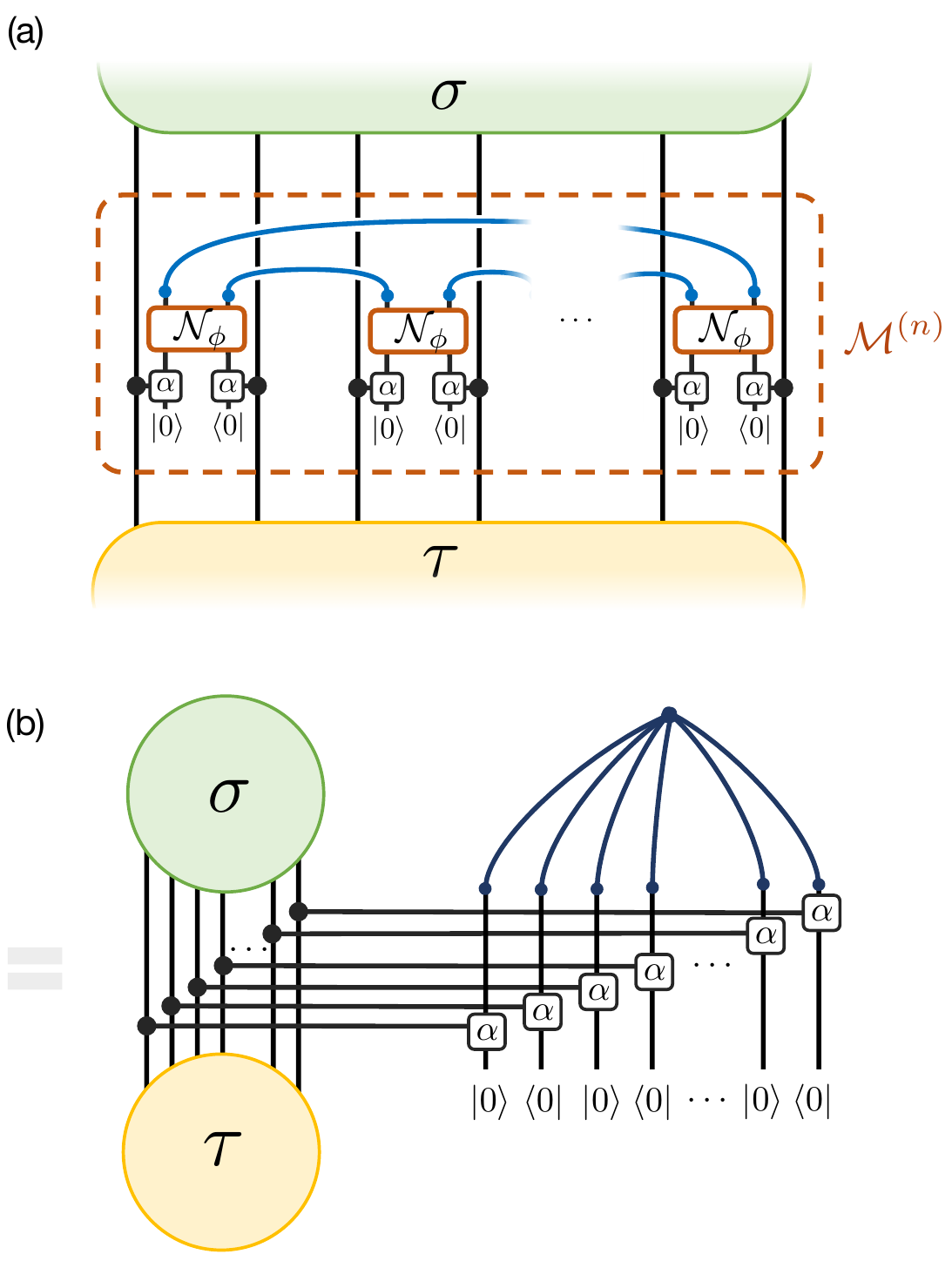}
	\caption{Tensor network representation of the diagonal weight $w_d^{(n)}$.
		(a) Every qudit is coupled to an ancilla qudit for a weak measurement.
		The ancilla qudits undergo the dephasing channel, $\mathcal{N}_\phi$, before contracted across different copies.
		(b) Rearrangement of the tensor network representation.
		Note that the diagrammatic representations of dephasing gates are simple in the computational basis of ancilla qudits.
	}
	\label{fig:two-body_weight_drv}
\end{figure}

As discussed in the main text, contracting a pair of diagonally neighboring $\hat{\sigma}$ and $\hat{\tau}$ tensors leads to a weight $w_d^{(n)}(\sigma,\tau)$ that depends on $q$ and $\alpha$.
Using the TN representation given in Fig.~\ref{fig:two-body_weight_drv}, a simple expression of $w_d^{(n)}$ can be written as
\begin{align}
    w_d^{(n)}(\sigma,\tau) = \sum_{\mathbf{aba'b'}}\hat{\sigma}_{\mathbf{ab}}\hat{\tau}_{\mathbf{a'b'}}\mathcal{M}_{\mathbf{ab},\mathbf{a'b'}}^{(n)},
\end{align}
where $\hat{\sigma}_{\mathbf{ab}}$ and $\hat{\tau}_{\mathbf{a'b'}}$ denotes the rank-$2n$ tensor in Eq.~\eqref{eqn:permutation_tensor} for a single qudit and $\mathcal{M}^{(n)}_{\mathbf{ab},\mathbf{a'b'}}$ denotes the tensor associated with the contraction of ancilla degrees of freedom.
We note that $\mathcal{M}^{(n)}_{\mathbf{ab},\mathbf{a'b'}}$ vanishes unless $\mathbf{a}=\mathbf{a}'$ and $\mathbf{b}=\mathbf{b}'$ due to our choice $\hat{R}_\alpha$, which does not have any off-diagonal element for system qudits in the computational basis [Fig.~\ref{fig:two-body_weight_drv}(b)].
Hence, we simplify our notation by using $\mathcal{M}^{(n)}_{\mathbf{ab}}$.
For given indices $\mathbf{ab}$, $\mathcal{M}^{(n)}_{\mathbf{ab}}$ takes the form
\begin{align}
    \mathcal{M}^{(n)}_{\mathbf{ab}} = \tr\left[ \prod_{k = 1}^n 
    \left(\mathcal{N}_\phi\left[ \rho^{(k)}_{a_kb_k}\right]\right) \right],
\end{align}
where $\rho^{(k)}_{a_kb_k}$ is the $k$-th copy of the density matrix of ancilla qudits defined as 
\begin{align}
    \rho^{(k)}_{a_kb_k} \equiv e^{-i \hat{X}_{a_k}\alpha}\ket{0}\bra{0} e^{i\hat{X}_{b_k}\alpha}.
\end{align}
We note that the subscript $a_k,b_k\in \{1,\dots q\}$ (c.f. not including $0$) are indices for system qudits and do not refer a matrix element, i.e., $\rho^{(k)}_{a_k,b_k}$ is a density matrix for an ancilla by itself.
The dephasing channel removes the off-diagonal elements of $\rho^{(k)}_{a_kb_k}$:
$\mathcal{N}_\phi[\rho^{(k)}_{a_kb_k}] = \cos^2\alpha \ket{0}\bra{0} + \sin^2\alpha\; \delta_{a_kb_k}\ket{a_k}\bra{a_k}$, leading to a simple expression
\begin{align}
     \mathcal{M}^{(n)}_{\mathbf{ab}} = \cos^{2n}\alpha + \sin^{2n}\alpha \prod_{k=1}^{n} \delta_{b_{k}a_{k+1}}\delta_{a_kb_k},
\end{align}
where $a_{n+1} \equiv a_1$, and the product of delta function is nonvanishing unless all $2n$ indices take the same value.
Using this expression for $\mathcal{M}^{(n)}_{\mathbf{ab}}$, we obtain 
\begin{align}
    &w_d^{(n)}(\sigma,\tau) \nonumber \\
    &= \sum_{\mathbf{ab}} \hat{\sigma}_{\mathbf{ab}} \hat{\tau}_{\mathbf{ab}} \left(\cos^{2n}\alpha + \sin^{2n}\alpha \prod_{k = 1}^n\delta_{a_kb_{k+1}}\delta_{a_kb_k} \right) \nonumber \\
    &= q^{\textrm{\#cycle}(\sigma\tau^{-1})}\cos^{2n}\alpha + q \sin^{2n}\alpha.
\end{align}
In the second equality, the tensor contraction of $\hat{\sigma}_{\mathbf{ab}}$ and $\hat{\tau}_{\mathbf{ab}}$ can be represented by a disconnected TN that consists of $\textrm{\#cycle}(\sigma\tau^{-1})$ loops. Each loop contributes a factor of $q$ to the summation and leads to $\sum_{\mathbf{ab}} \hat{\sigma}_{\mathbf{ab}} \hat{\tau}_{\mathbf{ab}} = q^{\textrm{\#cycle}(\sigma\tau^{-1})}$.
Hence, we derive $w_d^{(n)}$ in Eq.~\eqref{eq:weight_diag_n}. Equation~\eqref{eqn:w_d_2} can be obtained as a special case of $n = 2$. In the limit $\alpha = 0$, $w_d^{(n)} = q^{\textrm{\#cycle}(\sigma\tau^{-1})}$ reduces to the result in Ref.~\cite{nahum2018operator}.

\subsection{Derivation of \texorpdfstring{$\bar{w}^{(2)}$}{bar w^{(2)}}}
\label{app:three-body_weight}

In the case of $n = 2$, we have seen that the purity of a subsystem maps to the partition function of the classical Ising model on triangular lattice. Here, we derive and explicitly present the three-body weight $\bar{w}^{(2)}(\sigma_1,\sigma_2,\sigma_3)$ in Eq.~\eqref{eqn:Ising_spin_model} associated with three spin variables living on the vertices of a lower facing triangle, where $\sigma_1$, $\sigma_2$, and $\sigma_3$ refer to the spin variables located at the top left, top right, and bottom center of a triangle. 
The three-body weight $\bar{w}^{(2)}(\sigma_1,\sigma_2,\sigma_3)$ is given by
\begin{align}
    \bar{w}^{(2)}(\sigma_1,\sigma_2,\sigma_3) = \sum_{\tau=\pm1} w_d^{(2)}(\sigma_1,\tau)w_d^{(2)}(\sigma_2,\tau)w_g^{(2)}(\sigma_3,\tau).
\end{align}
Using $w_d^{(2)}$ in Eq.~\eqref{eqn:w_d_2} and $w_g^{(2)}$ in Eq.~\eqref{eqn:weingarten_coeff}, we can derive the full expression for $\bar{w}^{(2)}$. 
Considering that $\bar{w}^{(2)}$ is symmetric with respect to the exchange of $\sigma_1$ and $\sigma_2$, i.e., $\bar{w}^{(2)}(\sigma_1,\sigma_2,\sigma_3) = \bar{w}^{(2)}(\sigma_2,\sigma_1,\sigma_3)$, and that it has Ising symmetry, we only need to specify 
\begin{align}
    \bar{w}^{(2)}(\sigma,\sigma,\sigma) &= \frac{(q^2\cos^4\alpha + q\sin^4\alpha)^2 - (\cos^4\alpha + \sin^4\alpha)^2}{q^4-1}, \label{eqn:three-body_1}\\
    \bar{w}^{(2)}(\sigma,\sigma,\bar{\sigma}) &= \frac{(q\cos^4\alpha + q\sin^4\alpha)^2 - (q\cos^4\alpha + \sin^4\alpha)^2}{q^4-1}, \label{eqn:three-body_2}\\
    \bar{w}^{(2)}(\sigma,\bar{\sigma},\sigma) &= \frac{(q\cos^4\alpha + \sin^4\alpha)(\cos^4\alpha + \sin^4\alpha)}{q^2+1}, \label{eqn:three-body_3}
\end{align}
where $\bar{\sigma} = -\sigma$ represents the negation of $\sigma$.

We notice that, unlike in the absence of  measurements studied in Ref.~\cite{nahum2018operator}, $\bar{w}^{(2)}(\sigma,\sigma,\bar{\sigma}) \neq 0$ for $\alpha > 0$. We provide an explanation and discuss the implication of unitarity in Appendix~\ref{app:unitarity}.

\subsection{Derivation of Ising couplings $J_h$ and $J_d$}\label{app:Ising_coupling}
The three-body weight factorizes into pairwise contributions in the presence of Ising symmetry. Here, we derive the two-body Ising coupling $J_h$ and $J_d$ in Eq.~\eqref{eqn:Ising_Ham}. In terms of $J_h$ and $J_d$, $\bar{w}^{(2)}$ can be written as
\begin{align}
    \bar{w}^{(2)}(\sigma,\sigma,\sigma) &= Ce^{-2J_d-J_h},\\
    \bar{w}^{(2)}(\sigma,\bar{\sigma},\sigma) &= Ce^{J_h},\\
    \bar{w}^{(2)}(\sigma,\sigma,\bar{\sigma}) &= Ce^{-2J_d+J_h},
\end{align}
where $C$ is a constant prefactor. Using the expressions for $\bar{w}^{(2)}$, we have
\begin{align}
    J_d &= \frac{1}{4}\log \left( \frac{-u_2^2/q^2 + u_1^2}{u_2^2 - u_1^2/q^2} \right),\\
    J_h &= \frac{1}{4} \log \left( \frac{\left(u_1u_2(1-1/q^2)\right)^2}{(u_2^2 - u_1^2/q^2)(u_1^2 - u_2^2/q^2)} \right),
\end{align}
where
\begin{align}
    u_2 &= q^2\cos^4\alpha + q\sin^4\alpha,\\
    u_1 &= q\cos^4\alpha + q\sin^4\alpha.
\end{align}
Since $u_2 \geq u_1$, we can explicitly show that $J_d \leq 0$ is ferromagnetic, $J_h \geq 0$ is anti-ferromagnetic, and $J_d + J_h \leq 0$.

Using the explicit expressions for the Ising couplings, we can analyze the phase transition in the spin model. In the large $q$ limit, $J_d$ dominates over $J_h$ for any given $\alpha$ (i.e., $\abs{J_d} \gg \abs{J_h}$). The phase transition happens when $J_d \sim \BigO(1)$. For this reason, $\alpha_c$ is close to $\pi/2$, and it's reasonable to consider the physically interesting limit $q \gg 1$ and $\pi/2 - \alpha \ll 1$.

\subsection{A sufficient condition for nonnegative weights for $n\geq 3$}
\label{app:proof_validity}
In general, the negative weights in the expression of the $n$-th moment cannot be eliminated for arbitrary $q$ and $\alpha$ by simply integrating out $\tau$ variables.
Here, we derive a sufficient condition [Eq.~\eqref{eqn:validity}] for the weights being nonnegative, which allows the interpretation of the $n$-th moment as the partition function of a classical spin model.

After integrating out $\tau$ variables, the three-body weight in our classical spin model takes the form
\begin{align}
    &\bar{w}^{(n)}(\sigma_1,\sigma_2,\sigma_3) \nonumber \\
    &= \sum_{\tau\in\mathcal{P}_n} w_d^{(n)}(\sigma_1,\tau)w_d^{(n)}(\sigma_2,\tau)w_g^{(n)}(\sigma_3\tau^{-1};q^2).\label{eqn:three-body_w_n}
\end{align}
Here, we simplify the notation by using $w_g^{(n)}(\sigma_3\tau^{-1};q^2)$ for Weingarten function instead of  $w_g^{(n)}(\sigma_3,\tau;q^2)$ in the main text as it only depends on $\sigma_3\tau^{-1}$.
Also, we will often omit the dependence on $q^2$.
An exact formula for $w_g^{(n)}$ is known~\cite{collins2003moments,collins2006integration,zinn2010jucys,novak2010jucys,matsumoto2013weingarten,novaes2014elementary},
\begin{align}
    w_g^{(n)}(\sigma;d) = \frac{1}{n!}\sum_{\substack{\lambda \vdash n \\ \ell(\lambda)\leq d}}\frac{d_\lambda}{\mathcal{J}_\lambda^{(1)}(1^d)}\chi_\lambda(\sigma)
    \label{eq:weigngarten_full},
\end{align}
where $\lambda$ denotes a partition of $n$ elements and labels the irreducible representation (irrep) of the permutation group $\mathcal{P}_n$, $\ell(\lambda)$ is the length of the partition $\lambda$, $\chi_\lambda$ is the character of the irrep $\lambda$, $d_\lambda$ is the dimension of the irrep $\lambda$, and $\mathcal{J}_\lambda^{(1)}$ is the Jack Polynomial given by
\begin{align}
    \mathcal{J}_\lambda^{(1)}(1^d) = \prod_{j = 1}^{\ell(\lambda)}\frac{\Gamma(\lambda_j+d-j+1)}{\Gamma(d-j+1)}.
\end{align}
In this paper, we are generally interested in the large $q$ behavior in the model and consider the case $d = q^2 > n$. We can therefore ignore the criterion $\ell(\lambda) \leq d$ in the summation.

The three-body weight $\bar{w}^{(n)}$ can be negative owing to the existence of negative characters that appear in $w_g^{(n)}$. Here, we provide a lower bound of $\bar{w}^{(n)}$ and show that the lower bound becomes positive in a certain limit of $q$ and $\alpha$. 
The key idea is to realize that the leading order term in Eq.~\eqref{eqn:three-body_w_n} is positive and parametrically larger than the rest of the terms for a sufficiently large $q$ and fixed $\kappa$.
More specifically, the summation in $\bar{w}^{(n)}$ contains $n!$ terms, among which the term involving $w_g^{(n)}(\mathds{1};q^2) > 0$ provides the leading order contribution in the limit of a large $q$ and fixed $\kappa$.
To start with, we have a lower bound of the three-body weight:
\begin{align}
     \bar{w}^{(n)}  >& \left(u_1^{(n)}\right)^2 w_g^{(n)}(\mathds{1}) \nonumber \\
    &- (n!-1) \left(u_n^{(n)}\right)^2 \max_{\sigma \neq \mathds{1}}\Big( \abs{w_g^{(n)}(\sigma)}\Big),
    \label{eq:bound_w_n}
\end{align}
where $u_m^{(n)} \equiv q^m\cos^{2n}\alpha + q \sin^{2n}\alpha$.
We note that $u_m^{(n)} \geq u_1^{(n)}>0$.

Now, we provide a lower bound of $w_g^{(n)}(\mathds{1})$ and an upper bound of $|w_g^{(n)}(\sigma)|$ for $\sigma \neq \mathds{1}$.
The lower bound
\begin{align}
    w_g^{(n)}(\mathds{1}) > \frac{1}{n!}\sum_\lambda \frac{d_\lambda^2}{(q^2+n)^n} = \frac{1}{(q^2+n)^n} \label{eq:bound_wg1}
\end{align}
is obtained by considering the upper bound of the Jack Polynomial.
In order to obtain the upper bound of $|w_g^{(n)}(\sigma)|$ with $\sigma \neq \mathds{1}$, we first introduce $\mathcal{K}_\lambda = 1/\mathcal{J}^{(1)}_\lambda(1^d)$ and utilize the orthogonality relation between characters
$ \sum_{\lambda \vdash n} \chi_\lambda (\mathds{1}) \chi_\lambda (\sigma) = 0$. We note that $\chi_\lambda(\mathds{1}) = d_\lambda$. Multiplying both sides of this equation by $\mathcal{K}_{1^n}/n!$ and subtracting it from an expression for $w_g^{(n)}(\sigma)$, we obtain 
\begin{align}
    |w_g^{(n)}(\sigma)| &= \frac{1}{n!}\left| \sum_{\lambda \vdash n}(\mathcal{K}_\lambda - \mathcal{K}_{1^n})d_\lambda\chi_\lambda(\sigma)\right| \nonumber\\
    &\leq \frac{1}{n!} \sqrt{ \sum_{\lambda} (\mathcal{K}_\lambda - \mathcal{K}_{1^n})^2d_\lambda^2 \sum_{\lambda} \chi_\lambda(\sigma)^2} \nonumber\\
    &\leq \frac{1}{n!} \max\abs{\mathcal{K}_\lambda - \mathcal{K}_{1^n}} \sqrt{ \sum_{\lambda} d_\lambda^2 \sum_{\lambda} \chi_\lambda(\sigma)^2} \nonumber\\
    &< \max \abs{\mathcal{K}_\lambda -\mathcal{K}_{1^n}}. \label{eq:bound_wg2}
\end{align}
In the second line,  we used the Cauchy-Schwarz inequality, and in the the fourth line, we used the properties of irreps of a finite group $\sum_\lambda \chi_\lambda(\sigma)^2 < \sum_\lambda d_\lambda^2 = n!$.
$\max \abs{\mathcal{K}_\lambda -\mathcal{K}_{1^n}}$ has an upper bound
\begin{align}
    \max\abs{\mathcal{K}_\lambda -\mathcal{K}_{1^n}} < \frac{1}{(q^2-n)^n} - \frac{1}{(q^2+n)^n},\label{eq:bound_wg3}
\end{align}
when $q^2 > n$.
Combining these bounds in Eqs.~\eqref{eq:bound_wg1}, \eqref{eq:bound_wg2}, and \eqref{eq:bound_wg3} together and plugging them into Eq.~\eqref{eq:bound_w_n}, we obtain a lower bound
\begin{align}
    \bar{w}^{(n)}> \frac{u_1^2}{(q^2+n)^n} - n!  \left( \frac{u_n^2}{(q^2-n)^n} - \frac{u_n^2}{(q^2+n)^n} \right).
\end{align}

Finally, a sufficient condition of nonnegative  $\bar{w}^{(n)}(\sigma_1,\sigma_2,\sigma_3)$ for an arbitrary $ (\sigma_1,\sigma_2,\sigma_3)$ can be written as
\begin{equation}
    \frac{1}{n!}\frac{1}{(1+\kappa)^2} > \left( \frac{q^2+n}{q^2-n} \right)^n - 1.
\end{equation}
When $\kappa = q^{n-1}\cot^{2n}\alpha$ is fixed and $q\gg 1$, the leading order on the right-hand side is $2n^2/q^2$, and the inequality can be satisfied for a sufficiently large but finite $q$.
This completes the proof that there exists a finite region of $(q,\alpha)$ in which $\bar{w}^{(n)}$ is nonnegative and the $n$-th moment can be interpreted as the partition function of a classical spin model.

\section{Spin model for quantum relative entropy}\label{app:quantum}
We provide detailed derivations of the spin model for quantum relative entropy in Sec.~\ref{sec:quantum}.

\subsection{Derivation of two-body weight}
Here, we derive the two-body weight $v_d^{(n)}$ in the spin model description of the quantum relative entropy. Compared to the derivation of $w_d^{(n)}$ in Appendix~\ref{app:weights}, the only modification is the absence of the dephasing channel $\mathcal{N}_\phi$ applying to ancilla qudits.
The two-body weight is written as
\begin{align}
    v_d^{(n)}(\sigma,\tau) = \sum_{\mathbf{ab}} \hat{\sigma}_{\mathbf{ab}}\hat{\tau}_{\mathbf{ab}}\mathcal{M}^{(n)}_{\mathcal{Q}, \mathbf{ab}},
\end{align}
where $\mathcal{M}^{(n)}_{\mathcal{Q}, \mathbf{ab}}$ is the tensor associate with the tensor contraction of ancilla density matrices in the case without dephasing:
\begin{align}
    \mathcal{M}^{(n)}_{\mathcal{Q},\mathbf{ab}} = \tr\left(\prod_{k = 1}^{n} \rho_{a_kb_k}^{(k)} \right) = \prod_{k = 1}^{n} \left( \cos^2\alpha + \delta_{b_ka_{k+1}}\sin^2\alpha \right).\label{eqn:M_n}
\end{align}
An explicit expression for $v_d^{(n)}$ can be obtained using Eq.~\eqref{eqn:permutation_tensor} and the expression for $\mathcal{M}^{(n)}_{\mathcal{Q},\mathbf{ab}}$ above.
We should notice the two-body weight $v_d^{(n)}$ does not respect the permutation symmetry for any integer $n \geq 2$.
This is manifested in both the mathematical expression and diagrammatic representation of $\mathcal{M}^{(n)}_{\mathcal{Q},\mathbf{ab}}$.
Diagrammatically, without dephasing, $n$ copies of ancilla density matrices are contracted by $\mathcal{C}^{(n)}$ at the top illustrated in Fig.~\ref{fig:top_boundary_cond}(a), which explicitly breaks the symmetry $\mathcal{P}_n$ associated with  the permutation of $n$ replicated Hilbert spaces.
Mathematically, the permutation symmetry transformation maps $\sigma, \tau \mapsto \xi_1 \circ \sigma \circ \xi_2, \xi_1 \circ \tau \circ \xi_2$ for any pair $(\xi_1, \xi_2) \in \mathcal{P}_n \times \mathcal{P}_n$, and the invariance of $v_d^{(n)}$ requires $\mathcal{M}^{(n)}_{\mathcal{Q},\mathbf{ab}} =
\mathcal{M}^{(n)}_{\mathcal{Q},\xi_1(\mathbf{a})\xi_2^{-1}(\mathbf{b})}$, which is generally not satisfied.
Despite the absence of the symmetry, $v_d^{(n)}$ has interesting properties.
Particularly, $v_d^{(n)}$ has the maximally value $q^n$ when $\sigma = \tau = \mathcal{C}^{(n)}$. Furthermore, when either $\sigma$ or $\tau$ represents the cyclic permutation $\mathcal{C}^{(n)}$, $v_d^{(n)}$ becomes independent of $\alpha$ and reduces to $q^{\#\textrm{cycle}(\sigma\tau^{-1})}$, which is the result for the spin model description of a RUC without measurements~\cite{nahum2018operator,zhou2019emergent}.

Specifically, in the case of $n = 2$, $v_d^{(2)}$ takes the form
\begin{align}
    v_d^{(2)}(+, +) &= q^2\cos^2\alpha + q \sin^2\alpha,\\
    v_d^{(2)}(+, -) &= v_d^{(2)}(-, +) = q, \\
    v_d^{(2)}(-, -) &= q^2.
\end{align}

\subsection{Derivation of three-body weight}
The negative weights in the second moment due to negative Weingarten functions can be eliminated by integrating out $\tau$ variables. Hence, we obtain a Ising spin model on a triangular lattice with three-body interaction. The three-body weights $\bar{v}^{(2)}(\sigma_1,\sigma_2,\sigma_3)$ associated with three spins living on the vertices of lower-facing triangle can be explicitly written down. Considering the symmetry of exchange $\sigma_1$ and $\sigma_2$, i.e., $\bar{v}^{(2)}(\sigma_1,\sigma_2,\sigma_3) = \bar{v}^{(2)}(\sigma_2,\sigma_1,\sigma_3)$, all the independent $\bar{v}^{(2)}$ are written as
\begin{align}
\bar{v}^{(2)}(+,+,+) &= \frac{(q^2\cos^2\alpha + q\sin^2\alpha)^2 - 1}{q^4 - 1},\\
\bar{v}^{(2)}(+,+,-) &= \frac{q^2 - (q\cos^2\alpha + \sin^2\alpha)^2}{q^4 - 1},\\
\bar{v}^{(2)}(+,-,+) &= \frac{q^3\cos^2\alpha + q^2\sin^2\alpha-q}{q^4 - 1},\\
\bar{v}^{(2)}(+,-,-) &= \frac{q^3 - q\cos^2\alpha - \sin^2\alpha}{q^4 - 1},\\
\bar{v}^{(2)}(-,-,+) &= 0,\\
\bar{v}^{(2)}(-,-,-) &= 1.
\end{align}
The three-body weight $\bar{v}^{(2)}$ is nonnegative, which allows the interpretation of the second moment as the partition function of classical Ising model on a triangular lattice. Monte Carlo simulation in Fig.~\ref{fig:quantum_transient} is based on the Ising spin model derived here.
In this Ising spin model, we note that the spin variable with $\sigma = -1$ is preferred. 
More specifically, the average density of spins with $\sigma = -1$ is strictly increasing towards the bottom boundary (as $t$ decreases).
In the thermodynamic limit $T \rightarrow \infty$, the spin at the bottom layer will polarize to $\sigma = -1$, which implies $\langle m_1^-\rangle = 1$.

% Similar to the Ising spin model, in the generalized spin model with $n!$ internal states, the cyclic permutation $\mathcal{C}^{(n)}$ is preferred, and spins in the bottom layer are polarized to $\mathcal{C}^{(n)}$ in the thermodynamic limit, which leads to $\langle m_1^\downarrow\rangle = 1$.

\section{Implication of unitarity in classical spin models}
\label{app:unitarity}
The presence of weak measurements leads to a significant difference between weights from our mapping to spin models and those in Ref.~\cite{nahum2018operator}.
In what follows, we show that, in the absence of measurements, the three-body weight satisfies a special property that $\bar{w}^{(n)}(\sigma,\sigma,\sigma') = \delta_{\sigma\sigma'}$ when $q^2 \geq n$.
We note that this property is explicitly shown in Ref.~\cite{nahum2018operator} for a special case $n = 2$ without measurements.
This property is a consequence of the unitarity of the RUC and generally not satisfied in the presence of measurements.

The three-body weight $\bar{w}^{(n)}(\sigma,\sigma,\sigma')$ for three spins living on the vertices of lower-facing triangle can be written as
\begin{align}
    \bar{w}^{(n)}(\sigma,\sigma,\sigma') = \sum_{\tau\in\mathcal{P}_n} w_d^{(n)}(\sigma,\tau; q)^2 w_g^{(n)}(\tau,\sigma'; q^2),
\end{align}
where we put the explicit $q$ dependence in $w_d^{(n)}$ and $w_g^{(n)}$. In the case without measurements, $w_d^{(n)}(\sigma,\tau;q) = q^{\textrm{\#cycle} (\sigma\tau^{-1})}$, the three-body weight reduces to
\begin{align}
    \bar{w}^{(n)}(\sigma,\sigma,\sigma') = \sum_{\tau\in\mathcal{P}_n} \big(q^2\big)^{\textrm{\#cycle} (\sigma\tau^{-1})} w_g^{(n)}(\tau,\sigma'; q^2).
\end{align}
We notice that the Weingarten function $w_g^{(n)}(\sigma, \tau; d)$ can be interpreted as a matrix with two indices $\sigma$ and $\tau$ running over $n!$ elements of $\mathcal{P}_n$ when $d \geq n$. A simple expression is written as~\cite{roberts2017chaos}
\begin{align}
    w_g^{(n)}(\sigma, \tau; d) = \left(M^{-1}(d)\right)_{\sigma\tau}, \;\; (d\geq n)
\end{align}
where the matrix $(M(d))_{\sigma\tau} \equiv d^{\textrm{\#cycle}(\sigma\tau^{-1})}$.
Hence, the unitarity leads to the property of $\bar{w}^{(n)}$:
\begin{align}
    \bar{w}^{(n)}(\sigma,\sigma,\sigma') &= \sum_{\tau\in\mathcal{P}_n} \big(q^2\big)^{\textrm{\#cycle} (\sigma\tau^{-1})} \big(M^{-1}(q^2)\big)_{\tau\sigma'}, \nonumber \\
    &= \delta_{\sigma\sigma'}.
\end{align}
In our model, weak measurements explicitly break the unitarity, and this property does not hold.

\section{Duality in standard Potts models}\label{app:duality}
The phase transition point of the $2$D standard Potts model on square lattice given by Eq.~\eqref{eqn:standard_potts} is known to be exactly solvable using the Kramers-Wannier duality~\cite{potts1952some,kihara1954statistics}. Here, we briefly review the duality and derive the phase transition point.

In the low-temperature limit, $J < 0$ and $\abs{J} \gg 1$, we can perform the low temperature expansion of the partition function $\mathcal{Z} = \exp(-\beta H_{\textrm{Potts}})$:
\begin{align}
\mathcal{Z} &= Q e^{-2 N_{\textrm{sq}} J} \big[ 1 + e^{4J} N_{\textrm{sq}} (Q-1) + 2 e^{6J} N_{\textrm{sq}} (Q-1) \nonumber \\
&+ 2 e^{7J} N_{\textrm{sq}} (Q-1)(Q-2) + \BigO(e^{8J}) \big],
\end{align}
where $N_{\textrm{sq}}$ is the total number of sites on the square lattice, the Potts model contains $Q = n!$ local degrees of freedom.

In the high-temperature limit, $\abs{J} \ll 1$, we can perform the high temperature expansion of the partition function:
\begin{align}
\mathcal{Z} =& \sum_{\{\sigma_i\}} \prod_{\left<i,j\right>} \left(1 + \kappa\delta_{\sigma_i, \sigma_j}\right) = \sum_{\{\sigma_i\}} \prod_{\left<i,j\right>} \left(\frac{1 + r Q s(\sigma_i, \sigma_j)}{1-r}\right) \nonumber\\
=& \frac{Q^{N_{\textrm{sq}}}}{(1-r)^{2N_{\textrm{sq}}}}\Big[1 + r^4 N_{\textrm{sq}} (Q - 1) + 2 r^6 N_{\textrm{sq}} (Q - 1) \nonumber\\
&+ 2 r^7 N_{\textrm{sq}} {(Q - 1)(Q - 2)} + \BigO(r^8) \Big],
\end{align}
where we recall that $\kappa = e^{-J}-1$, the small parameter $r \equiv {\kappa}/({{Q} + \kappa})$ and
$s(\sigma_i, \sigma_j) \equiv \delta_{\sigma_i\sigma_j} - {1}/{Q}$.

A duality is observed in the low- and high-temperature expansion. This connects the partition function at high and low temperature through the relation $e^J = r$:
\begin{equation}
\frac{1}{\tilde{\kappa}+1} = \frac{\kappa}{{Q} + \kappa},
\end{equation}
where $\kappa$ and $\tilde{\kappa}$ parametrize the high- and low-temperature expansion.
The phase transition point, where the free energy exhibits a singular behavior, satisfies the self-dual relation $\tilde{\kappa}=\kappa$.
Hence, the transition point $\kappa_c$ is given by
\begin{equation}
\kappa_c = \sqrt{Q}.
\end{equation}
At $Q = 2$, the result reduces to the well-known Kramers-Wannier duality in $2$D classical Ising model on a square lattice.

\section{Monte-Carlo simulation of the transition in $\mathcal{F}^{(2)}$}\label{app:mc_distinguishability}
\begin{figure}[t!]
	\includegraphics[width=0.48\textwidth]{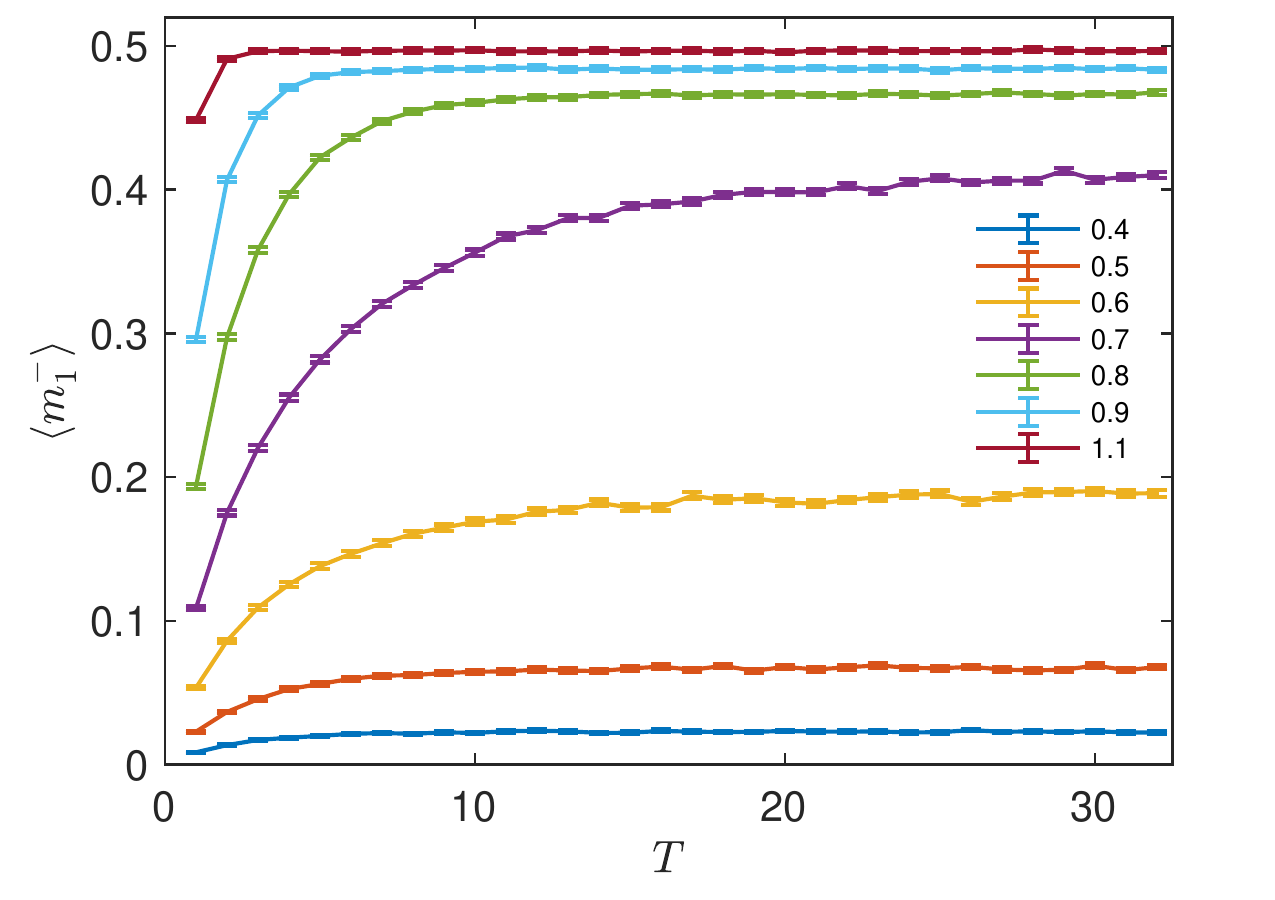}
	\caption{Density of spins with $\sigma = -1$ in the bottom layer $\langle m_1^-\rangle$ as a function of time $T$. Different curves represent different measurement strength $\alpha$ given on the right. The Monte-Carlo simulation is done with $q = 2$, $N = 32$. The data are averaged over $72000$ Monte Carlo samples.
	}
	\label{fig:classical_transient}
\end{figure}

We present the numeric simulation of the phase transition using the Monte Carlo algorithm. Here, we focus on the case of $n = 2$, where the second divergence and Fisher information have an Ising spin model description.
In the spin model description, the second Fisher information $\mathcal{F}^{(2)}$ is given by the density of $\sigma = -1$ spins at the bottom through the relation $\mathcal{F}^{(2)} = 2\langle m_1^-\rangle$.

In Fig.~\ref{fig:classical_transient}, we show $\langle m_1^- \rangle$ as a function of time for various measurement strengths $\alpha$ computed from Monte-Carlo simulations. 
$\langle m_1^- \rangle$ generally grows at early time, indicating that the measurement device gains more information about the initial state.
As $T$ is further increased, $\langle m_1^- \rangle$ eventually saturates to a value determined by $\alpha$ and $q$.
When $\alpha>\alpha_c^{(2)}$, $\langle m_1^- \rangle$ saturates to $1/2$, which implies the equal numbers of $\sigma =\pm1$ spins at the bottom, corresponding to the paramagnetic phase.
In contrast, when $\alpha<\alpha_c^{(2)}$, the saturation value of $\langle m_1^- \rangle$ is less than $1/2$, implying long range correlations between spin variables at top and bottom boundaries.
This phase corresponds to the ferromagnetic phase.

\begin{figure}[t!]
	\centering
	\includegraphics[width=0.48\textwidth]{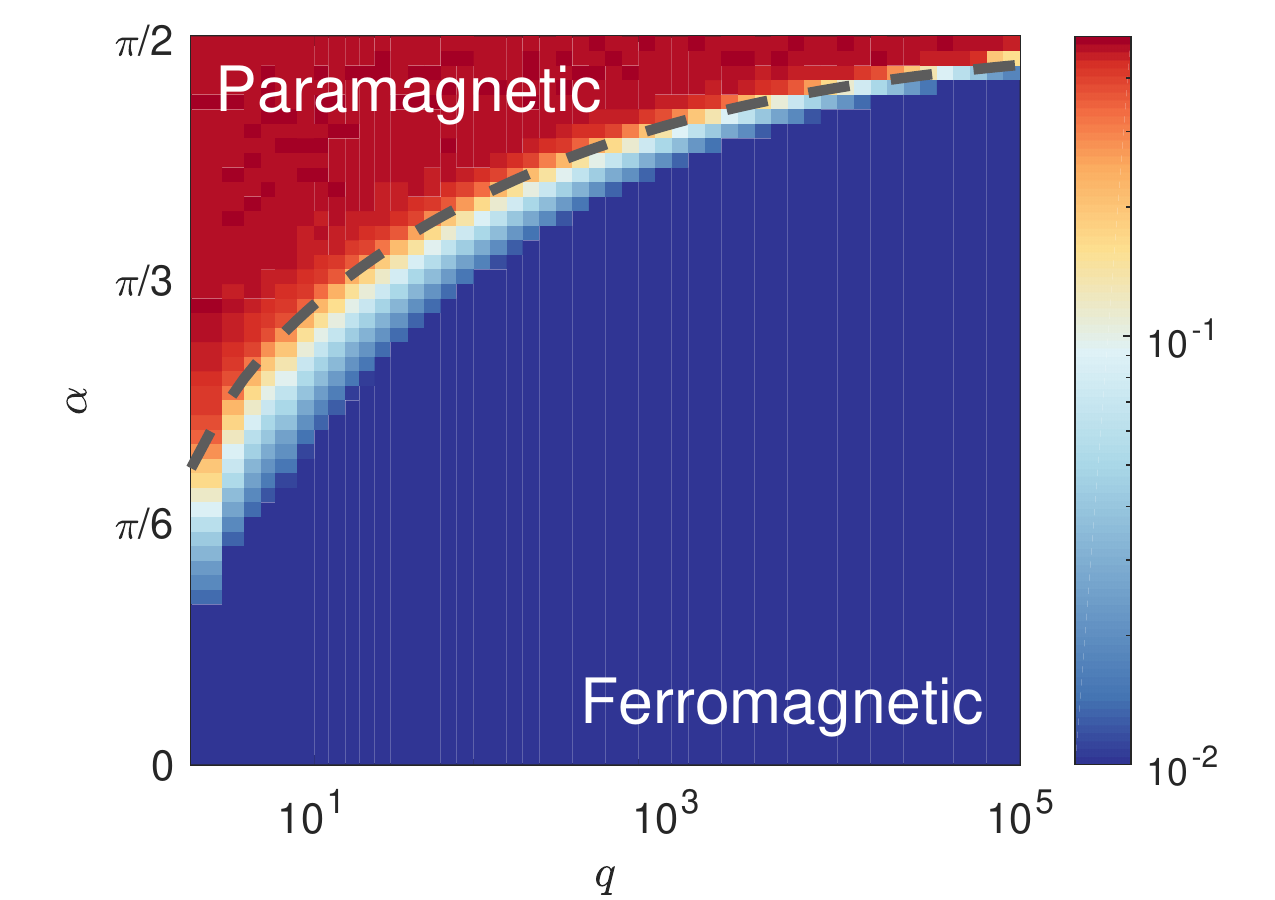}
	\caption{Phase diagram of the  transition in the second Fisher information $\mathcal{F}^{(2)}$, a proxy of the ability of the measurements to distinguish between two different initial states. Horizontal and vertical axes represent $q$ and $\alpha$, respectively. The gray line represents the exact solution of critical point in the classical Ising model in Eq.~\eqref{eq:alphac_2}. The color-coded background displays the density of spins with $\sigma = -1$ in the bottom layer $\langle m_1^-\rangle$. The simulation is done in a system of $N = 20, T = 40$. 
	}
	\label{fig:distinguishability_pd}
\end{figure}

Figure~\ref{fig:distinguishability_pd} shows the phase diagram of the transition in the second Fisher information $\mathcal{F}^{(2)}$ as a function of $\alpha$ and $q$.
Two distinct phases are seen in the color-coded background representing $\langle m_1^-\rangle$.
Large $\alpha$ and small $q$ correspond to the high-temperature limit in the classical Ising model and lead to the paramagnetic phase.
On the contrary, small $\alpha$ and large $q$ correspond to the low-temperature limit and lead to the ferromagnetic phase.
We find that the phase boundary between two phases matches with the exact solution of critical point in classical Ising model in Eq.~\eqref{eq:alphac_2} as expected.

\section{Exact Numeric simulation for the entanglement phase transition}
\subsection{Algorithm}\label{app:ED_numerics}
Here, we provide the details of the numerical simulations presented in Sec.~\ref{sec:pt}.
In order to efficiently store exact many-body wave functions for as large as $N=30$ qubits, we leverage the fact that a fraction of qudits are disentangled in every time step.
More specifically, we define a single discrete time step at even (odd) time $t$ as the applications of nearest neighboring two-qubit unitary gates at site $i$ and $i+1$ for every even (odd) $i$ and the projective measurement of each qubit in the computational basis with probability $p$.
The measurement outcomes are probabilistically determined according to the Born rules, and the wave function is normalized after each measurement.
At any given time $t$, on average $pN$ qubits are projected and remain disentangled from the rest of the system.
In our numerical algorithm, we only store the wave function for the entangled spins and separately keep track of the configuration of the disentangled ones.
In order to simulate the time evolution, we update the many-body wave function at every time step in the following way: (1) generate $N/2$ (or $N/2-1$) random two-qubit gates to apply and determine which qubits to measure projectively, and (2) sequentially update the wave function by applying a unitary gate at site $i$ and $i+1$ followed by the projection of the qubit(s) at $i$, $i+1$, or both, if necessary.
Crucially, our sequential update of the wave function and the sampling of the projective measurement outcomes are equilvalent to applying the entire unitary gates first and then performing projective measurements, since different unitary gates within a single time step have disjoint supports. 
The order of the sequential updates within a single time step can be further optimized in order to minimize the number of entangled spins during the time evolution.
Our simulation involves only even $N$ with open boundary conditions.
Half chain entanglement entropies are computed after time step $t = 3N$ (or $t=3N-1$) when $N/2$ is even (odd), in order to minimize the even/odd effect associated with the layout of our unitary circuits.

\subsection{Finite-size scaling}\label{app:ED_numerics_fss}
Here, we present the details of finite size scaling in Sec.~\ref{sec:pt_numerics}. We use the scaling ansatz proposed in Ref.~\cite{skinner2018measurement}:
\begin{align}
    S(p, L) - S(p_c, L) = g\left( (p-p_c) N^{1/\nu} \right).
\end{align}
The critical measurement probability $p_c$ and critical exponent $\nu$ are extracted by numerically optimizing the quality function of data collapse introduced in Refs.~\cite{kawashima1993critical,houdayer2004low}:
\begin{align}
    K(p_c, \nu) = \frac{1}{\mathcal{N}}\sum_{i,j}\frac{(y_{ij} - Y_{ij})^2}{dy_{ij}^2 + dY_{ij}^2},
\end{align}
where $i$ and $j$ run over the set of measurement probabilities and system sizes in the simulation, respectively.
$y_{ij}$ and $\rd y_{ij}$ are the data points $S(p_i, L_j) - S(p_c, L_j)$ and its error, $Y_{ij}$ and $\rd Y_{ij}$ are the estimated value of the master curve and its corresponding error.
The master curve is obtained by linear fit with weighted least squares using the data points nearby, and its error is obtained from the fit as described in Refs.~\cite{houdayer2004low,strutz2010data}.
Specifically, the optimization is done in the following way: (1) choose a $p_c \in [0.2, 0.35]$ and compute $S(p_c, L)$ by polynomial fitting of $S(p, N)$ using the seventh-th order polynomials; (2) find $\nu$ that optimize to the quality function for the given $p_c$, which gives the corresponding minimum $K_{\min}(p_c)$; (3) find the global minimum of $K_{\min}(p_c)$ for $p_c \in [0.2, 0.35]$ in order to extract the optimal $p_c$ and $\nu$.
The error bars of $p_c$ and $\nu$ are estimated using the bootstrapping method.
More specifically, out of $15$ measurement probabilities for $p \in [0.2, 0.35]$ in our simulation, we randomly choose $10$ data points and perform the aforementioned finite size scaling analysis to obtain $p_c$ and $\nu$. We repeat the analysis for $100$ times with independently randomly chosen data points and use the standard deviation of $p_c$ and $\nu$ as the estimated error bars.

\section{The Fisher information for nonlocal measurements}\label{app:Fisher_info}
Here, we review the connection between the KMB Fisher information and measurements in an optimal nonlocal basis.
The most general form of projective measurement can be characterized by the positive-operator valued measure (POVM) $\Pi$~\cite{nielsen2002quantum}, which consists of a complete set of observables $\{\Pi_i\}$ that need not be mutually orthogonal. For a given $\Pi$, the measurement outcome $i$ is drawn from a  probability distribution $p_i^\Pi \equiv \tr{\rho \Pi_i}$.
Hence, analogous to local projective measurements, we can compute the KL divergence
\begin{align}
    D_{\Pi}(\theta) \equiv 
    \sum_{i} p^{\Pi}_{0,i} \left( \log p^{\Pi}_{0,i} -\log p^{\Pi}_{\theta,i} \right),
\end{align}
where $p^\Pi_{\theta,i} = \tr[ \rho_{\theta}\Pi_{i} ]$ is the measurement probability distribution when the system is initialized in $\ket{\Psi_\theta}$.
We note that our original simple local measurements in the main text correspond to a special choice of $\Pi$ that consists of a set of $(q')^{NT}$ projectors in the computational basis of ancilla qudits. 
The amount of information carried in the measurement outcomes is quantified by the Fisher information
\begin{align}
    \mathcal{F}_{\Pi} = \partial_\theta^2 D_{\Pi}(\theta)\Big|_{\theta = 0}.
\end{align}

Most importantly, the KMB Fisher information $\mathcal{F}_{\textrm{KMB}}$ provides an attainable upper bound to $\mathcal{F}_\Pi$~\cite{uhlmann1977relative,hiai1991proper,ogawa1999strong,hayashi2001asymptotics,nagaoka2005fisher}:
\begin{align}
    \mathcal{F}_{\textrm{KMB}} = \max_{\Pi} \mathcal{F}_{\Pi},
\end{align}
where $\Pi$ is optimized over all possible POVM. In our work, we compute $\mathcal{F}_{\textrm{KMB}}$ averaged over unitary gates $\mathcal{U}$, and the optimal $\Pi$ for each $\mathcal{U}$ may not be the same.
\end{document}